\documentclass[11pt]{article}

\usepackage{graphicx, amsmath, amssymb,cancel,multirow,pdflscape,authblk,float,morefloats,hyperref}
\usepackage{cite}
\usepackage[small,normal,up]{caption2}

\newcommand{\brac}[2]{\ensuremath{\left< #1\left|#2 \right. \right>}}
\newcommand{\avg}[1]{\ensuremath{\left< #1 \right>}}

\newcommand{\h}{\hat}
\newcommand{\be}[0]{\begin{equation}}
\newcommand{\ee}[0]{\end{equation}}
\newcommand{\bea}[0]{\begin{eqnarray}}
\newcommand{\eea}[0]{\end{eqnarray}}

\def\hyph{-\penalty0\hskip0pt\relax}

\newcommand{\op}[1]{\hat{#1}}
\newcommand{\ket}[1]{\lvert #1\rangle}
\newcommand{\bra}[1]{\langle #1 \rvert}
\newcommand{\abs}[1]{\lvert #1 \rvert}

\newcommand{\braket}[2]{\langle #1 \vert #2 \rangle}

\hypersetup{
    colorlinks=true,
    citecolor=blue,
    linkcolor=black,
    urlcolor=blue
  }

\title{Quantum Imaging Technologies}
\author[1,2]{Mehul Malik\thanks{mehul.malik@univie.ac.at}}
\author[1,3]{Robert W.~Boyd}
\affil[1]{The Institute of Optics, University of Rochester,
Rochester, New York 14627 USA}
\affil[2]{Institute for Quantum Optics and Quantum Information (IQOQI), Austrian Academy of Sciences, Boltzmanngasse 3, A-1090, Austria}
\affil[3]{Department of Physics, University of Ottawa, Ottawa, ON K1N 6N5 Canada}

\begin{document}

\maketitle
\tableofcontents

\begin{abstract}
\normalsize
Over the past three decades, quantum mechanics has allowed the development of technologies that provide unconditionally secure communication. In parallel, the quantum nature of the transverse electromagnetic field has spawned the field of quantum imaging that encompasses technologies such as quantum lithography, quantum ghost imaging, and high-dimensional quantum key distribution (QKD). The emergence of such quantum technologies also highlights the need for the development of accurate and efficient methods of measuring and characterizing the elusive quantum state itself.

In this document, we describe new technologies that use the quantum properties of light for security. The first of these is a technique that extends the principles behind QKD to the field of imaging and optical ranging. By applying the polarization-based BB84 protocol to individual photons in an active imaging system, we obtained images that are secure against any intercept-resend jamming attacks. The second technology presented in this article is based on an extension of quantum ghost imaging, a technique that uses position-momentum entangled photons to create an image of an object without directly obtaining any spatial information from it. We used a holographic filtering technique to build a quantum ghost image identification system that uses a few pairs of photons to identify an object from a set of known objects.

The third technology addressed in this document is a high-dimensional QKD system that uses orbital-angular-momentum (OAM) modes of light for encoding. Moving to a high-dimensional state space in QKD allows one to impress more information on each photon, as well as introduce higher levels of security. We discuss the development of two OAM-QKD protocols based on the BB84 and Ekert protocols of QKD.  The fourth and final technology presented in this article is a relatively new technique called direct measurement that uses sequential weak and strong measurements to characterize a quantum state. We use this technique to characterize the quantum state of a photon with a dimensionality of $d=27$, and measure its rotation in the natural basis of OAM.

\end{abstract}

\section{Key Concepts}
\label{chap:Key}
\subsection{Introduction}

Here we introduce certain key concepts that are essential to understanding current research in quantum information and especially the experiments that are described in later sections. While the field of quantum mechanics (QM) is vast, there are certain ideas that underlie almost all quantum technologies today. For example, the quantum no-cloning theorem \cite{Wootters:1982ex} states that one cannot create a perfect copy of an arbitrary single quantum state. This seemingly simple theorem has led to applications such as quantum cryptography \cite{Bennett:1984wv}, which offers encryption with unconditional security --- a feat considered impossible with classical physics. Similarly, quantum entanglement, long considered one of the ``spookier" concepts in quantum mechanics, has allowed the development of quantum technologies such as quantum lithography \cite{Boto:2000eg,Nagata:2007jo} --- the ability to make measurements more sensitive and lithographic patterns finer than those allowed by classical physics. 

While it is important to understand the formal theory behind these concepts, it is perhaps more important to gain intuition for what really is going on in quantum mechanics. The chief difficulty in understanding or explaining QM is that of language. Modern day English (or any other language) is steeped in the language of classical mechanics. This makes sense of course, as we experience the world through our five senses, which are essentially classical detectors. It would be absurd to describe the iconic apple that supposedly fell on Newton's head in terms of the apple's wavefunction. How do we, then, go about trying to explain the highly counter-intuitive aspects of QM using our everyday Newtonian language?

In the beginning of book VII of Plato's \textit{The Republic} \cite{Plato:Republic}, Socrates describes a cave whose inhabitants are chained and forced to look upon a wall since they were born, not knowing anything else. Behind them, a fire burns and casts shadows of objects moving in front of it upon the cave wall. These people, having been forced to gaze only at the wall, can see just these shadows, and not the objects themselves or the fire. Plato uses this allegory to describe the nature of the philosopher as a person who has been freed from these chains and can see the true nature of reality. One could argue that the quantum mechanical world is like Plato's cave. Limited by our classical senses, we can only see the ``classical" shadows of the wavefunction. The quantum physicist then rises as the freed philosopher, empowered to see the ``true nature" of reality! 

\begin{figure}[b!]
\centering\includegraphics[scale=0.5]{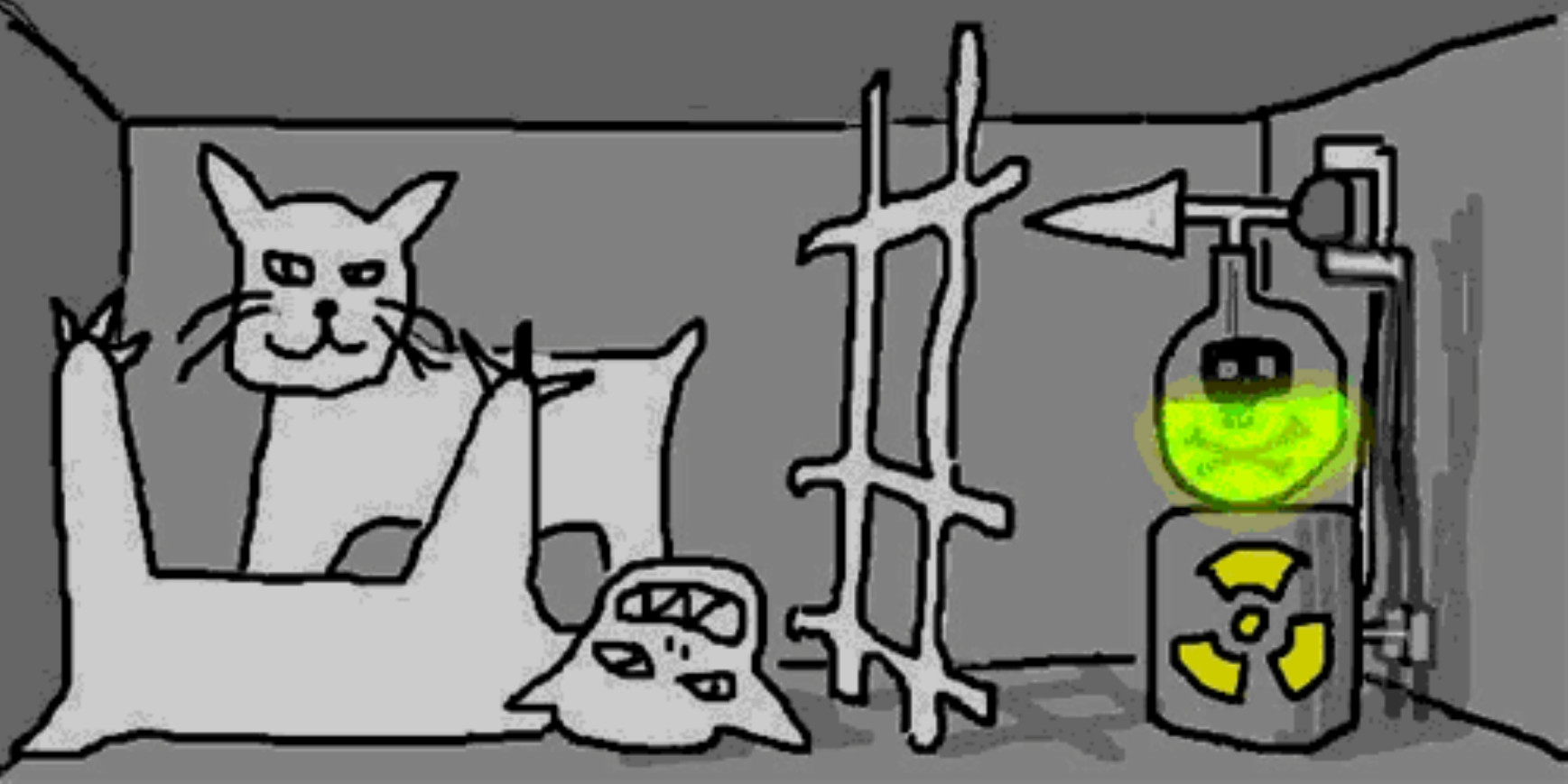}
  \caption{A cartoon depicting the ``Schr\"{o}dinger's cat" thought experiment \cite{cat_website}.}\label{schro_cat}
\end{figure}

The broad and fantastical imagery of Plato's cave allegory allows one to visualize the divide between the classical and quantum world using classical language. However, it is perhaps too broad to describe individual concepts in QM. In order to do that, one needs a metaphor for a quantum state itself. In 1935, Erwin Schr\"{o}dinger proposed a thought experiment that has come to be called ``Schr\"{o}dinger's cat." In it, Schr\"{o}dinger describes a cat that has been put in a box that contains a tiny amount of radioactive substance that has an equal probability of decaying and not decaying. If it decays, it sets off a geiger counter that triggers a hammer that breaks a vial of poison that kills the cat (see Fig.~\ref{schro_cat}). If the box is closed, the tiny amount of radioactive substance can be expressed for a certain instant in time as simultaneously being in a state of decay and not having decayed, putting the cat in a similar state of being alive and dead. On one hand, this thought experiment raises many deeper issues in QM such as the macroscopic limits of superposition and the role of the observer in collapsing the wavefunction. On the other hand, it is extremely useful for illustrating many fundamental concepts in QM using a simple, visual metaphor. Throughout this section, we will refer to Schr\"{o}dinger's cat whenever possible in an effort to provide some intuition for the topic at hand.

\subsection{Superposition and No-Cloning}

Schr\"{o}dinger's cat is an archetypal metaphor for quantum superposition. In the language of quantum mechanics, the state of the cat is written as

\be\ket{\textrm{cat}}=\frac{1}{\sqrt{2}}\big[\ket{\textrm{dead}}+\ket{\textrm{alive}}\big].\ee

\begin{figure}[t!]
\centering\includegraphics[scale=0.5]{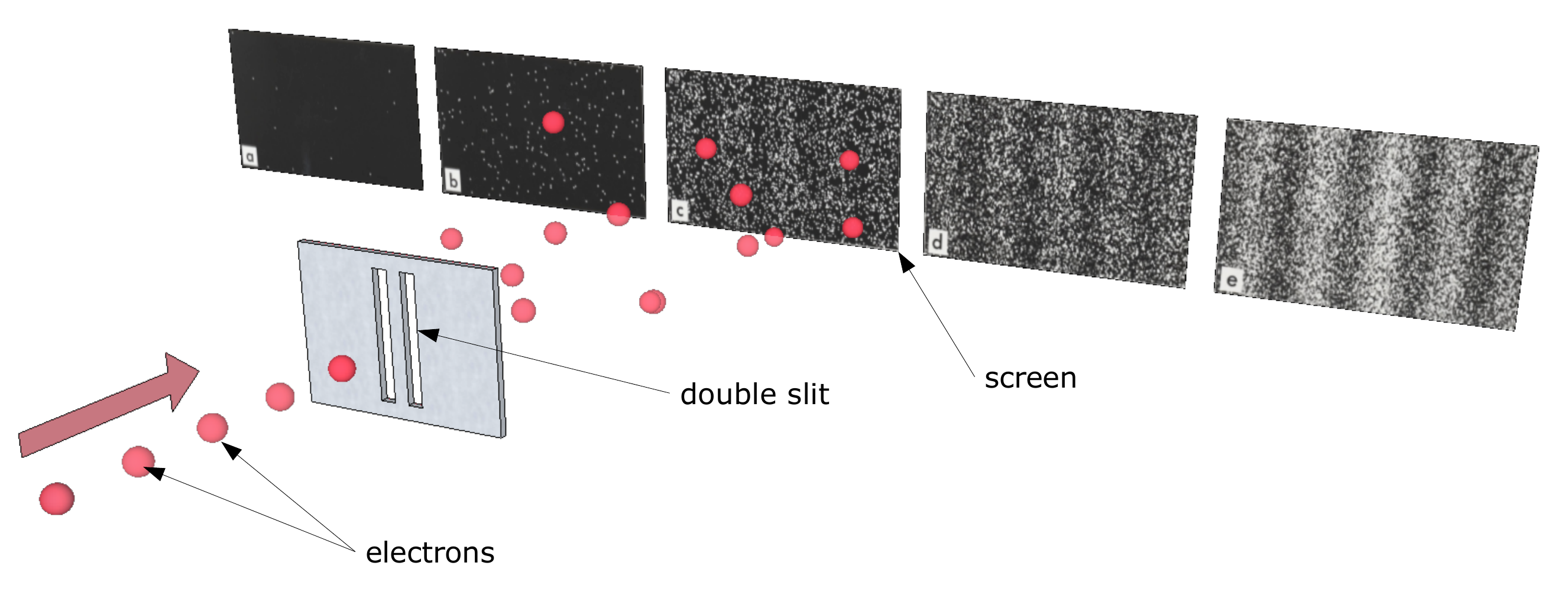}
  \caption{A schematic of Young's double slit experiment performed with electrons. Plates (a)-(e) show the buildup of an interference pattern at the single electron level. Thus, each electron only interferes with itself.}\label{doubleslit}
\end{figure}

The act of opening the box constitutes a measurement, which \textit{collapses} the wavefunction of the cat into one of the two states, dead or alive. Here we are using language associated with the Copenhagen interpretation of quantum mechanics, which describes reality in terms of probabilities associated with observations or measurements. While Schr\"{o}dinger's cat lies in the domain of \textit{gedankenexperiments}, realistic proposals have been made to put small living objects such as viruses into a quantum superposition \cite{RomeroIsart:2009iv}. Perhaps the simplest real world example of superposition is found in Young's famous double slit experiment when applied to particles. Many electrons are fired one at a time through a set of narrow slits. The resulting pattern measured on a screen on the other side of the slits shows distinct peaks and valleys (Fig.~\ref{doubleslit}). Care is taken to ensure that only one electron is present in the setup at any given time. How, then, can each electron independently know where to land in order to create an interference pattern usually associated with waves?

The answer lies in interpreting each electron as being in a probabilistic mixture of going through both slits at the same time. To paraphrase Dirac, ``each electron only interferes with itself." This experiment was the first to illustrate the principle of wave-particle duality, which states that all matter has a wavelength equal to $h/p$, where $h$ is Planck's constant and $p$ the momentum of any particle. Thus, one can see how particles with a very small mass (such as electrons) would have a measurable wavelength. The double slit experiment has been performed with molecules as large as Buckyballs (with a diameter of about 0.7 nm), steadily bringing the idea of quantum superposition into the macroscopic domain.

Current technology allows us to perform Young's double slit experiment with light as originally intended, but at the single photon level. As expected, one sees the familiar buildup of interference fringes on an electron-multiplying CCD camera, one photon at a time (as seen for electrons in Fig.~\ref{doubleslit}). Besides being in a superposition of two slits or positions, photons can be easily put into many other types of superpositions. For this reason, they form the building blocks for many proof-of-principle experiments in quantum optics and quantum information. Perhaps the simplest quantum superposition of a photon is that of polarization, where the state of the photon is written as

\be\ket{\psi}=\frac{1}{\sqrt{2}}\big[a\ket{H}+b\ket{V}\big],\label{super}\ee

\noindent where $\ket{H}$ and $\ket{V}$ refer to the horizontal and vertical quantum states of polarization. The coefficients $a$ and $b$ are, in general, complex coefficients of probability known as probability amplitudes. This type of superposition state is known as a \textit{qubit}, as it serves as a fundamental unit of quantum information. Just like a classical bit can have a value 0 or 1, a qubit can be in a superposition of 0 \textit{and} 1. The modulus-squares of the probability amplitudes $\abs{a}^2$ and $\abs{b}^2$ dictate the probability of finding the photon in either state $\ket{H}$ or $\ket{V}$. The relative phase between $a$ and $b$ governs the phase relationship between the $H$ and $V$ components, which can be interpreted as a measure of the ellipticity of polarization. A diagonally or anti-diagonally polarized photon can be written as superposition of horizontally and vertically polarized states as

\bea \ket{D} &=& \frac{1}{\sqrt{2}}\big[\ket{H}+\ket{V}\big]\nonumber\\
\ket{A} &=& \frac{1}{\sqrt{2}}\big[\ket{H}-\ket{V}\big].\label{DAHV}\eea

\noindent Thus, in this case, the coefficients $a$ and $b$ are unity, except for state $\ket{A}$, where the coefficient of $\ket{V}$ has a minus sign. The states $\ket{D}$ and $\ket{A}$ are ``mutually unbiased" with respect to the $\ket{H}$ and $\ket{V}$ states, as measuring one of them in the $H/V$ basis is equally likely to give an outcome of $H$ and $V$. Throughout this article, we will deal with superpositions of other properties of a photon, such as its position, momentum, and orbital angular momentum.

The no-cloning theorem, postulated by Wootters and Zurek in 1982 \cite{Wootters:1982ex}, states that one cannot create a perfect copy of an arbitrary quantum state. When applied to our metaphor of Schr\"{o}dinger's cat, this means that one cannot create a second ``cat superposition" that is identical to the first, without opening the box and destroying the first superposition. In their simple proof, Wootters and Zurek used an example of a device that perfectly clones a polarization qubit. In order to do so, they assume the device can independently clone a $\ket{H}$ photon as well as a $\ket{V}$ photon. However, when an arbitrary superposition state such as that shown in Eq.~\ref{super} is fed into this device, it creates a two-photon state that, in general, cannot be a replica of the original. Despite its simplicity, the ramifications of the quantum no-cloning theorem were huge. Just two years later, Bennett and Brassard applied this theorem to create the field of quantum cryptography \cite{Bennett:1984wv}. If one cannot perfectly clone a quantum state, then why not use it to securely send information? In this manner, two parties using quantum states for communication could detect any tampering, as an eavesdropper could not create copies of their communication states without introducing some error in the protocol.

\subsection{Entanglement}

\begin{figure}[t!]
\centering\includegraphics[scale=0.65]{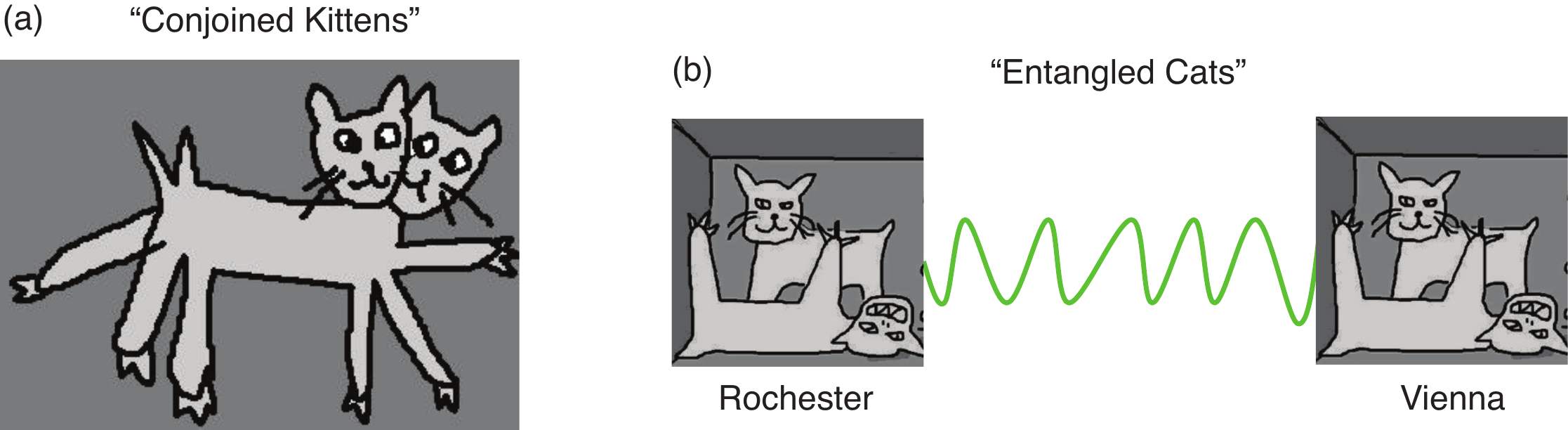}
  \caption{(a) A pair of conjoined kittens are separated by skilled veterinarians at birth and grow up to strangely feel each others pain (b) When put into identical Schr\"{o}dinger boxes, these cats constitute an entangled state. When one cat is measured to be alive, one knows immediately that the other cat is alive (and vice-versa), no matter how far apart they are.}\label{entcats}
\end{figure}

Since it was first proposed in a seminal paper by Schr\"{o}dinger in 1935 \cite{Schrodinger:1935ub}, the phenomenon of entanglement has captured the
imaginations of physicists and philosophers alike, and has made itself manifest as one of the most counterintuitive aspects of quantum mechanics. We can use a variation of the Schr\"{o}dinger's cat metaphor to illustrate the concept of entanglement. Imagine the freak occurrence of a birth of a pair of conjoined kittens (Fig.~\ref{entcats}(a)). These kittens are separated at birth by the finest veterinarians in the land. However, due to some unexplained unnatural phenomenon, the kittens grow up to share each others feelings and pain. Now, imagine putting both of these cats into identical Schr\"{o}dinger boxes. One box is kept in Rochester, while the other is sent to Vienna (Fig.~\ref{entcats}(b)). If the Rochester box is opened and the cat is found to be alive, we know instantly that the cat in the Vienna box is also alive (and vice versa)! The state of the cats can be written as

\be\ket{\textrm{cats}}=\frac{1}{\sqrt{2}}\big[\ket{\textrm{dead}}_R\ket{\textrm{dead}}_V+\ket{\textrm{alive}}_R\ket{\textrm{alive}}_V\big].\label{entcateq}\ee

\noindent This state indicates the cats share a highly correlated, or entangled state. By measuring the state of one of the cats, one non-locally collapses the state of the other cat. It is crucial to point out that this nonlocal relationship is not causal, i.e., one cannot kill the Vienna cat by shooting the Rochester cat. Only the different outcomes of measurement are perfectly correlated.

Of course, it is not realistic to imagine such a situation in real life. However, the phenomenon of entanglement is readily seen in the laboratory setting in the form of entangled photons. The strong correlations found in entangled photons have allowed great headway in experimental quantum mechanics, facilitating experiments ranging from the most fundamental to the very applied. Polarization-entangled photons have been used to obtain some of the most exacting experimental violations of Bell's inequality \cite{Bell:1964wu,Clauser:1969ff,Aspect:1982ja}. Time-energy entangled photons have found large application in various nonclassical techniques such as quantum cryptography \cite{Ekert:1991kl} and quantum teleportation \cite{Bouwmeester:1997wk}. The strong spatial correlations found in position-momentum entangled photons have given rise to the field of quantum imaging and have allowed the development of techniques such as quantum lithography \cite{Boto:2000eg,Boyd:2011eh} and ghost imaging \cite{Pittman:1995jb}.

In direct analogy with the above entangled Schr\"{o}dinger cats state, one can entangle a pair of photons in their polarization to create the state

\be\ket{\psi}=\frac{1}{\sqrt{2}}\bigg[\ket{H}_1\ket{H}_2+e^{i\phi}\ket{V}_1\ket{V}_2\bigg],\ee

\noindent where $\phi$ is the phase between the $H$ and $V$ states. If one were to measure the polarization of one these photons to be horizontal, one would know immediately that the polarization of the other is horizontal, and vice versa. Interestingly, when $\phi=0$ or some multiple of $2\pi$, this state can be written in the diagonal-anti-diagonal ($D/A$) basis by substituting Eqs. \ref{DAHV} into the above equation and simplifying to:

\be\ket{\psi}=\frac{1}{\sqrt{2}}\bigg[\ket{D}_1\ket{D}_2+\ket{A}_1\ket{A}_2\bigg].\ee

\noindent The correlations seen in $H$ and $V$ polarizations are perfectly preserved in the mutually unbiased basis of $D$ and $A$ polarization. This aspect of entanglement is crucial for distinguishing it from classically correlated states. Photons similarly entangled in position will retain these correlations when observed in the conjugate basis of momentum. However, in this case the correlations are opposite, i.e. position-entangled photons are anti-correlated in momentum. In the momentum representation, the state of position-momentum entangled photons is written as \cite{Monken:1998vy,Walborn:2007ix}

\be\ket{\psi}=\int\int d\bf{q}_1d\bf{q}_2\Phi(\bf{q}_1,\bf{q}_2)\ket{\bf{q}_1}_1\ket{\bf{q}_2}_2,\ee

\noindent where the vector $\bf{q}_i$ is the transverse component of the wave vector $\bf{k}_i$, and $\ket{q_i}$ represents the state of a single photon in momentum space (i.e. a plane wave mode). This definition uses the paraxial approximation $\abs{\bf{q}}<<\abs{\bf{k}}$ and the normalized function $\Phi(\bf{q}_1,\bf{q}_2)$ is defined as

\be\Phi(\bf{q}_1,\bf{q}_2)=\frac{1}{\pi}\sqrt{\frac{2L}{K}}v(\bf{q}_1+\bf{q}_2)\gamma(\bf{q}_1-\bf{q}_2).\ee

\noindent Here, $\gamma(\bf{q})$ is a phase-matching function and $\bf{v}(\bf{q})$ is the angular spectrum of the pump beam. For a pump beam with a narrow angular spectrum, this function is large only when the argument is zero, i.e. $\bf{q}_1=-\bf{q}_2$. This describes a state strongly anti-correlated in momentum. The same state can be written in position space, which we do later in Section \ref{chap:QGI}. 

Another property of photons that can be entangled is their orbital angular momentum (OAM). While the topic of OAM is discussed in detail in the next section, it is worth mentioning some key points here. Just as momentum entanglement manifests as momentum anti-correlations, OAM entangled photons also exhibit OAM anti-correlations. The state of these photons is written as

\be\ket{\psi} = \sum_\ell c_\ell \ket{\ell}\ket{-\ell}.\ee

\noindent As OAM exists in a discrete, infinite dimensional space, the entangled state is written in a simpler form as a sum over all possible $\ell$ modes. The range of $\ell$ modes in an OAM-entangled state usually depends on experimental considerations such as aperture sizes and the pump beam waist.

Entangled photons are readily produced via the process of spontaneous parametric down-conversion (SPDC). In this process, one photon conventionally known as the \textit{pump (p)} is annihilated in a second-order ($\chi^{(2)}$) nonlinear crystal and two photons are created, known as the \textit{signal (s)} and \textit{idler (i)} photons. This process is governed by the conservation of energy and momentum, and the frequencies and wave vectors of the photons involved are related as follows:

\bea \omega_p&=&\omega_s+\omega_i\nonumber\\
\bf{k}_p&=&\bf{k}_s+\bf{k}_i.\eea

\noindent Entanglement will play a strong role in Sections \ref{chap:QSS} and \ref{chap:QGI}, where we utilize it for techniques such as quantum-secured surveillance and quantum ghost image identification.

\subsection{Orbital Angular Momentum}

It is well known that light carries both spin and orbital angular momentum (OAM). The spin angular momentum of light is associated with its circular polarization. Such circularly polarized light was shown to exert a torque on a suspended wave plate by Beth in 1936 \cite{Beth:1936vh}. In the language of quantum mechanics, each circularly polarized photon carries a spin angular momentum of $\hbar$. Subsequently, Allen et al. extended this idea to OAM and showed that light also carries an angular momentum of $\ell\hbar$, where $\ell$ is the azimuthal mode index of the Laguerre-Gaussian mode solution to the paraxial wave equation \cite{Allen:1992by}. 

By making a paraxial approximation to the Helmholtz equation $(\Delta^2+k^2)E(k)=0$, we can write the paraxial wave equation

\be \bigg(\frac{\partial^2}{\partial x^2}+\frac{\partial^2}{\partial y^2}+2ik\frac{\partial}{\partial z}\bigg)E(x,y,z)=0.\ee

\noindent This equation is satisfied by the cylindrically symmetric Laguerre-Gaussian modes, whose normalized amplitude is described as

\bea LG_{\ell}^p(\rho,\theta,z)&=&\sqrt{\frac{2p!}{\pi(\abs{\ell}+p)!}}\frac{1}{w(z)}\bigg[\frac{\sqrt{2}\rho}{w(z)}\bigg]^{\abs{\ell}}L_p^{\ell}\bigg[\frac{2\rho^2}{w^2(z)}\bigg]\nonumber\\
&\times& \exp\bigg[-\frac{\rho^2}{w^2(z)}\bigg]\exp\bigg[-\frac{ik^2\rho^2z}{2(z^2+z_R^2)}\bigg]\nonumber\\
&\times& \exp\bigg[i(2p+\abs{\ell}+1)\tan^{-1}\bigg(\frac{z}{z_R}\bigg)\bigg]e^{i\ell\theta},
\eea

\noindent where $z_R$ is the Rayleigh range, $w(z)$ is the beam radius, $k$ is the wave vector magnitude, and $L^{\ell}_p$ is the associated Laguerre polynomial. The quantities $\ell$ and $p$ are the azimuthal and radial quantum numbers respectively. Allen et al. showed that each photon in such a Laguerre-Gaussian beam carries a well defined OAM of $\ell\hbar$ in vacuum.

\begin{figure}
\centering\includegraphics[scale=0.85]{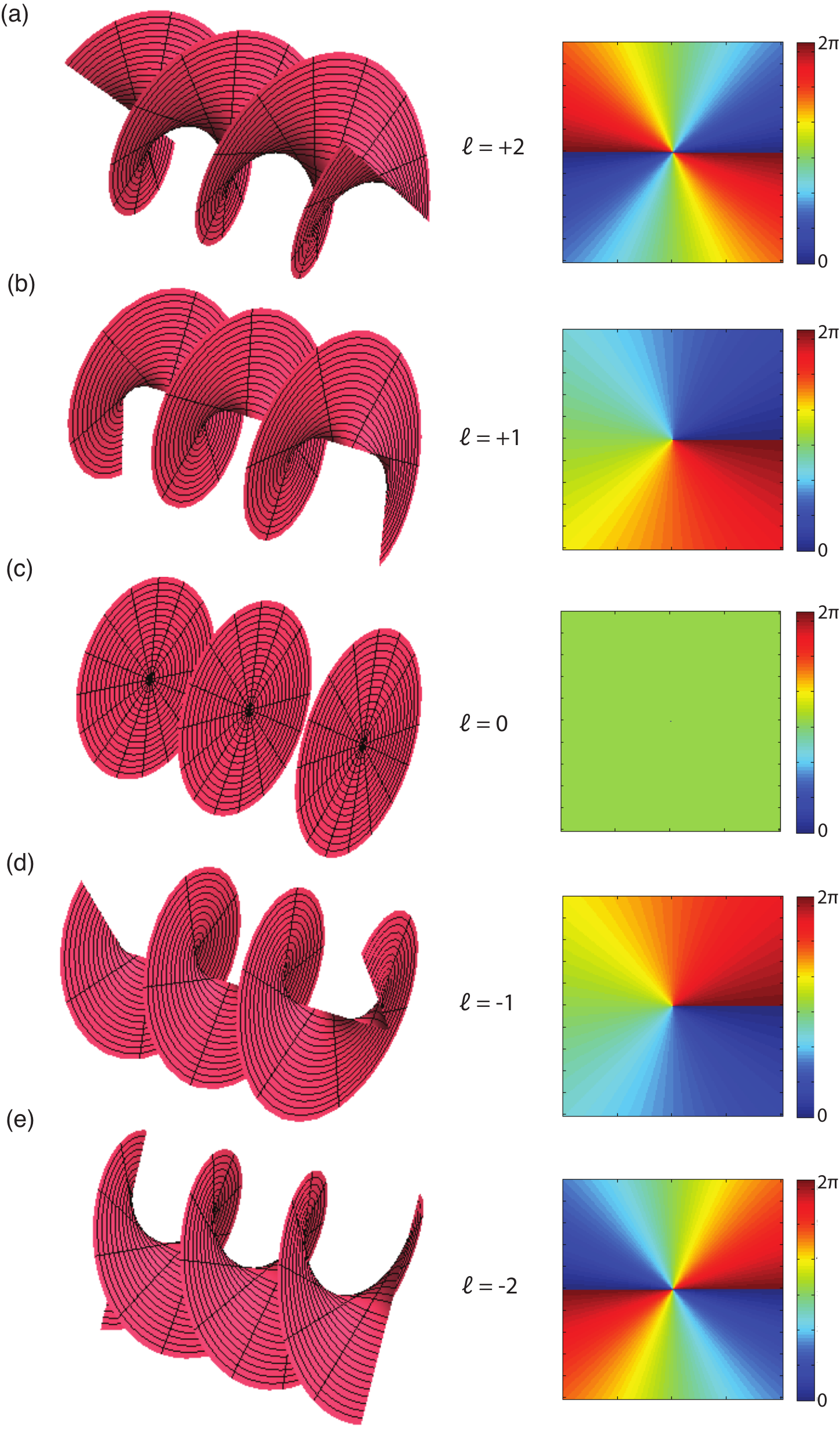}
  \caption{The wavefronts and transverse phase structure of five vortex modes with azimuthal quantum numbers (a) $\ell=+2$, (b) $\ell=+1$, (c) $\ell=0$, (d) $\ell=-1$, and (e) $\ell=-2$.}\label{ellmodes}
\end{figure}

\begin{figure}
\centering\includegraphics[scale=0.85]{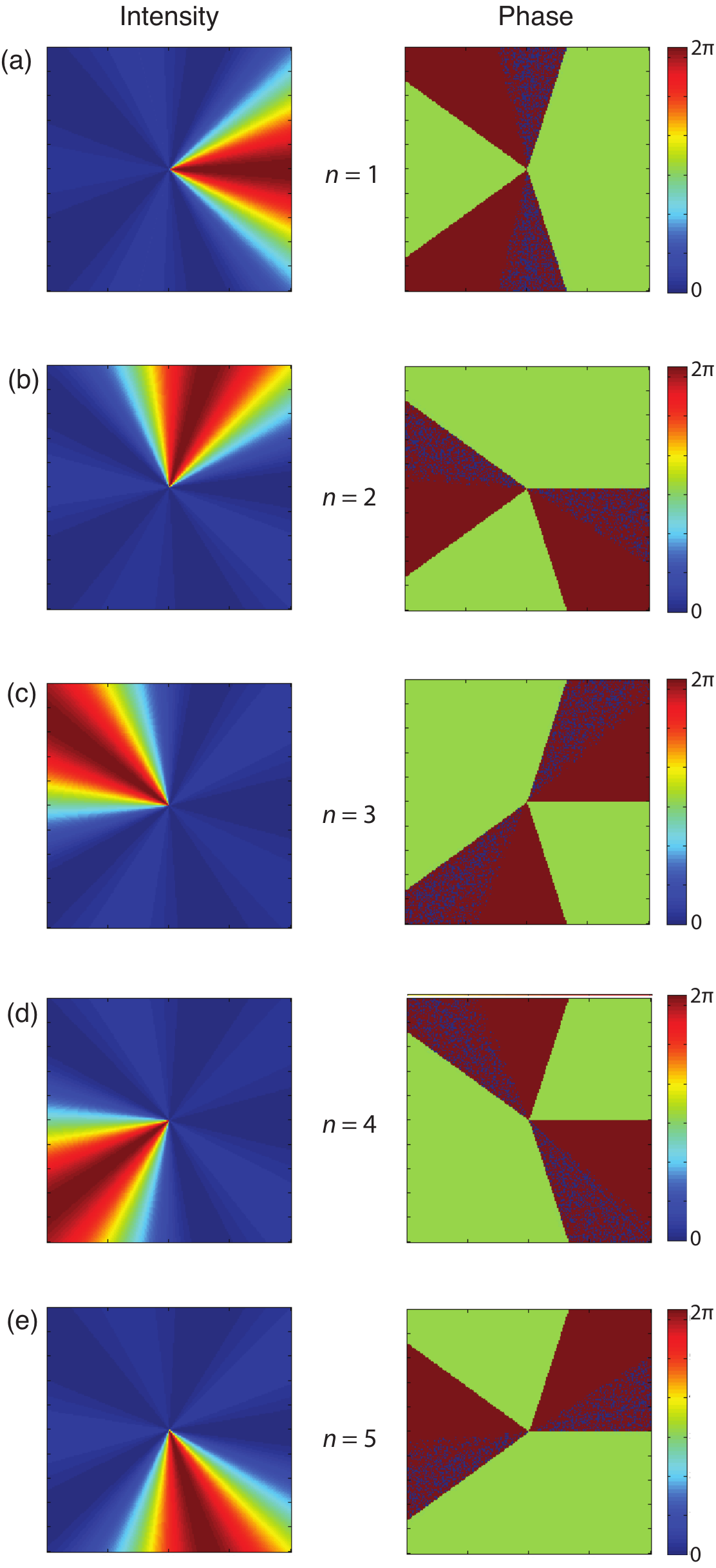}
  \caption{Simulated intensity and phase for the set of five angular position (ANG) modes composed of OAM modes with $\ell=0,\pm1,\pm2$.}\label{angmodes}
\end{figure}

The ramifications of the quantized nature of OAM modes are huge. Theoretically, OAM modes reside in a discrete, infinite dimensional Hilbert space. The dimensionality of this space is only limited by the physical size of apertures. The quantum nature of the spin angular momentum, or polarization, has allowed the use of photons as carriers of quantum information, or \textit{qubits}. The ability to use the OAM of photons for encoding quantum information opens up the potential to encode vastly more amounts of information per photon. Further, it also provides increased security in quantum information protocols. These key points are discussed further in Section \ref{chap:OAMQKD}. Photons carrying more than two bits of quantum information are referred to as \textit{qudits}, where $d$ refers to the dimensionality of the Hilbert space. 

In this article, we will focus on pure vortex modes, which are OAM modes with a spatially uniform amplitude. This allows for a simpler theoretical treatment, and lets one use the full aperture of the transmitter. Vortex modes can be written as

\be\Psi_{\ell}=A_0W(r/R)\exp{(i\ell\theta)},\ee

\noindent where $A_0$ is the spatially uniform field amplitude, $W(x)$ is an aperture function such that $W(x)=1$ for $\abs{x}\leq 1$ and zero otherwise, $r$ and $\theta$ are the radial and azimuthal coordinates, and $\ell$ is the azimuthal quantum number. The radial quantum number $p$ is zero in these modes. This allows us to isolate the azimuthal phase dependance for further study. The wavefronts of five vortex modes ($\ell=0,\pm1,\pm2$) are shown in Fig.~\ref{ellmodes}. The helical nature of the phase fronts is apparent in this figure, and shows why they are referred to as \textit{vortex} modes. Fig.~\ref{ellmodes} also shows X-Y cross-sections of the phase profiles. These cross-sections clearly show how the phase winds $\ell$ times from 0 to $2\pi$ in the azimuthal direction for a mode with azimuthal quantum number $\ell$. There is a phase singularity at the very center of these modes, which results in the intensity having a null at the center. This is the reason that these modes are sometimes referred to as ``donut" modes. Throughout this article, a reference to an ``OAM mode" implies a vortex mode as defined above.

As the set of OAM modes is complete and orthonormal, one can express any spatial mode with rotational symmetry in terms of its component OAM modes. Another set of orthonormal modes can be formed by taking a finite number of OAM modes $N=2L+1$ and adding them coherently according to the relationship 

\be \Theta_n=\frac{1}{\sqrt{2L+1}}\sum_{\ell=-L}^L\Psi_{\ell}\exp{\bigg(\frac{i2\pi n\ell}{2L+1}\bigg)}.\label{ANGeq}\ee

This second set modes is called the angular position, or ANG basis. These modes are so named because their intensity profile looks like an angular slice that moves around the center of the beam as one changes the relative phases of the component OAM modes. It is important to note that the number of ANG modes in the basis will be equal to the number of OAM modes used to form the basis. As an example, we show simulated intensity and phase for the set of five ANG modes composed of the OAM modes with $\ell = 0,\pm1,\pm2$ in Fig.~\ref{angmodes}. These modes are given by the formula

\be \Theta_n=\frac{1}{\sqrt{5}}\sum_{\ell=-2}^2\Psi_{\ell}\exp{\bigg(\frac{i2\pi n\ell}{5}\bigg)},\ee

\noindent which is simply a special case of Eq.~\eqref{ANGeq}. As the set of ANG modes is composed of an equal superposition of OAM modes, they form a basis that is mutually unbiased (i.e.~a MUB) with respect to the OAM basis. This point is important for application in quantum key distribution and will be discussed in the next section.

\subsection{Weak Values}
\label{sec:WV}

Weak values were first introduced in a seminal paper by Aharanov, Albert, and Vaidman in 1988 \cite{Aharonov:1988fk}. In general, weak values are complex numbers that one can assign to the powers of a quantum observable operator $\op{A}$ using two states: an initial, preparation state $\ket{i}$, and a final, post-selection state $\ket{f}$.  The $n$\textsuperscript{th} order weak value of $\op{A}$ has the form
\begin{align}\label{eq:awv}
  A^n_w &= \frac{\bra{f}\op{A}^n\ket{i}}{\braket{f}{i}}
\end{align}
where the order $n$ corresponds to the power of $\op{A}$ that appears in the expression. 

Weak values have long been considered an abstract concept. However, since their introduction 25 years ago, they have gradually transitioned from a theoretical curiosity to a practical laboratory tool. In a recent review, we show how these peculiar complex expressions appear naturally in laboratory measurements \cite{Dressel:2014ks}. In order to do so, we derive them in terms of measurable detection probabilities. 

In a standard prepare-and-measure experiment, the probability of detecting an event is given by $P=\abs{\braket{f}{i}}^2$ where $\ket{i}$ corresponds to the initial and $\ket{f}$ to the final state. If we introduce an intermediate unitary interaction $\op{U}(\epsilon)=\exp(-i\epsilon \hat A)$ that modifies the initial state, the detection probability also changes to $P_\epsilon=\abs{\braket{f}{i'}}^2=\abs{\bra{f}\op{U}(\epsilon)\ket{i}}^2$. If $\epsilon$ is small enough, we can consider $\op{U}(\epsilon)$ to be ``weak.'' In this case, the operator $\op{U}$ can be expanded in a Taylor series. The detection probability above can then be written as (shown here to first order):

\begin{align}\label{eq:pepsilongen}
  P_\epsilon &= |\bra{f}\op{U}(\epsilon)\ket{i}|^2 = |\bra{f}(1-i\epsilon\hat A+\dots)\ket{i}|^2\nonumber\\
  &=P  + 2 \epsilon\, \text{Im}\braket{i}{f}\bra{f}\op{A}\ket{i} + O(\epsilon^2).
\end{align}
Assuming $\ket{i}$ and $\ket{f}$ are not orthogonal (i.e. $P\neq0$), we can divide both sides of the previous equation by $P$ to obtain:
\begin{align}\label{eq:singlewv}
  \frac{P_\epsilon}{P} &= 1 + 2 \epsilon\, \text{Im}A_w - \epsilon^2\left[ \text{Re}A^2_w - |A_w|^2\right] + O(\epsilon^3),
\end{align}

\begin{figure}[b!]
\centering\includegraphics[scale=.55]{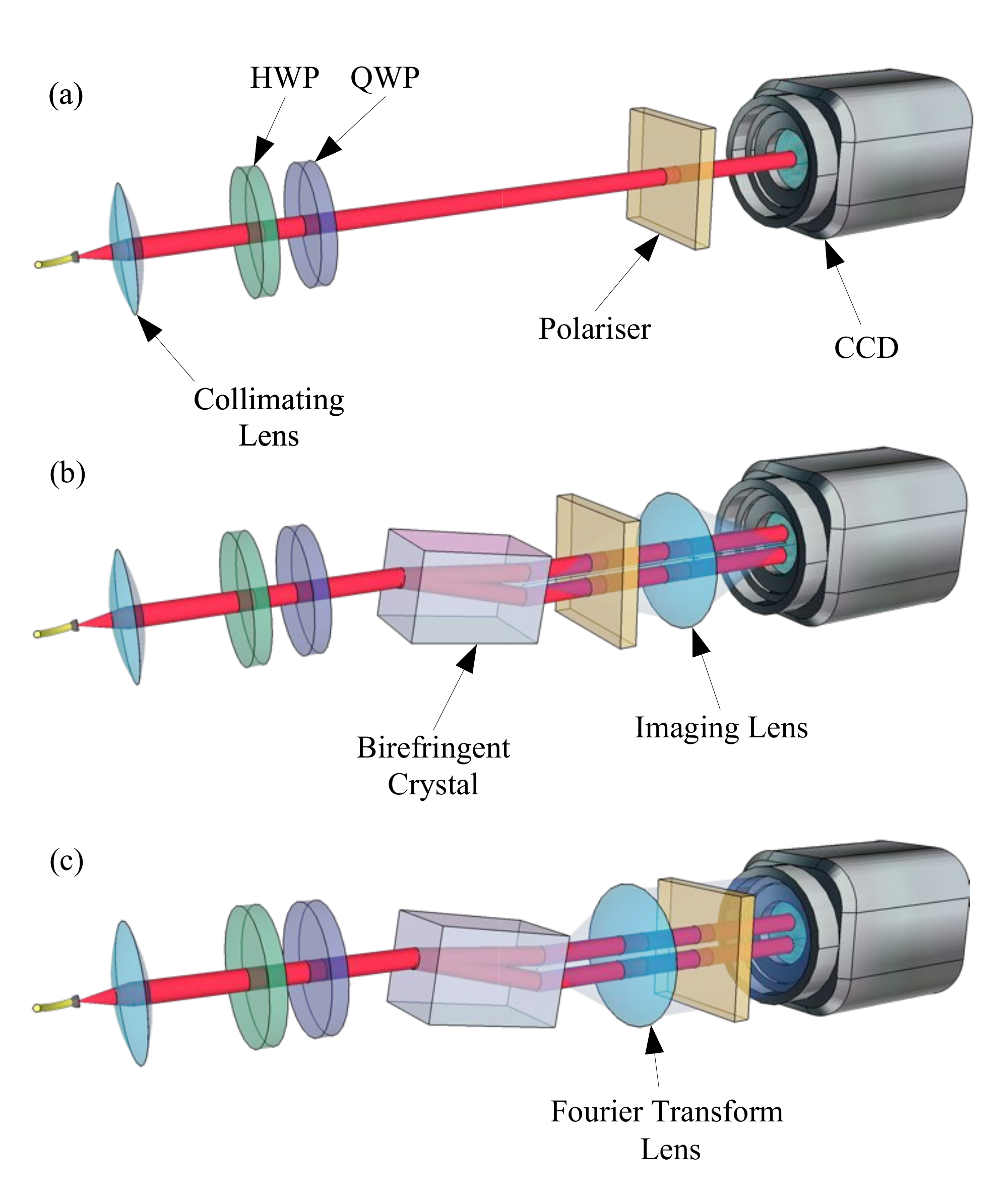}
\caption{An experiment illustrating how weak values are measured by using a coupled system of two observables. (a) A collimated Gaussian beam from a single mode fiber (SMF) is prepared in an initial polarization state by a quarter-wave plate (QWP) and half-wave plate (HWP). A polarizer is used for post-selection into a final polarization state. A CCD is used for measuring the position-dependent beam intensity. (b) A birefringent crystal inserted between the wave plates and polarizer displaces the beams by a small amount. A lens is used for imaging the output face of the crystal onto the CCD in order to measure the real part of the polarization weak value. (c) A lens is used for imaging the far-field of the crystal face onto the CCD in order to measure the imaginary part of the polarization weak value (Figure redrawn from Ref. \cite{Dressel:2014ks}).}\label{crystal}
\end{figure}

\noindent where $A_w$ is the first order weak value and $A^2_w$ is the second order weak value as defined above in Eq.~\eqref{eq:awv}. Here, we arrive at our operational definition: weak values characterize the relative correction to a detection probability $\abs{\braket{f}{i}}^2$ due to a small intermediate perturbation $\op{U}(\epsilon)$ that results in a modified detection probability $\abs{\bra{f}\op{U}(\epsilon)\ket{i}}^2$. When the higher order terms in the expansion given in Eq.~\eqref{eq:singlewv} can be neglected, one has a linear relationship between the probability correction and the first order weak value, which we call the \emph{weak interaction regime}. The conditions under which the higher order terms cannot be neglected are discussed in detail in Ref. \cite{Dressel:2014ks}.

In general, weak values are complex quantities. In order to measure a weak value, one has to measure both its real and imaginary parts. In most laboratory measurements of the weak value, one uses a coupled system of observables in order to do so. Such a system is composed of two parts---a ``system" observable, whose weak value we are interested in measuring, and a ``pointer" observable, which provides us with information about the system observable. For example, in the experiment of Ritchie et al. \cite{Ritchie:1991gh}, the authors use a coupled system of photon polarization and position. This system is illustrated in Fig. \ref{crystal} by an experimental setup where the polarization of a photon is weakly coupled to its position by a thin birefringent crystal. In this setup, the polarization state is prepared by a half-wave and quarter-wave plate, and post-selected by a polarizer. The position state is prepared in a gaussian mode by collimating light from a fiber, and post-selected by either imaging or Fourier-transforming the mode onto a CCD detector. The real and imaginary parts of the polarization weak value are obtained by making appropriate post-selections on the photon position or momentum, which result in the real or imaginary parts of the polarization weak value being isolated. 

Such a system is described by a symmetric combination of the real and imaginary parts of the weak values of polarization ($S_w$) and momentum ($p_w$):
\begin{align}\label{eq:spwv}
  \frac{P_{\epsilon}}{P} - 1 &\approx 
   \frac{2\epsilon}{\hbar}\left[\text{Re}S_w\text{Im}p_w + \text{Im}S_w\text{Re}p_w\right].
\end{align}
\noindent Here, the real and imaginary parts of the polarization weak value are isolated by using two experimental configurations. In one configuration, a post-selection of the photon position is performed such that the momentum weak value is purely imaginary. This corresponds to imaging the crystal face onto the detector, as shown in Fig. \ref{crystal}(b). In this case, the real part of the momentum weak value in Eq. \eqref{eq:spwv} goes to zero, which effectively isolates the real part of the polarization weak value. In the second configuration, a post-selection of the photon momentum is performed, which results in the momentum weak value being purely real. This corresponds to looking at the far-field of the crystal face with a Fourier-transform lens, as shown in Fig. \ref{crystal}(c). In this case, the imaginary part of the momentum weak value in Eq. \eqref{eq:spwv} goes to zero, which effectively isolates the imaginary part of the polarization weak value. This example is analyzed in more detail in Ref.  \cite{Dressel:2014ks}.

In this manner, any system of two coupled observables can be used to measure the weak value of one of the observables. For example, by making appropriate post-selections on the polarization degree of freedom, one could isolate and measure the real and imaginary parts of the momentum weak value. This is indeed the technique used in the direct measurement method that is explained in section \ref{DM}.


\section{Quantum Technologies Today}
\label{chap:QTT}
\subsection{Introduction}

The Heisenberg uncertainty principle lies at the heart of most modern quantum technologies. The ultimate limits of measurement precision are set by this principle and have been reached through the use of quantum resources such as entanglement and squeezed light \cite{Giovannetti:2004jg,Boto:2000eg,Dutton:2010fj}. The uncertainty principle also bounds the probability of simultaneously measuring two complementary observables, such as position and momentum. By extending this idea to discrete properties of a photon such as its polarization, the field of quantum secure communication was developed \cite{Bennett:1984wv,Ekert:1991kl}. Quantum concepts such as superposition and entanglement have expanded the fields of information theory and computing to remarkable frontiers \cite{Steane:1999fn,Bennett:2000kl}. Clearly, quantum mechanics has had a profound impact on modern technology. However, most of these technologies still live in the domain of proof-of-principle experiments on the lab bench. Given the rate of technological progress today, it won't be long before we see technologies such as practical quantum computing and long-distance quantum communication become a reality. In this section, we  briefly describe three quantum technologies that form the backbone of this article---quantum key distribution, quantum ghost imaging, and direct measurement. 

\subsection{Quantum Key Distribution}

Quantum key distribution (QKD) was first proposed by Bennett and Brassard in 1984 as a method by which two parties, Alice and Bob, could share a random string of bits with one another with unconditional security \cite{Bennett:1984wv,Gisin:2002zz}. A third party, Eve, with the intention of eavesdropping on Alice and Bob's communication channel, would be unable to do so without introducing a certain amount of statistical error in the channel. This is best illustrated by the use of a simple example. Let's say Alice wants to convey a bit value of 1 or 0 to Bob. She will then choose at least two mutually unbiased bases in which to represent this bit value. For polarization, two such bases are the horizontal-vertical ($H/V$) basis and the diagonal-anti-diagonal ($D/A$) basis. These are known as ``mutually-unbiased bases'' (MUBs) because one can express each state in one basis as an equally weighted sum of the states in the other. We can invert Eqs. \ref{DAHV} from Section \ref{chap:Key} to write states in the $H/V$ basis as a sum of states in the $D/A$ basis:

\bea \ket{V} &=& \frac{1}{\sqrt{2}}\big[\ket{D}-\ket{A}\big]\nonumber\\
\ket{H} &=& \frac{1}{\sqrt{2}}\big[\ket{D}+\ket{A}\big].\eea

Let's say Alice picks the $H/V$ basis for encoding. Then, state $\ket{H}$ corresponds to 0 and state $\ket{V}$ corresponds to 1. Alice sends a bit value of 0 encoded as an $\ket{H}$ state. If Eve intercepts the communication channel and measures this state in the correct $H/V$ basis, she will obtain the correct bit value. However, if she measures the state in the incorrect $D/A$ basis, she will have an error half the time. This is because when measuring an $\ket{H}$ state in the $D/A$ basis, Eve has an equal probability of measuring a $\ket{D}$ or an $\ket{A}$ state, which in turn correspond to 0 or 1. This results in an error of 50\% when Eve measures in the wrong basis. Combined with the 50\% chance of Eve picking the wrong basis leads to a total error probability of 25\%.

For the protocol to work, Alice randomly picks between the $H/V$ and $D/A$ bases for encoding. She then transmits these states to Bob. Bob measures these states in the same manner as Eve, by also randomly picking between the $H/V$ and $D/A$ bases. Just like Eve, Bob will also get an error with 25\% probability. To remove these errors, Alice shares her basis choices through a public channel \textit{after} Bob has made all his measurements. Bob then discards all the measurements that he made where his basis choice did not match Alice's. This procedure is known as ``sifting.''

\begin{figure}[t!]
\centering\includegraphics[scale=.7]{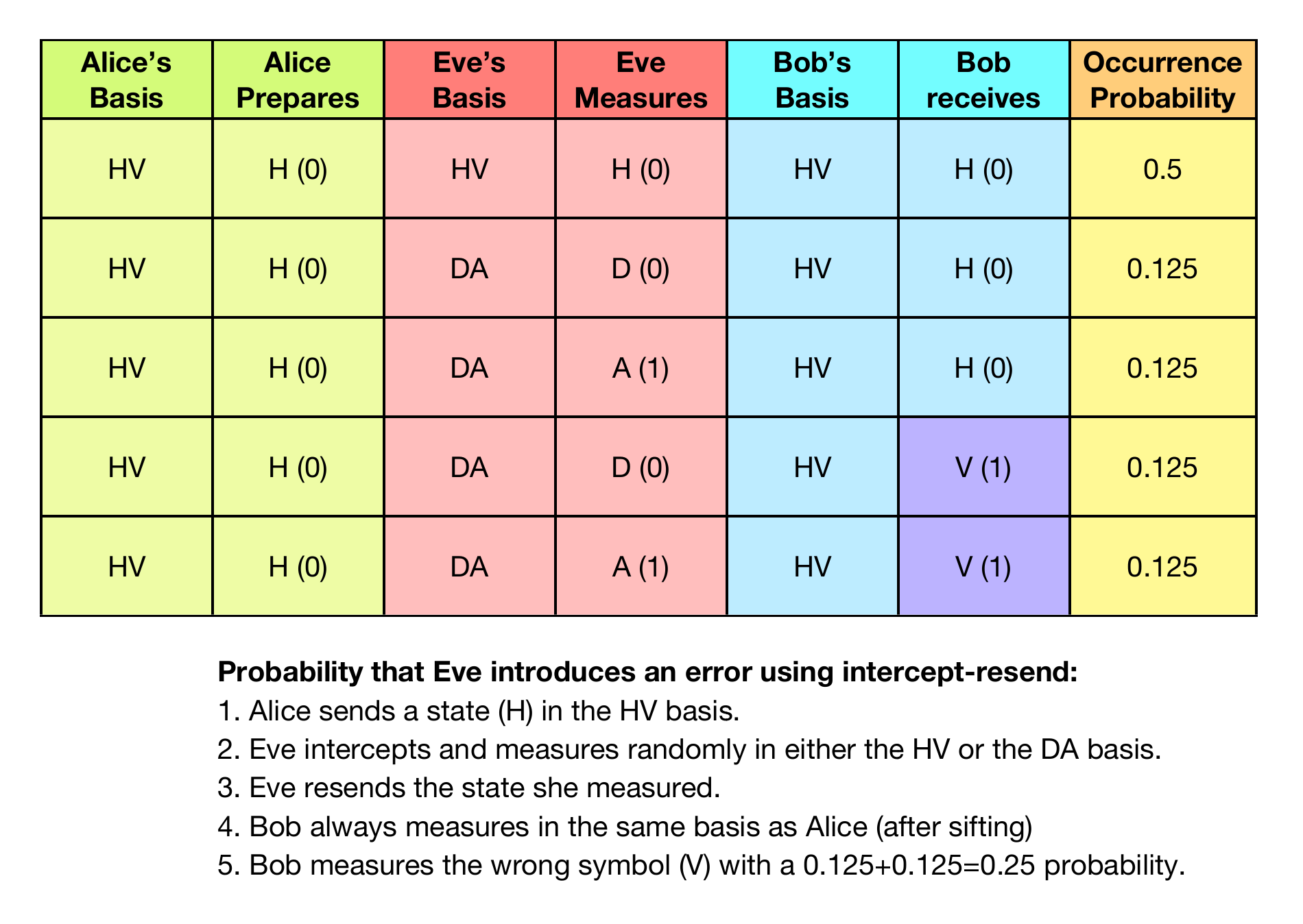}
  \caption{A table showing the different possible outcomes in an intercept-resend eavesdropping attack when Alice and Bob use a 4 state, 2 MUB protocol. As shown, Eve introduces a 25\% error between Alice and Bob after the sifting procedure (i.e. Alice and Bob discard all mismatched basis measurements).}\label{QKDTable}
\end{figure}

After the sifting procedure, Alice and Bob ideally share an error-free string of bits. Now let's reinsert Eve, who intercepts and resends all of Alice's states to Bob. Like Bob, Eve also measures these states by randomly picking between the $H/V$ and $D/A$ bases. Again, like Bob, Eve gets an error rate of 25\%. She then resends her measured states to Bob in the basis she measured them in. By doing so, she introduces errors than cannot be removed by the sifting process. For example, after sifting, Alice and Bob both have a bit value measured in the $H/V$ basis. If Eve also measured and resent that bit in the $H/V$ basis, she would have introduced no error. However, if she measured and resent that bit in the $D/A$ basis, she would have introduced an error half the time, leading to a total error rate of 25\%. 

Thus, Alice and Bob can determine if Eve had intercepted and resent their states by sacrificing a small part of the key and checking the error rate. If they obtain an error rate less than 25\%, they can assume their protocol was secure. If they obtain an error rate greater than or equal to 25\%, they assume an eavesdropper was present and abandon the protocol. In this manner, Alice and Bob can generate a secure key using the method of QKD. 

The intercept-resend attack explained above is illustrated by means of a table in Fig.~\ref{QKDTable}. All possible outcomes (post-sifting) are shown for the case when Alice picks an $\ket{H}$ photon to encode a 0. The cases where Bob has an error, i.e. he registers a $\ket{V}$ photon, are shown as purple cells. These occur with a probability of $0.125+0.125=0.25$. It is clear from this table that an intercept-resend eavesdropping attack by Eve will introduce an error of 25\% between Alice and Bob.

\begin{figure}[t!]
\centering\includegraphics[scale=.75]{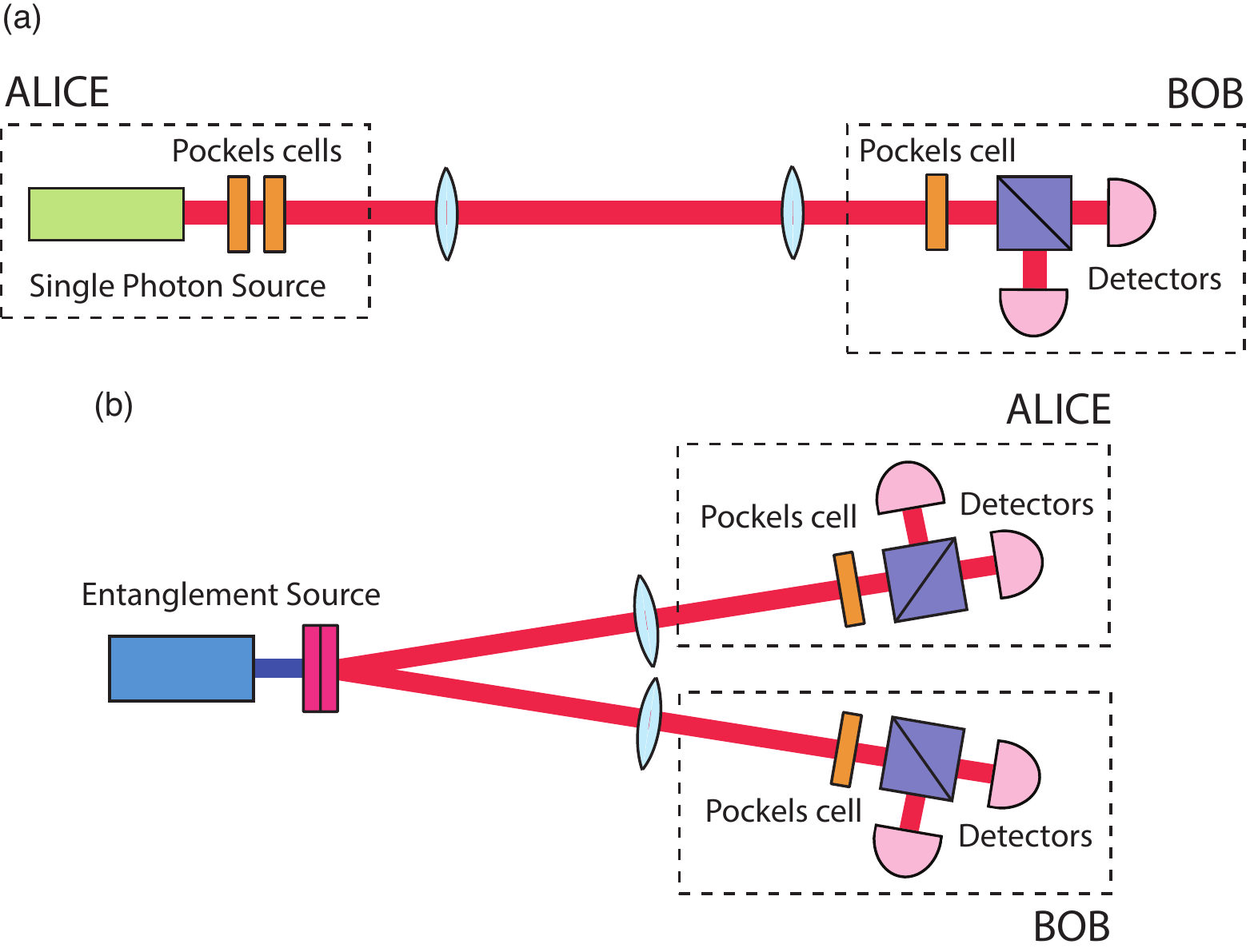}
  \caption{Schematics for the (a) prepare and measure BB84 QKD protocol and the (b) entanglement-based Ekert QKD protocol.}\label{EkertvBB84}
\end{figure}

The two main protocols used to perform QKD are known as the BB84 protocol \cite{Bennett:1984wv} and the Ekert protocol \cite{Ekert:1991kl}, named after their founders Charles Bennett, Gilles Brassard, and Artur Ekert. The procedure described above is known as the BB84 protocol and relies on the impossibility of cloning single photons for security \cite{Wootters:1982ex}. Recent work has cleverly extended this protocol for use with weak coherent pulses which are susceptible to a photon number splitting eavesdropping attack. In this extension known as the decoy state protocol, Alice randomly modulates the mean photon number of her pulses and later shares this information with Bob \cite{Lo:2005cd}. The decoy state protocol is discussed further in Section  \ref{chap:OAMQKD}. The second main QKD protocol invented by Artur Ekert relies on the quantum correlations in entanglement for security. Alice and Bob initially share correlated photons from a common entanglement source. Any eavesdropper intercepting and resending either Alice or Bob's photons disturbs the fragile entanglement, which introduces errors as before. This loss of entanglement can also be checked via other means such as a test of Bell's inequalities \cite{Clauser:1969ff} and entanglement witnesses \cite{Agnew:2012gs}. Schematics for both these protocols are shown in Fig.~\ref{EkertvBB84}.

The security analysis in the Ekert protocol is identical to that of BB84. The main difference appears in the passive selection of states by Alice \textit{and} Bob and the location of the source. In BB84, the source is located at Alice and she actively picks states to send to Bob. In Ekert, both Alice and Bob make passive measurements in their chosen measurement bases. In addition, the source of entangled photons can be spatially separated from both Alice and Bob. The lack of active preparation can be considered a technological advantage of Ekert over BB84, especially since entangled sources are available rather easily today.

\subsection{Quantum Ghost Imaging}
\label{Ghost}

\begin{figure}[t!]
\centering\includegraphics[scale=.75]{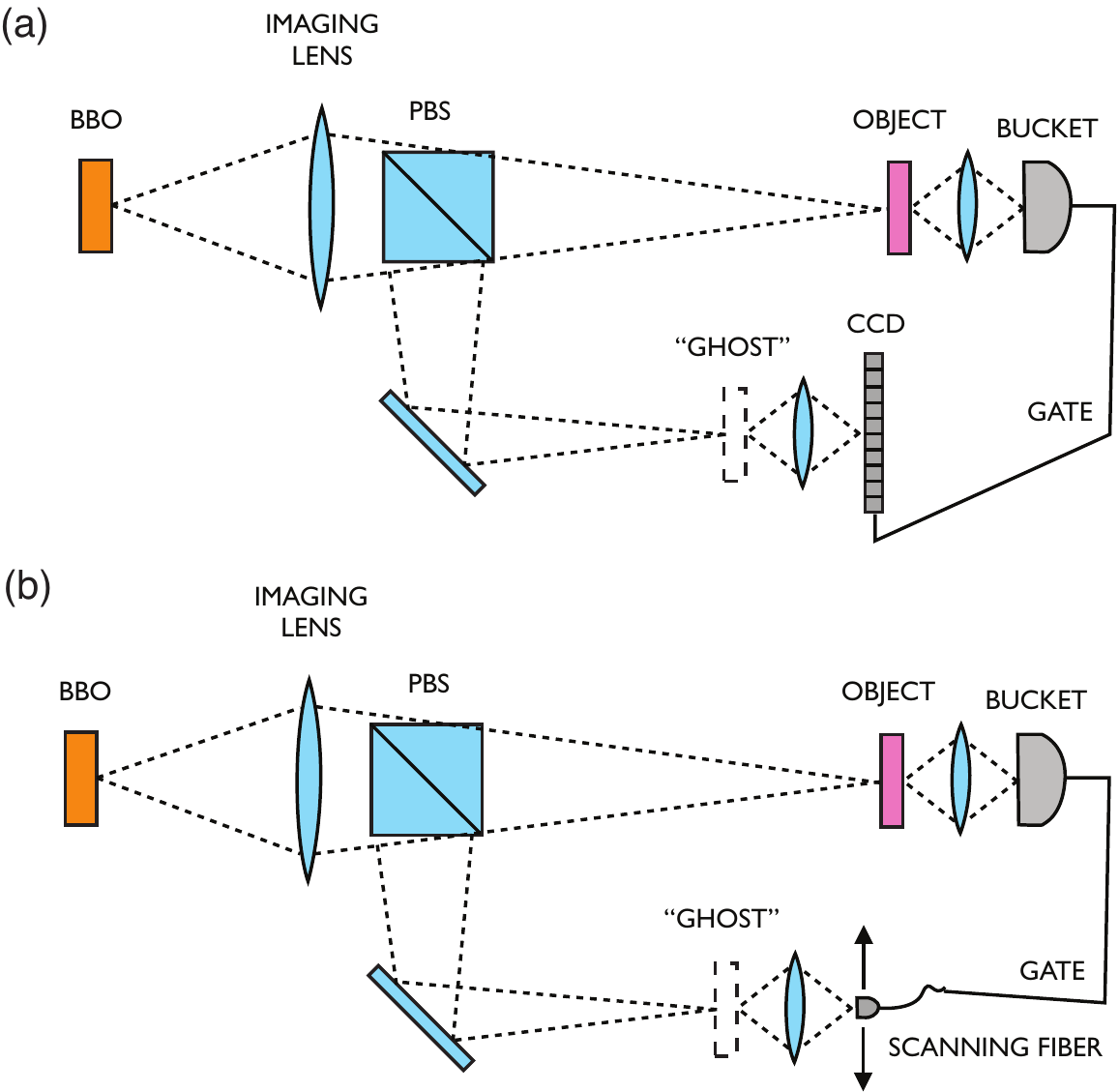}
  \caption{(a) An ideal quantum ghost imaging scheme that uses a CCD gated by a bucket detector for creating a ``ghost" image of an object without directly gaining any spatial information from it. (b) In reality, CCDs that can detect single photons are not commonplace. Hence, a scanning fiber tip is used in place.}\label{ghostsetup}
\end{figure}

Ghost imaging, also known as coincidence imaging, was first implemented with position-momentum entangled photons \cite{Pittman:1995jb,Strekalov:1995ub}. The strong position and momentum correlations shared by such photons allow one to perform imaging without a spatially resolving detector. This process is depicted in Fig.~\ref{ghostsetup}(a). The entangled photons are generated by pumping a $\beta$-Barium Borate (BBO) crystal with a pump laser (not shown). The entangled signal and idler photons generated in the type II downconversion process are orthogonally polarized, and can be separated with a polarizing beamsplitter (PBS). The crystal face is imaged both onto an object and onto a ``ghost" image plane through the PBS. The signal photon, as it has come to be called, is then allowed to fall onto a spatially non-resolving bucket detector. As its name implies, the bucket detector collects all the signal photons that make it past the object. The idler photons, on the other hand, are imaged from the ghost image plane onto a spatially resolving detector (CCD). A sharp image is obtained in the coincidence counts of the CCD and the bucket detector. The term ``ghost image" was coined for this phenomenon based on the fact that the image was formed without directly obtaining any spatially resolved image information from the object itself \cite{Pittman:1995jb}.

It was soon shown that ghost imaging relied solely on the spatial correlations of the two light fields. The same effect was reproduced by using randomly but synchronously directed twin beams of classical light \cite{Bennink:2002jr}. A benefit of using entangled photons was found to be that imaging could be performed both in the near and far fields, without having to change the source \cite{Gatti:2004ec,Bennink:2004df}. This is a direct consequence of the fact that entangled photons have strong correlations in both position and momentum, which correspond to correlations in the near and far fields respectively. In the case of ghost imaging with an entangled source, the choice of whether to measure in  the image plane or the diffraction pattern is left to the observer, instead of being determined by the source. Subsequently, even this property of ghost imaging with an entangled source was mimicked by using a pseudothermal source \cite{Ferri:2005hq}. The twin speckle patterns created by shining an intense beam of light through a ground glass plate and a beamsplitter were found to have strong spatial correlations in both the near and far fields \cite{Gatti:2004ec}.

In practice, both the quantum and thermal ghost imaging methods require the use of many single-photon pairs or random speckle patterns to obtain an image. Also, long processing times are needed for scanning an avalanche photodiode in the quantum case \cite{Pittman:1995jb} (Fig.~\ref{ghostsetup}(b)) or averaging many speckle patterns on a CCD camera in the thermal case \cite{Ferri:2005hq}. These requirements have made the practical applicability of such schemes difficult. Due in part to this, ghost imaging has steadily inhabited the domain of proof-of-principle experiments. Studies of ghost imaging through turbulence \cite{Meyers:2011cv,Dixon:2011wq} and ghost imaging experiments using compressive sensing \cite{Katz:2009fv,Bromberg:2009ca,MaganaLoaiza:2013jq} have been performed. More recent efforts to exploit the quantum correlations of spatially-entangled photons have led to new techniques for sub-shot-noise imaging \cite{Brida:2010jw}. A very recent experiment was able to use induced coherence between entangled photons from two separate sources in order to image an object with photons that never interacted with it \cite{Lemos:2014uw}. It is clear that the field of quantum imaging still has many new insights to offer and unexplored ideas yet to be discovered. In Section \ref{chap:QGI}, we extend the ghost imaging technique described in this section into a ghost image identification scheme. Instead of using a scanning fiber tip to measure the image, we use a hologram as an ``image sorter." In this manner, a known set of objects can identified by a pair of entangled photons, one of which interacts with the object and the other with the image sorting hologram. This technique is much faster than building a ghost image pixel by pixel, and maximizes the amount of image information carried by each photon.

\subsection{Direct Measurement}
\label{DM}

Weak values first aroused interest in the context of amplifying a small detector signal. By making an appropriate post-selection, the weak value can be made very large, which allows an experimenter to easily estimate the unknown small parameter $\epsilon$. However, this amplification is achieved at the cost of a large loss due to the post-selection process. Due to this, the benefits of weak value amplification as compared to standard statistical techniques have been studied in detail \cite{Ferrie:2013gf, Jordan:2014jv}. More recently, weak values have been used in an alternative technique for measuring a quantum state. Conventionally, a quantum state is measured through the indirect process of quantum tomography \cite{Altepeter:2005dc}. Like its classical counterpart, quantum tomography involves making a series of projective measurements in different bases of a quantum state. This process is indirect in that it involves a time consuming post-processing step where the density matrix of the state must be globally reconstructed through a numerical search over the many allowed alternatives. Due to this, tomography is prohibitive for measuring high-dimensional multipartite quantum states such as those of orbital angular momentum. 

Recent work has shown that a quantum state can be expanded into sums and products of complex weak values, which are proportional to the probability amplitudes of the state \cite{Lundeen:2011isa,Lundeen:2012db,Salvail:2013bo}. As these weak values are measurable quantities, a quantum state can thus be determined directly without the need for the complicated post-processing step involved in tomography. A particularly notable application of such an expansion is the direct determination of the complex components of a pure quantum state $\ket{\psi}$ expanded in a particular measurement basis $\{\ket{a}\}$ \cite{Lundeen:2011isa,Lundeen:2012db,Salvail:2013bo}. This is accomplished by the insertion of the identity and multiplication by a strategically chosen constant factor $c = \braket{b}{a}/\braket{b}{\psi}$, where the auxilliary state $\ket{b}$ must be unbiased with respect to the entire basis $\{\ket{a}\}$ such that $\braket{b}{a}$ is a constant for all $a$.  With this choice we have,
\begin{align}
  c \ket{\psi} = c \sum_a \ket{a}\braket{a}{\psi} = \sum_a \ket{a} \frac{\braket{b}{a}\braket{a}{\psi}}{\braket{b}{\psi}}.
\end{align}
That is, each scaled complex component $c \braket{a}{\psi}$ of the state $\ket{\psi}$ can be directly measured as a complex weak value of the projection operator $\op{\Pi}_a = \ket{a}\bra{a}$ using the unbiased auxilliary state $\ket{b}$ as a post-selection.  After determining these complex components experimentally, the state can be renormalized to eliminate the constant $c$ up to a global phase.  For mixed states, one can additionally vary the auxilliary state $\ket{b}$ within a mutually unbiased basis to determine the Dirac distribution for the state directly using the same technique \cite{Lundeen:2012db,Salvail:2013bo}. 

\begin{figure}[t!]
\centering\includegraphics[scale=1.3]{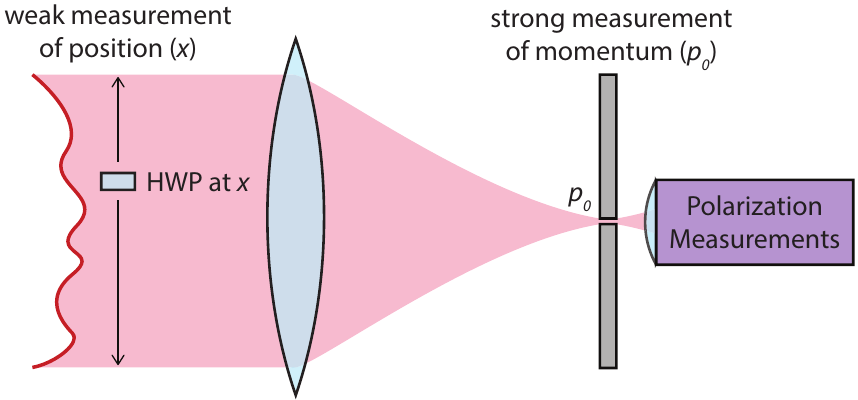}
  \caption{A schematic outlining the direct measurement experiment of Lundeen et al. \cite{Lundeen:2011isa} where the authors measured the wavefunction in the position basis by using polarization as a pointer.}\label{Lundeen}
\end{figure}

We can use our previous example of a polarized beam going through a birefringent crystal (Fig.~\ref{crystal}) to illustrate this idea. We showed earlier how we can isolate and measure both the real and imaginary parts of the polarization weak value $S_w$ in this experiment. In order to apply our setup to characterize the polarization quantum state, we must perform the post-selection in a basis mutually unbiased with respect to the weak measurement basis. Specifically, as the birefringent crystal in our example performs a weak measurement in the $H/V$ basis, we must orient the post-selecting polarizer in the $D/A$ basis (which is mutually unbiased with respect to the $H/V$ basis). This is the exact procedure used in the polarization state characterization experiment of Ref. \cite{Salvail:2013bo} and discussed in further detail in Ref. \cite{Dressel:2014ks}.

In the pioneering experiment of Lundeen et al. \cite{Lundeen:2011isa}, the authors first used this technique to measure the wavefunction of an ensemble of identically prepared photons in the position basis. Following the theoretical treatment above, one can expand the wavefunction $\ket{\Psi}$ in the position basis in terms of weak values. By inserting the identity $\ket{x}\bra{x}$ and multiplying by a constant factor $c=\brac{p}{x}/\brac{p}{\Psi}$, one can write the wavefunction as
\be
  c \ket{\Psi} = c \int dx \ket{x}\braket{x}{\Psi} = \int dx \ket{x} \frac{\braket{p}{x}\braket{x}{\Psi}}{\braket{p}{\Psi}} = \int dx \avg{\pi_x}_{\tiny \textrm{W}}\ket{x}.
\ee

\noindent In this manner, the wavefunction at a particular position $x$ is found to be proportional to the weak value at that position
\be c \Psi(x) =  \avg{\pi_x}_{\tiny \textrm{W}}.\ee

In order to measure the position weak value, Lundeen et al. used polarization as a pointer. As explained in section \ref{sec:WV}, the weak value of a particular observable can be measured by coupling that observable to a pointer observable. As shown in Fig.~\ref{Lundeen}, the authors performed a weak measurement of the position $x$ of a photon by rotating the polarization at $x$ by a small angle with a sliver of half-wave plate (HWP). A strong measurement of the conjugate variable of momentum was performed by Fourier-transforming with a lens and post-selecting value $p=0$ of momentum. By measuring the rotation of the polarization vector in the linear and circular polarization bases, the authors were able to measure the real and imaginary parts of the position weak value. From this, they obtained the real and imaginary parts of the wavefunction, which then gave them the amplitude and phase as a function of $x$. They verified this measurement by comparing it to a regular measurement with a Shack-Hartmann wavefront sensor. We use this direct measurement technique in Section \ref{chap:DM} to measure the wavefunction of a high-dimensional quantum state in the OAM basis.

The primary benefit of this tomographic approach is that minimal post-processing is required to construct the state from the experimental data. The real and imaginary parts of each state component directly appear in the linear response of the measurement device up to appropriate scaling factors. The downside of this approach is that the auxiliary state $\ket{b}$ must be chosen carefully so that the denominator $\braket{b}{\psi}$ or $\braket{p}{\Psi}$ (and hence the detection probability) does not become too small and break the weak interaction approximation used to determine the weak values \cite{Haapasalo:2011en}. This restriction limits the generality of the technique for faithfully determining a truly unknown $\ket{\psi}$. For a more comprehensive review of recent work on this topic, see Ref. \cite{Dressel:2014ks}.


\renewcommand{\abs}[1]{\ensuremath{\left| #1 \right|}}
\renewcommand{\bra}[1]{\ensuremath{\left< #1 \right|}}
\renewcommand{\ket}[1]{\ensuremath{\left| #1 \right>}}

\section{Quantum-Secured Surveillance}
\label{chap:QSS}
\subsection{Quantum-Secured Imaging}

\begin{figure}[t!]
\centering\includegraphics[scale=.8]{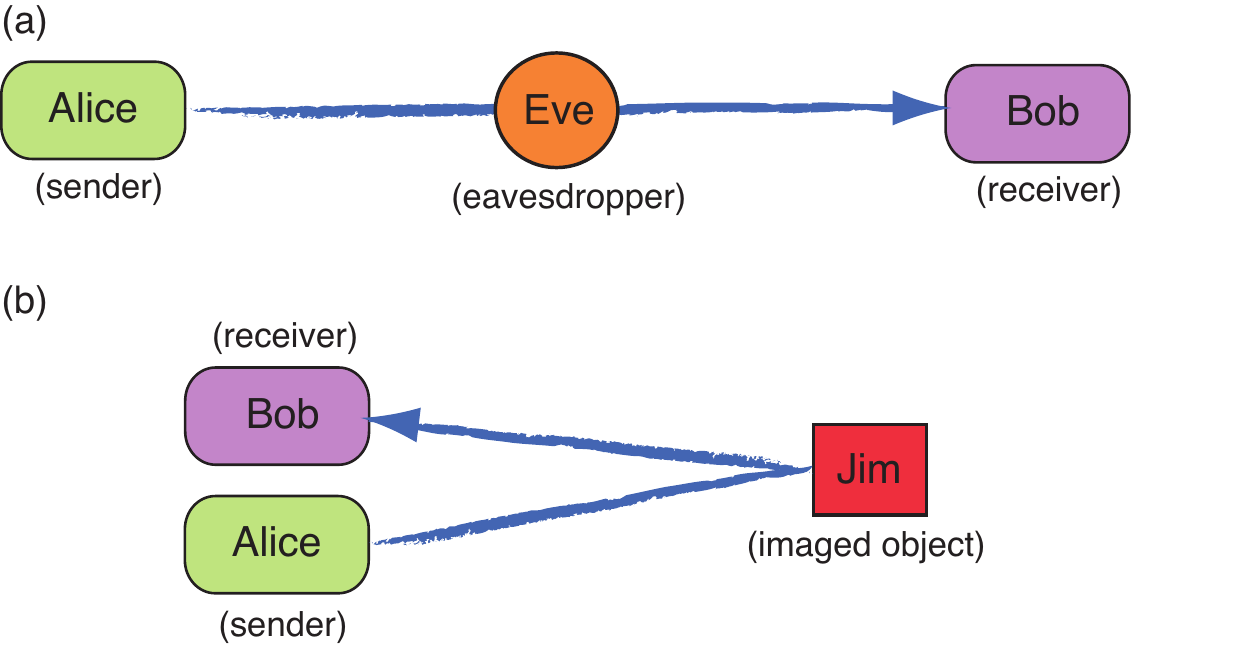}
  \caption{A sketch comparing the (a) quantum key distribution and (b) quantum-secured imaging protocols (Figure redrawn from Ref. \cite{Malik:2012js}, copyright 2012 American Institute of Physics).}\label{SQIvQKD}
\end{figure}

Active imaging systems such as radar and lidar are susceptible to intelligent jamming attacks, where the light used for querying an object is intercepted and resent. In this manner, an object can send false information to the receiver belying its true position or velocity, or even creating a false target \cite{Roome:1990th}. In this section, we show how one can detect such jamming attacks by using quantum states of light modulated in polarization. The BB84 protocol of quantum key distribution (QKD) uses such quantum states to generate a random key with unconditional security \cite{Bennett:1984wv}. By randomly modulating the polarization of single photons in two mutually unbiased polarization bases, the sender (Alice) can send a stream of 1s and 0s to the receiver (Bob). Any eavesdropper trying to gain information about the polarization of a photon will have to perform a measurement, thus unalterably changing the polarization state and introducing errors that Alice and Bob can detect. In this manner, the eavesdropper will reveal herself (Fig.~\ref{SQIvQKD}a).

\begin{figure}[t!]
\centering\includegraphics[scale=.8]{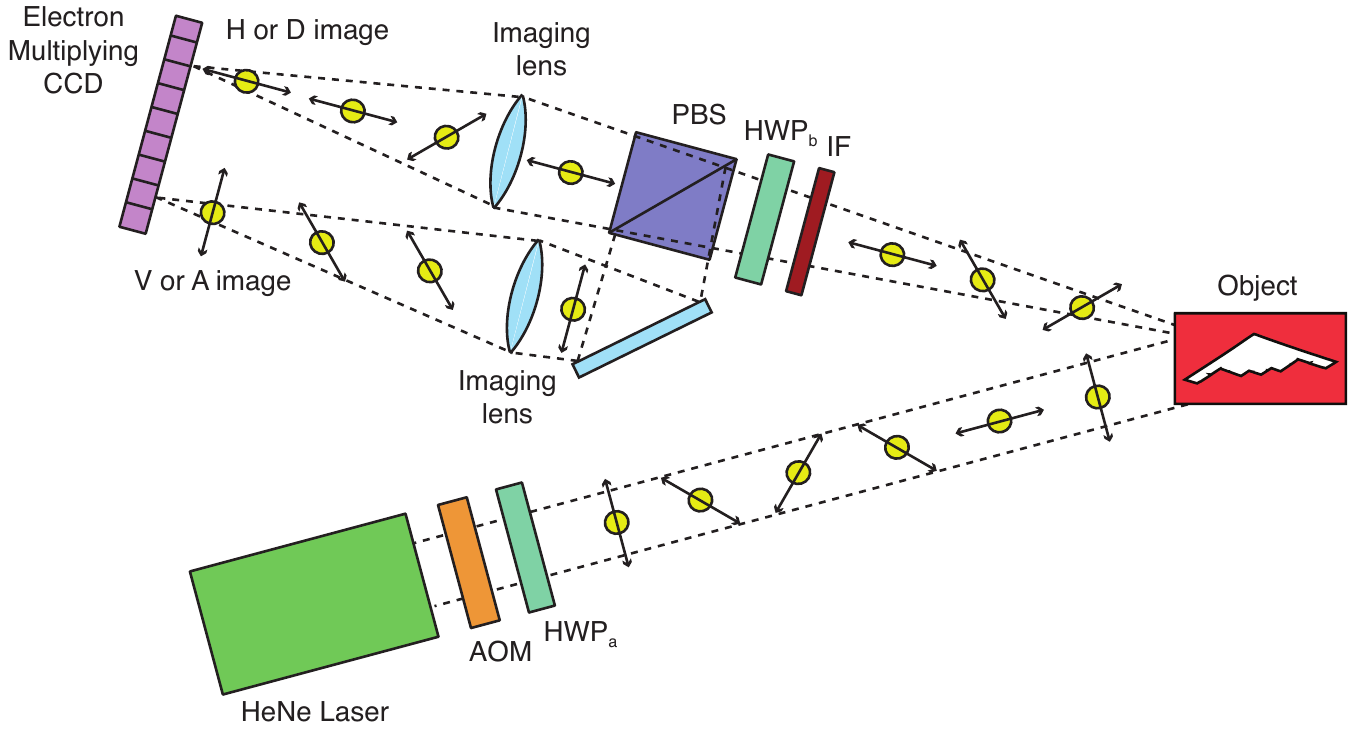}
  \caption{Schematic of our quantum-secured imaging experiment. An object is securely imaged with single-photon pulses modulated in polarization. Security is verified by measuring the error between sent and received polarizations (Figure redrawn from Ref. \cite{Malik:2012js}, copyright 2012 American Institute of Physics).}\label{setup}
\end{figure}

We extend this idea to an imaging system, where the object to be imaged now plays the role of the eavesdropper. Alice and Bob constitute the imaging system, and are hence located in the same place (Fig.~\ref{SQIvQKD}b). If the object intercepts and resends any of the imaging photons, it will introduce errors in the polarization encoding that can be detected by Alice and Bob. In a two-dimensional polarization-based QKD system, the minimum error introduced by an eavesdropper using an intercept-resend attack is equal to 25\%. We apply this same error bound to our imaging system. As shown in Fig.~\ref{setup}, a HeNe laser is intensity modulated by an acousto-optic modulator (AOM) to create pulses with less than one detected photon on average. A half-wave plate (HWP$_\textrm{a}$) mounted on a motorized rotation stage randomly switches the polarization state of the photon among horizontal, vertical, diagonal, and anti-diagonal (\ket{H}, \ket{V}, \ket{D}, and \ket{A}). The single-photon pulses are incident on the object, which consists of a stealth aircraft silhouette on a mirror. They are then specularly reflected from the object towards our detection system. In Fig.~\ref{setup}, a non-zero reflectance angle is shown for clarity. An interference filter (IF) is used to eliminate the background. A second rotating half-wave plate (HWP$_\textrm{b}$) and a polarizing beam-splitter (PBS) perform a polarization measurement in either the horizontal-vertical ($H/V$) or diagonal-anti-diagonal ($D/A$) basis. Two lenses are used after the PBS to create four images corresponding to the four measured polarizations on an electron-multiplying CCD camera (EMCCD), which serves as a spatial single-photon detector.

\begin{figure}[b!]
\centering\includegraphics[scale=.7]{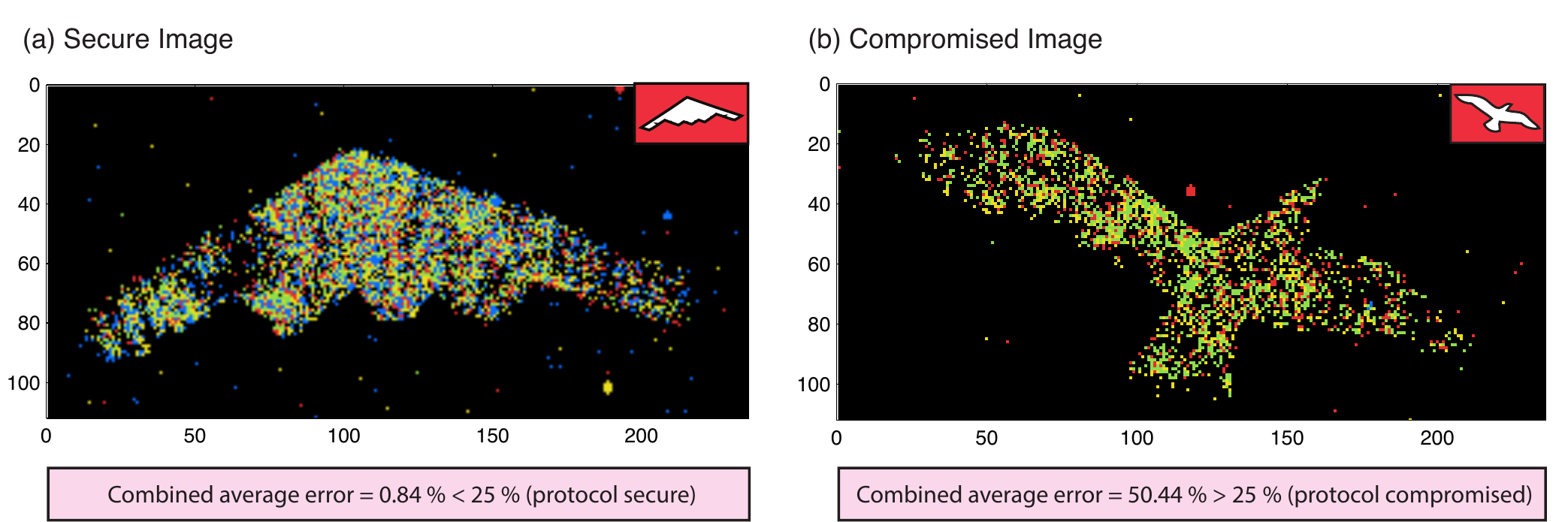}
  \caption{Laboratory demonstration of quantum-secured imaging. (a) When there is no jamming attack, the received image faithfully reproduces the actual object, which is shown in the inset. (b) In the presence of an intercept-resend jamming attack, the received image is the ``spoof" image of a bird. However, the imaging system can always detect the presence of the jamming attack, because of the large error rate in the received polarization. In (a) the error rate is 0.84\%, while in (b) it is 50.44\%. A detected error rate of $>25\%$ indicates that the image received has been compromised (Figure redrawn from Ref. \cite{Malik:2012js}, copyright 2012 American Institute of Physics).}\label{image_data}
\end{figure}

Figure \ref{image_data}(a) and (b) show two images formed using this system. The first image of a stealth aircraft has a measured polarization error of 0.84\%. This means that 0.84\% of the time in the image, the measured polarization of a received photon was different from the photon polarization sent. Since this is less than the error bound of 25\%, this image can be considered secure. The second image shows results from a simulated intercept-resend attack where the object simply blocked all the incoming photons, and resent photons in state \ket{H} with modified information indicating the image of a bird. The polarization error measured in this case is 50.44\%, which is very close to the expected value of 50\%. Since this is greater than the error bound of 25\%, it indicates that the object was actively jamming the imaging system.

\subsection{Quantum-Secured LIDAR}

We can apply this quantum-secured protocol to a lidar system that uses the time-of-flight information of a light pulse to measure the velocity of a moving object. In Ref. \cite{Malik:2012js}, we propose such a system based on the Ekert protocol of QKD. In our system, one photon from an entangled pair is kept locally. The second photon travels to the object and back. Any intercept-resend attack by the object can be detected by testing the presence of entanglement in the system via a test of Bell's inequalities. The loss of entanglement would indicate that the lidar system system is being actively jammed. This lidar system is discussed in more detail in Ref. \cite{Malik:2012js}. While our quantum-secured protocols are certainly limited by cloaking techniques that don't modify the polarization state of a photon, they can be easily integrated into modern optical ranging systems given the current state of QKD technology.


\section{Quantum Ghost Image Identification}
\label{chap:QGI}

\subsection{Introduction}

Holograms are a hallmark of science fiction movies, often used to instill a sense of futuristic reality where one can create three-dimensional objects made entirely of light. While realistic 3D holograms are indeed a technology of the future, simpler holograms are more commonplace than one may think. For example, holograms are used as security marks on identity and credit cards. Holograms are also used for recording large amounts of information onto a medium. A simple hologram can be formed by interfering two mutually coherent waveforms, which can be written in terms of their amplitude and phase as \cite{Goodman:Fourier}

\bea A(x,y) &=& \abs{A(x,y)}e^{i\phi_A(x,y)}\nonumber\\
R(x,y) &=& \abs{R(x,y)}e^{i\phi_R(x,y)}
\eea

\noindent Here, $A$ is an arbitrary unknown field and $R$ is a known reference field. In the developing process, the interference pattern written onto the hologram is given by the intensity $I(x,y)=\abs{A(x,y)+R(x,y)}^2$ as follows

\bea I(x,y) &=& \abs{A(x,y)}^2+\abs{R(x,y)}^2+A(x,y)^\dag R(x,y)+A(x,y)R(x,y)^\dag\nonumber\\
&=&\abs{A(x,y)}^2+\abs{R(x,y)}^2+2\abs{A(x,y)}\abs{R(x,y)}\cos\big[\phi_R(x,y)-\phi_A(x,y)\big]\nonumber\\
\eea

\noindent Here, the first two terms depend only on the intensities of each field while the third term depends on their relative phases. In this manner, a hologram stores information about both, the amplitude and the phase of the unknown waveform $A$. By sending the reference waveform $R$ through the developed hologram, the unknown waveform $A$ can be reconstructed. In fact, we use this technique in Section \ref{chap:OAMQKD} to create fields with a helical phase structure, also known as orbital angular momentum modes.

\begin{figure}[t!]
\centering\includegraphics[scale=0.7]{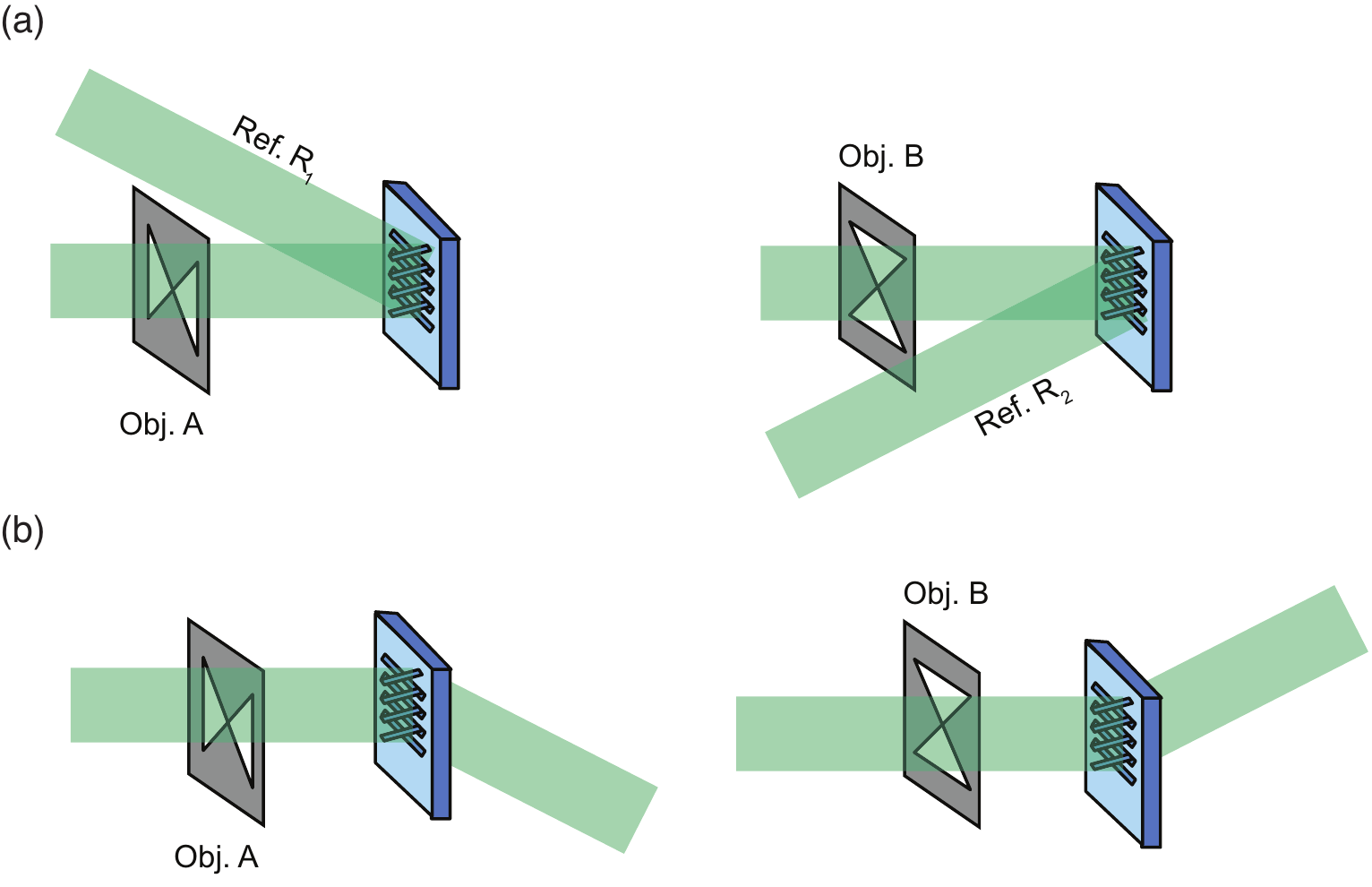}
  \caption{A cartoon showing (a) the construction and (b) the operation of a two-object multiplexed hologram as an image sorter.}\label{2holograms}
\end{figure}

\subsection{Holograms as Image Sorters}

The wavefront reconstruction process described above can be employed in reverse. Instead of recreating the waveform $A$ by sending reference $R$ into the hologram, one can send waveform $A$ into the hologram to recreate reference $R$. In this manner, a hologram can be used to ``identify" an image from a set of images. The resulting output can be written as a product of the original waveform with the transmission function given by the intensity of the hologram:

\bea &A(x,y)&I(x,y)=\nonumber\\
&&A(x,y)\abs{A(x,y)}^2+A(x,y)\abs{R(x,y)}^2+\abs{A(x,y)}^2 R(x,y)+A^2(x,y)R(x,y)^\dag\nonumber\\ \eea

\noindent If a plane wave is used as the reference beam, the first two terms in the above equation are simply an attenuated version of the original unknown waveform propagating in the normal direction. We refer to this as the ``zero-order" output from the hologram. The third term gives us the recreated reference beam $R$, which also carries the intensity distribution of the original waveform $\abs{A}^2$. The fourth term refers to a virtual reference beam $R^\dag$. A similar hologram can be constructed for two arbitrary fields $A$ and $B$, with two difference reference fields $R_1$ and $R_2$. Such a hologram is referred to as a \textit{multiplexed} hologram, and can be used an image sorter for distinguishing two images for each other.

\begin{figure}[t!]
\centering\includegraphics[scale=0.8]{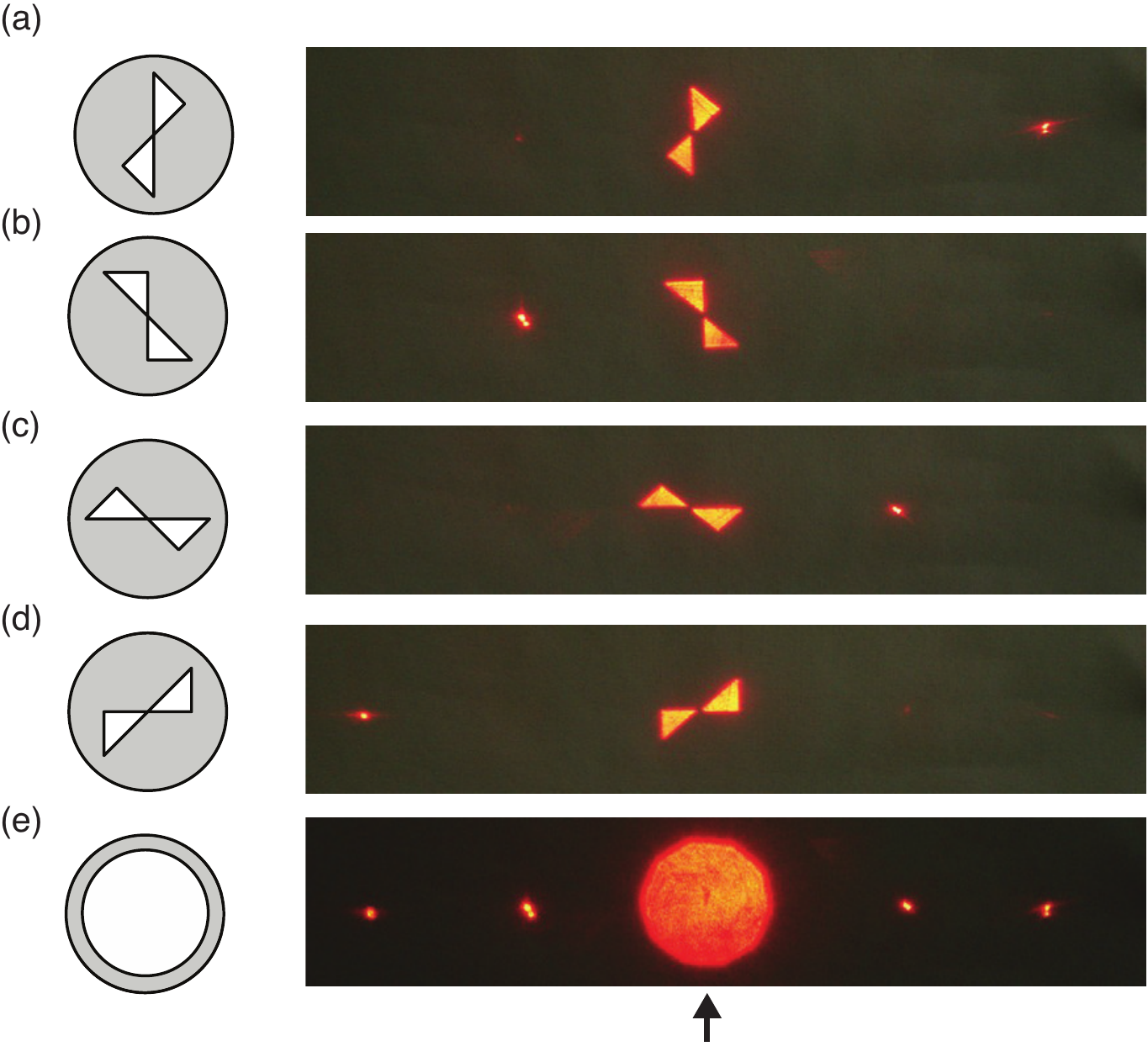}
  \caption{CCD images showing the output from a four-object multiplexed hologram sorting objects (a)-(d). (e) shows the output when a collimated field is sent into the hologram. Notice that a strong zero-order field (indicated by the vertical black arrow) is present in all cases.}\label{4holograms}
\end{figure} 

A cartoon showing the construction and operation of such a multiplexed hologram is shown in Fig.~\ref{2holograms}(a) and (b). First, the holographic material is exposed with an interference pattern formed by interfering field $A$ originating from object A with a reference plane wave $R_1$. Then, this procedure is repeated for object B with a reference plane wave $R_2$ from a different direction (Fig.~\ref{2holograms}(a)). The hologram is then fixed (or developed) to make the interference pattern permanent. This multiplexed hologram then acts as an image sorter, converting a field from object A into reference $R_1$ and one from object B into reference $R_2$ (Fig.~\ref{2holograms}(b)). The output from such a hologram made for 4 different objects is shown in Fig.~\ref{4holograms}. CCD images of the output for each object (a-d) show a strong zero-order output in the normal direction (indicated by a black arrow), and a strong diffracted component in the direction of the original plane wave associated with object (a-d). In Fig.~\ref{4holograms}(e), the output obtained when a collimated beam with uniform amplitude is input into the hologram is shown. As can be seen, all four diffracted orders are present, as image information for all four objects is present in the input. While these images show the hologram operating at high light levels, we use the same hologram in the next section to perform ghost image identification with single photons. Many of these predictions have been verified in recent publications \cite{Broadbent:2009tq,Malik:2010cr}.  

\subsection{Ghost Image Identification with Correlated Photons}

\begin{figure}[t!]
\centering\includegraphics[scale=0.7]{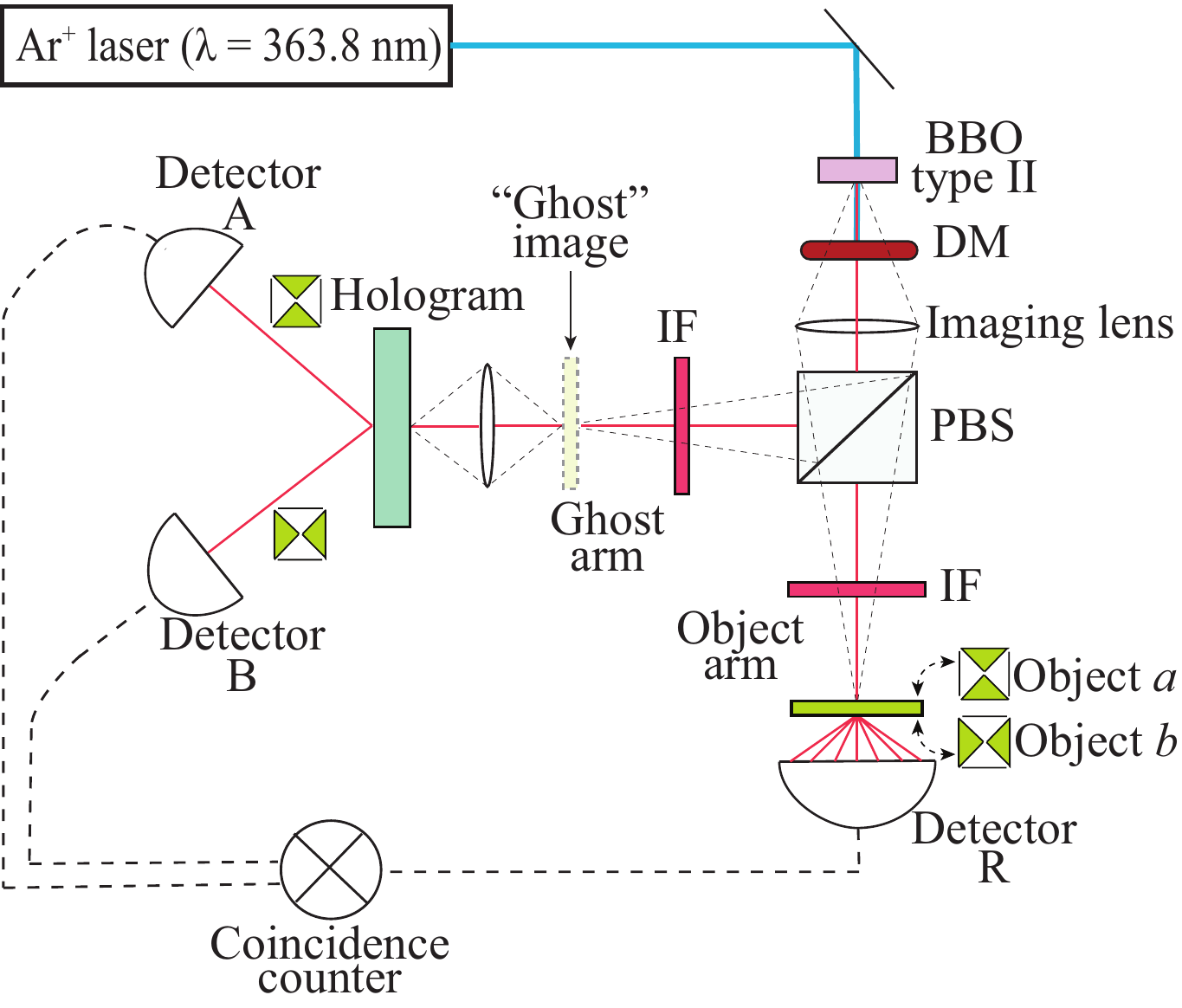}
  \caption{Experimental setup for quantum ghost image identification.
  DM is a dichroic mirror for blocking the pump laser; IF is an interference filter
  with 10 nm bandwidth, centered at 727.6 nm. The dotted lines indicate the imaging process for a point object. (Figure adapted from Ref. \cite{Malik:2010cr}.)}\label{ghostexp}
\end{figure}

In this section, we describe a quantum ghost imaging scheme that
uses the aforementioned holographic filtering technique to identify an object from a large basis set of objects \cite{Malik:2010cr}. As a proof-of-principle experiment, we demonstrate this method for both a set of two and a set of four spatially non-overlapping objects. We do so by replacing the CCD in the idler arm of the standard ghost imaging setup described in Section \ref{Ghost}
with a holographic sorter. The ghost image is
obtained from the coincidence counts of the bucket detector and
the beams diffracted by the hologram. In this manner, we are able
to determine which object from our pre-established set is in the
signal arm without directly acquiring any spatial information about it. In our analysis, we change the naming system by referring to the signal arm as the ``object'' arm and the idler arm as the ``ghost'' arm. The object arm is the path taken by the object photon and contains the object followed by the bucket detector. The ghost arm is the path taken by the ghost photon and contains a holographic sorter and single-photon detectors (Fig. \ref{ghostexp}).

Let us first describe the two-object case. The measurement in the object arm is carried out by the object-bucket detector combination. The object photon is either transmitted into the bucket detector, labelled R in Fig. \ref{ghostexp}, or is blocked by the object. The measurement in the ghost arm is carried out by the hologram-detectors combination. The ghost photon is diffracted into either detector A or B. If object $a$ is present in the object arm and transmits an object photon into bucket detector R, the corresponding ghost photon will always be diffracted by the hologram into detector A. This is due to the strong position correlations between the two photons. A similar explanation holds for object $b$. 

Our experimental setup for the two-object case is sketched in Fig. \ref{ghostexp}. The holographic sorter is created by multiplexing the two spatially non-overlapping objects $a$ and $b$ with reference beams incident at different angles. It is recorded with a collimated HeNe laser at 633 nm on a silver-halide
plate. The entangled photon pairs are created by degenerate SPDC in a collinear type-II phase
matched BBO crystal pumped by a cw beam from an argon-ion laser operating at a wavelength of 363.8 nm. The pump beam is well collimated with a divergence of less than 0.31 mrad and a beam waist of 3 mm. A dichroic mirror placed
after the crystal blocks the pump laser light. A polarizing beam splitter separates the object photon from the ghost photon. The distance between the
crystal and the object (and the ghost image plane) is 45 cm.
The imaging condition is met by placing a 10-cm-focal-length lens 15 cm after the crystal. The ghost image plane then acts as a ``virtual object'' for the hologram. This imaging process is
illustrated in Fig. \ref{ghostexp} with dotted lines for a point object at the
crystal. In the unfolded or Klyshko interpretation of the setup \cite{Strekalov:1995ub}, one can understand the object as being imaged onto the crystal face, which is then imaged onto the ghost image plane, and consequently imaged onto the hologram. A more detailed theoretical analysis of the imaging process for entangled two-photon fields can be found in Ref. \cite{Abouraddy:2002vi}. Perkin-Elmer
avalanche photodiodes and a coincidence circuit with a window of
approximately 12 ns are
used for the detection.

\begin{figure}[t!]
\centering\includegraphics[scale=0.6]{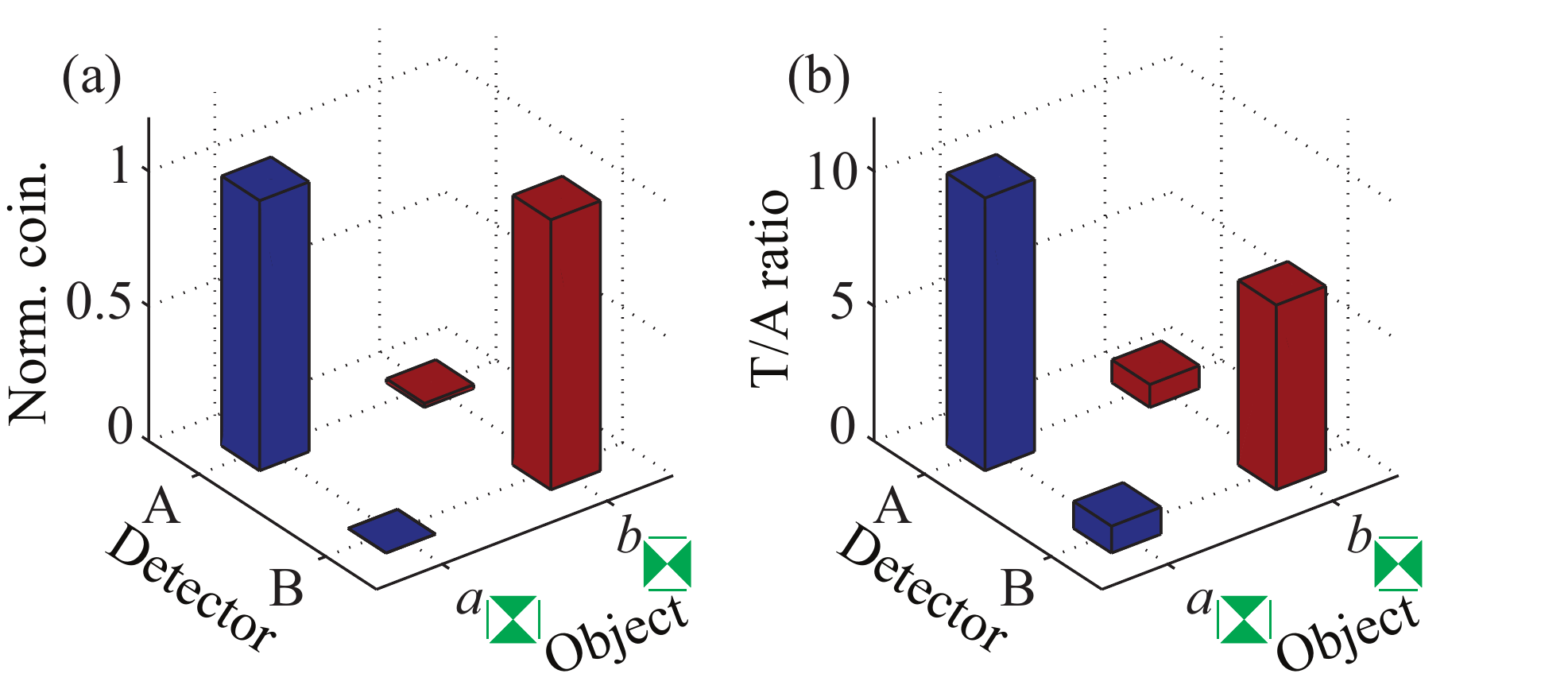}
  \caption{Image-identification results for the two-object case. (a) Data
  for each object-detector combination is normalized by the
  maximum coincidence count for the corresponding object.
  (b) T/A ratio is calculated by dividing the total coincidences by the
  accidental coincidences for each object-detector combination. (Figure adapted from Ref. \cite{Malik:2010cr}.)
  }\label{2objresults}
\end{figure}

When a coincidence count correctly identifies the object, we refer to it as a true case (A-$a$ or B-$b$), and the opposite as a false case (A-$b$ or B-$a$). The normalized experimental results for each object are graphed in Fig.~\ref{2objresults}(a). The data for each detector are normalized by the number of coincidence counts recorded by that detector for a true case. It is clear from this figure that our experimental system has high contrast between true and false cases. The ratio between total and accidental coincidence counts (T/A ratio) for each object-detector combination serves as a measure of the system fidelity \cite{Broadbent:2009tq} and is graphed in Fig.~\ref{2objresults}(b).

\begin{figure}[t!]
\centering\includegraphics[scale=0.7]{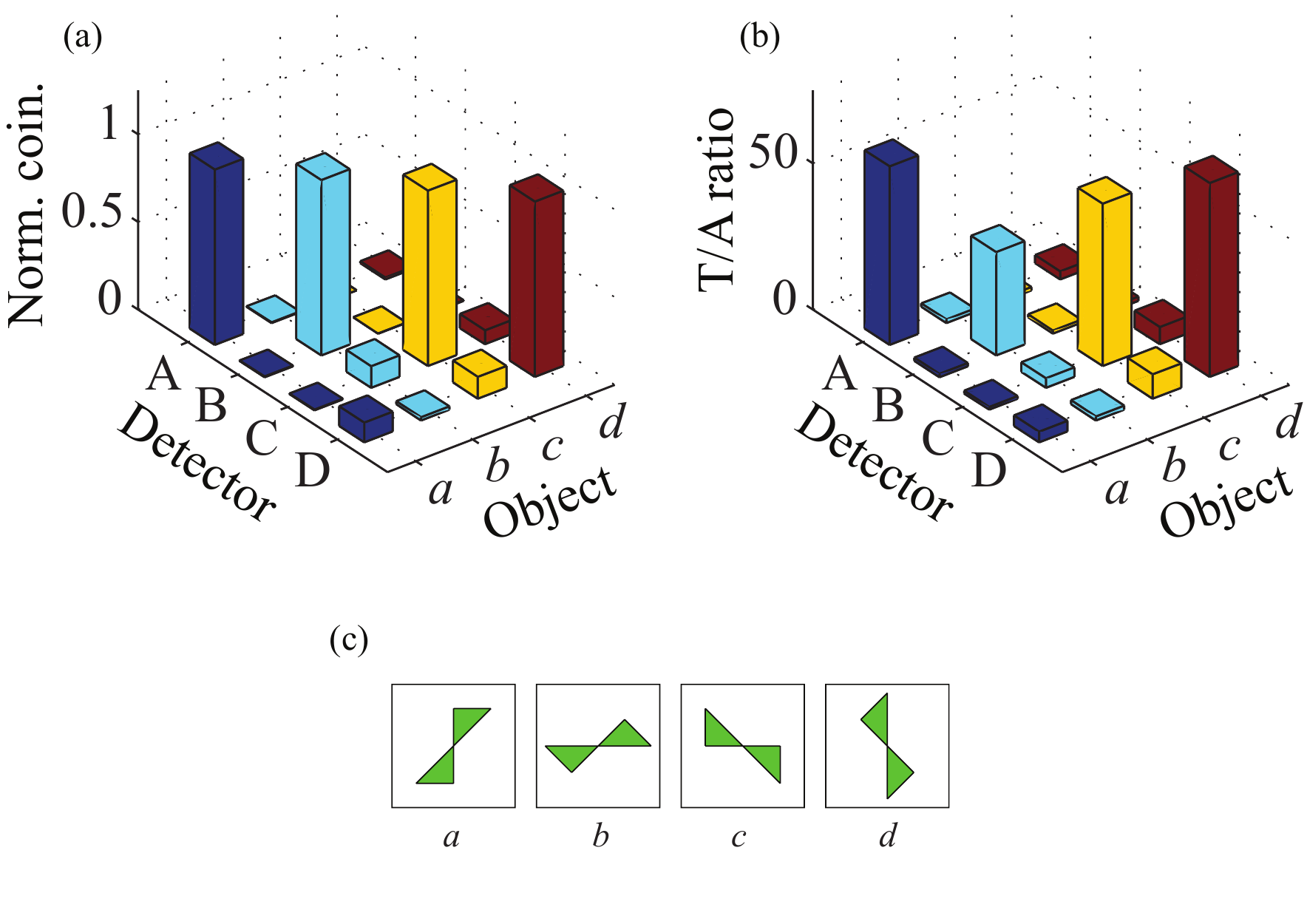}
  \caption{(a) and (b) Graphs of ghost-image-identification results for the four-object case. (c) The four spatially non-overlapping objects used in our experiment. (Figure adapted from Ref. \cite{Malik:2010cr}.)
   }\label{4objresults}
\end{figure}

We repeat this experiment for an object space of four objects.  An angularly multiplexed hologram was created for the four spatially non-overlapping objects shown in Fig. \ref{4objresults}(c) using the same method as before. The image sorting operation of this hologram at high light levels is discussed in the previous section. The normalized coincidence counts and the T/A ratios for each object-coincidence combination are plotted in Fig. \ref{4objresults}.

Ghost image identification using a holographic sorter clearly
has many advantages over other ghost imaging schemes. First, a hologram provides an all-optical method of sorting images that can overcome the limitations of slow CCD frame rates \cite{He:2008ud}. Second, distinguishing among objects of a known set is much faster than building an image pixel by pixel. This approach has practical applications in situations where the objects to be distinguished fall into a relatively small class of objects. Third, an advantage of using quantum ghost image identification appears in the applicability of this method when extremely low light levels are required. One can classify this as a type of ``stealth imaging," where a minimum number of photons is used in order to avoid optical eavesdropping or letting the object become aware of its detection. The small number of photons used in quantum ghost image identification make it an excellent candidate for such imaging schemes. When combined with the quantum-secured imaging technique that is discussed in Section \ref{chap:QSS}, quantum ghost image identification could prove especially valuable for securely identifying an object while economizing the number of photons used.

Matched filters have been used for pattern recognition for many years \cite{Turin:1960tw}. Highly overlapping objects can be sorted with a high confidence factor using matched filters made with holograms \cite{Goodman:Fourier}. While our experiment addresses only non-overlapping amplitude objects, in principle it is possible to construct matched filters that distinguish among complicated and overlapping objects. However, the efficiency of the identification process is reduced for such sets of objects, and more than one photon pair is needed to distinguish unambiguously among them \cite{Morris:1984vk}. 

In conclusion, we have shown that it is possible to discriminate
among non-overlapping objects using a small number of
correlated photon pairs, without gaining any spatially-resolved
information about the objects themselves. Although we have performed
this experiment for object spaces of two and four objects, it is
possible to expand the size of the object space markedly.
Multiplexed holograms have been designed to store as many as
10,000 images \cite{An:1999ui}. However, as the object
space increases, limitations on coincidence counts will be imposed
by large cross talk and low diffraction efficiency. The
possibility of using thick holograms to remedy such problems is a topic worth exploring in the future. 


\section{High-Dimensional Quantum Key Distribution}
\label{chap:OAMQKD}

\medskip

\subsection{Introduction}

Since Bennett and Brassard introduced the first quantum key distribution (QKD) protocol in 1984 \cite{Bennett:1984wv}, the field of QKD has rapidly developed to the extent that QKD systems are commercially available today. Secure transmission of a quantum key has been performed over 148.7 km of fiber \cite{Hiskett:2006kt} as well as over 144 km of free space \cite{Ursin:2007jj}. One of the limiting aspects of these key distribution systems is that they use the polarization degree of freedom of the photon to encode information \cite{Hiskett:2006kt,Ursin:2007jj}. The use of polarization encoding limits the maximum amount of information that can be encoded on each photon to one bit. In addition, it places a low bound on the amount of error an eavesdropper can introduce without compromising the security of the transmission \cite{Bourennane:2002uo}. Because of these limitations, there has been great interest in exploring other ways to encode information on a photon that would allow for higher data transmission rates and increased security \cite{Groeblacher:2005ec,Cerf:2002fp}.

In this section, we report results on the use of orbital angular momentum (OAM) modes of a photon in QKD. The motivation for doing so is that OAM modes span a discrete, infinite-dimensional basis. Hence, there is no limit to how much information one can send per photon in such a system. The large dimensionality of this protocol also provides a much higher level of security than the two-state approach \cite{Bourennane:2002uo}. However, in a practical communication system using OAM modes, the maximum number of modes that can be used is limited by the size of the limiting aperture in the system. This occurs because the radius of an OAM mode increases with the mode number. In this section, we discuss the advantages of increasing the dimensionality of the Hilbert space for a QKD system in detail. Then, we explain how we use holograms to generate the high-dimensional modes we use in our system. A significant part of any quantum communication system is the efficient sorting of single photons carrying information. we describe the approach we use to sort photons carrying OAM. Finally, we describe two high-dimensional QKD systems that use OAM modes for encoding, which we are currently in the process of building. Since our system uses spatial modes, it is highly susceptible to turbulence. Recently, there have been several theoretical studies on how atmospheric turbulence affects OAM modes \cite{Paterson:2005wj,Tyler:2009wz,Smith:2006ek,Gbur:2008fb,Roux:2011dr}. In addition, many recent experiments have been performed that study the effects of atmospheric turbulence on the channel capacity of an OAM communication channel at high light levels \cite{Malik:2012ka,Rodenburg:2013dn,Krenn:2014wj}.

\subsection{Advantages of High-Dimensionality}

\subsubsection{Channel Capacity of an Ideal Channel}

The amount of information that can be carried by a channel is related to the concept of entropy. Generally, entropy is understood as a measure of disorder in a system. For example, in the context of thermodynamics, the entropy of a glass of ice water increases as its reaches room temperature and the ice melts. Similarly, in information theory, entropy is understood as a measure of randomness of a variable. First applied to the field of communication systems by Claude E. Shannon \cite{Shannon:1948zz,Nielsen:2004}, the \textit{Shannon entropy} of a random variable can be thought of as a measure of its uncertainty $before$ we learn its value. Another way of understanding this is in terms of the information gained $after$ learning the value of the variable. The Shannon Entropy is defined as:

\begin{figure}[t!]
\centering\includegraphics[scale=0.6]{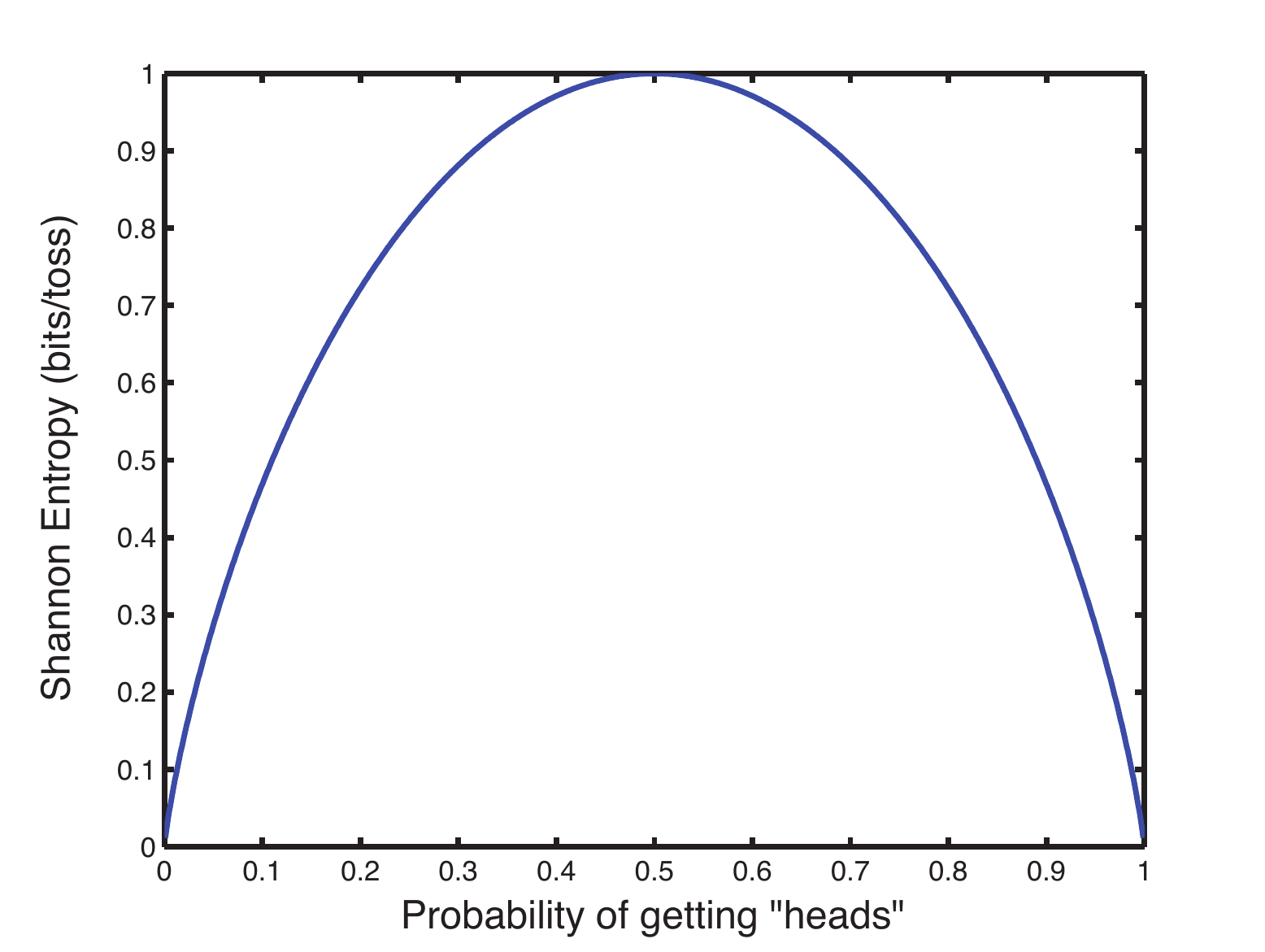}
  \caption{Shannon entropy of a coin toss as a function of fairness.}\label{cointoss}
\end{figure}

\be H(X) = H(p_1,...,p_N)=-\sum_n p_n \log_2(p_n)\label{shannon}\ee

\noindent where $p_n$ is the probability of the $n^{th}$ outcome of the variable $X$. A simple example is a coin toss. A fair coin has equal probability of resulting in a ``heads" or ``tails" outcome. This results in a maximum entropy of 1 as follows:

\be H(X) = H(p_{heads},p_{tails}) = -(0.5\times\log_2(0.5)+0.5\times\log_2(0.5))=1\textrm{ bit}\ee

\noindent If the coin we are using is unfair such that it has a $3/4$ probability of resulting in ``heads" and $1/4$ probability of resulting in ``tails," its shannon entropy is reduced to 0.81 as follows:

\be H(X) = -(0.75\times\log_2(0.75)+0.25\times\log_2(0.25))=0.8113\textrm{ bit}\ee

The plot in Fig.~\ref{cointoss} shows the entropy of a coin toss as a function of the fairness of the coin (the probability of getting ``heads"). A fair coin has a 0.5 probability of getting ``heads" in a toss. In the limits of a completely unfair coin, the entropy goes to zero. This makes sense if you think of the entropy in terms of the \textit{information gained}. If the coin toss always results in the same outcome, no net information is gained. Another way of understanding this definition of entropy is in terms of the resources needed to store information. For a 50-50 fair coin, we need at least 1 bit per toss to store this information. For an unfair coin as shown in eq. (3), we need at least 0.8113 bit per toss to store the information.

\begin{figure}[b!]
\centering\includegraphics[scale=0.6]{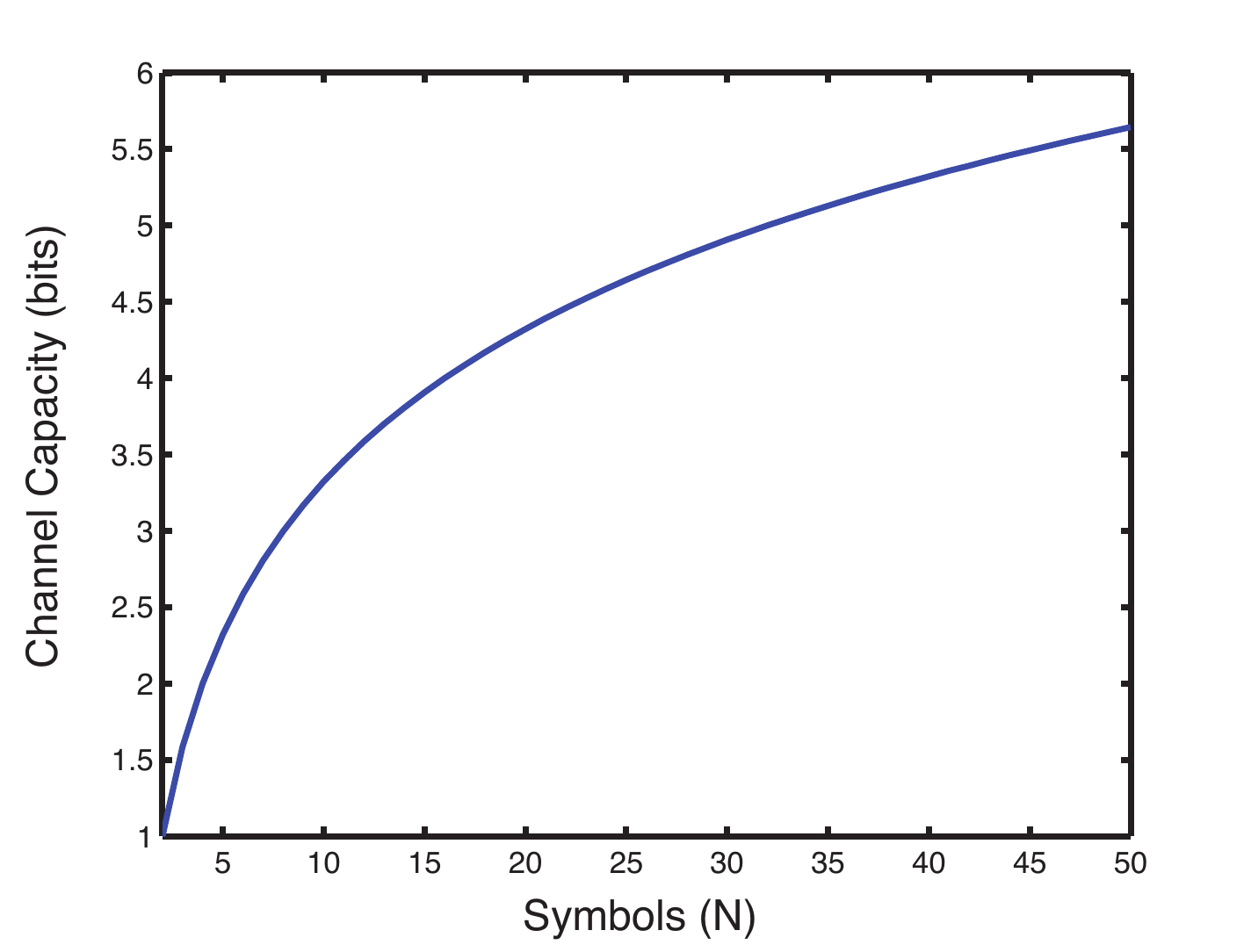}
  \caption{Channel capacity of a communication channel as a function of the number of symbols $N$.}\label{channelcap}
\end{figure}

Now lets extend this idea to a communication channel. Imagine a channel where the sender encodes a message by picking ``heads" or ``tails" on a coin and then sending the coin to the receiver. A completely random message can be thought of as the result of many tosses of a fair coin. A communication channel employing such a 2-symbol encoding can then at most carry 1 bit of information. If the channel is biased towards one result (i.e. the coin is unfair), the amount of information that can be carried by this channel is reduced from 1 bit. For now, lets consider an ideal channel that employs an N-symbol encoding, with each symbol being equally likely to occur ($p_n=1/N$). The maximum amount of information that such a channel can carry is given by simplifying Eq.~\ref{shannon}:

\be H(X) = -\sum_N \frac{1}{N} \log_2\bigg(\frac{1}{N}\bigg) = \log_2(N)\ee

As shown in Fig.~\ref{channelcap}, the channel capacity increases logarithmically as a function of the number of symbols, or the channel dimension, $N$. As mentioned in the introduction above, QKD systems conventionally use the polarization degree of freedom of a photon for encoding. Polarization is inherently a two-dimensional state space, as there are only two orthogonal polarizations in any given polarization basis (for e.g., horizontal and vertical, or left-circular and right-circular). For polarization, the maximum channel capacity is then limited to $\log_2(2) = 1$ bit/photon. However, for an OAM-based QKD system employing 25 OAM modes, the channel capacity is increased to $\log_2(25) = 4.64$ bits/photon, which is almost 5 times the capacity of the polarization-based system!

\begin{figure}[b!]
\centering\includegraphics[scale=0.6]{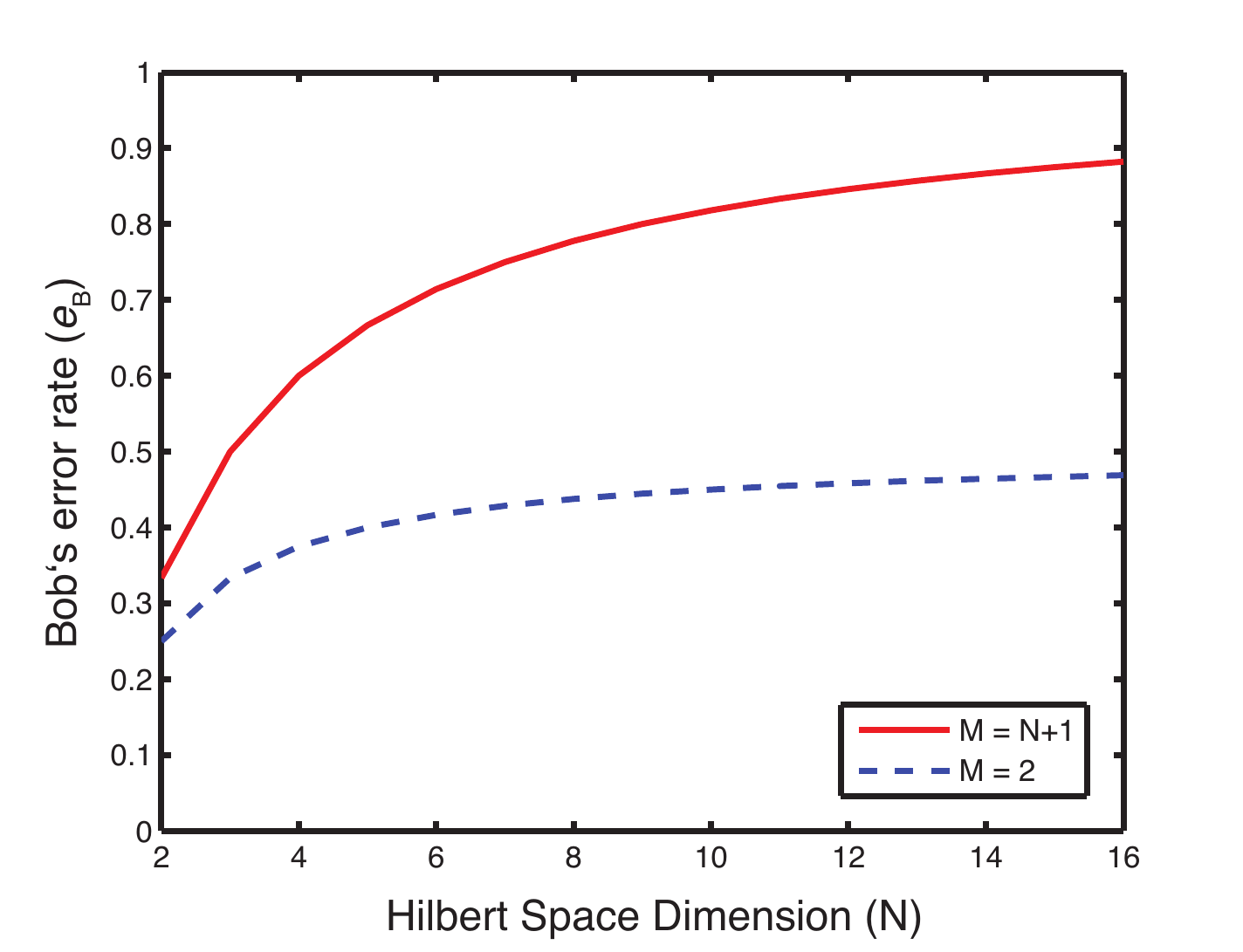}
  \caption{Bob's allowed error rate for an intercept-resend eavesdropping attack as a function of system dimension $N$ for $M=2$ MUBs (dashed blue line) and the maximum of $M=N+1$ MUBs (solid red line).}\label{errorrateIR}
\end{figure}

\subsubsection{Enhanced Security in QKD}

As explained in Section \ref{chap:QTT}, a QKD link between two parties (Alice and Bob) is susceptible to eavesdropping. However, due to the quantum no-cloning theorem \cite{Wootters:1982ex}, an eavesdropper (Eve) cannot perfectly replicate a quantum system without destroying it. Thus, an eavesdropper using the simplest form of eavesdropping---intercept and resend---will introduce statistical errors in the channel that can be measured by Alice and Bob. For this reason, Alice and Bob must attribute all errors in their channel to Eve. If their measured error rate is equal to or higher than that expected from an eavesdropper using a known method of eavesdropping, their protocol is no longer secure and they must abandon it.

In Section \ref{chap:QSS}, we discuss the error bound for a polarization-based quantum-secured imaging system in detail ($e_{\textrm{B}}<25\%$). Similarly in polarization-based QKD, if Alice and Bob measure an error rate greater than or equal to 25\%, they must abandon their protocol. However, it is important to note that this error rate was derived for a QKD system using two mutually unbiased bases (MUBs). In general, the maximum number of MUBs in an $N$-dimensional QKD system is equal to $N+1$, for when $N$ is a prime number \cite{Wootters:1989ba}. For the polarization-based implementation of the BB84 protocol \cite{Bennett:1984wv}, $N$ is equal to 2 and there are three available MUBs --- the horizontal-vertical (HV) basis, the diagonal-anti-diagonal (DA) basis, and the left-circular-right-circular (LR) basis. A polarization-based QKD system can use all three MUBs for encoding. Such a protocol is referred to as the ``six-state protocol" \cite{Bruss:1998gg}. The use of three instead of two MUBs has two effects. First, the data rate drops by 50\%. This is because Alice and Bob will now prepare and measure in the same MUB only 1/3 of the time (as opposed to 1/2 the time) and will discard 2/3 of the data in the sifting process. Second, the error bound increases from 25\% to 33\%. This is because Eve has a higher probability of measuring in the wrong MUB now that there are three MUBs, and hence has a higher probability of introducing errors in the transmission.

\begin{figure}[t!]
\centering\includegraphics[scale=0.6]{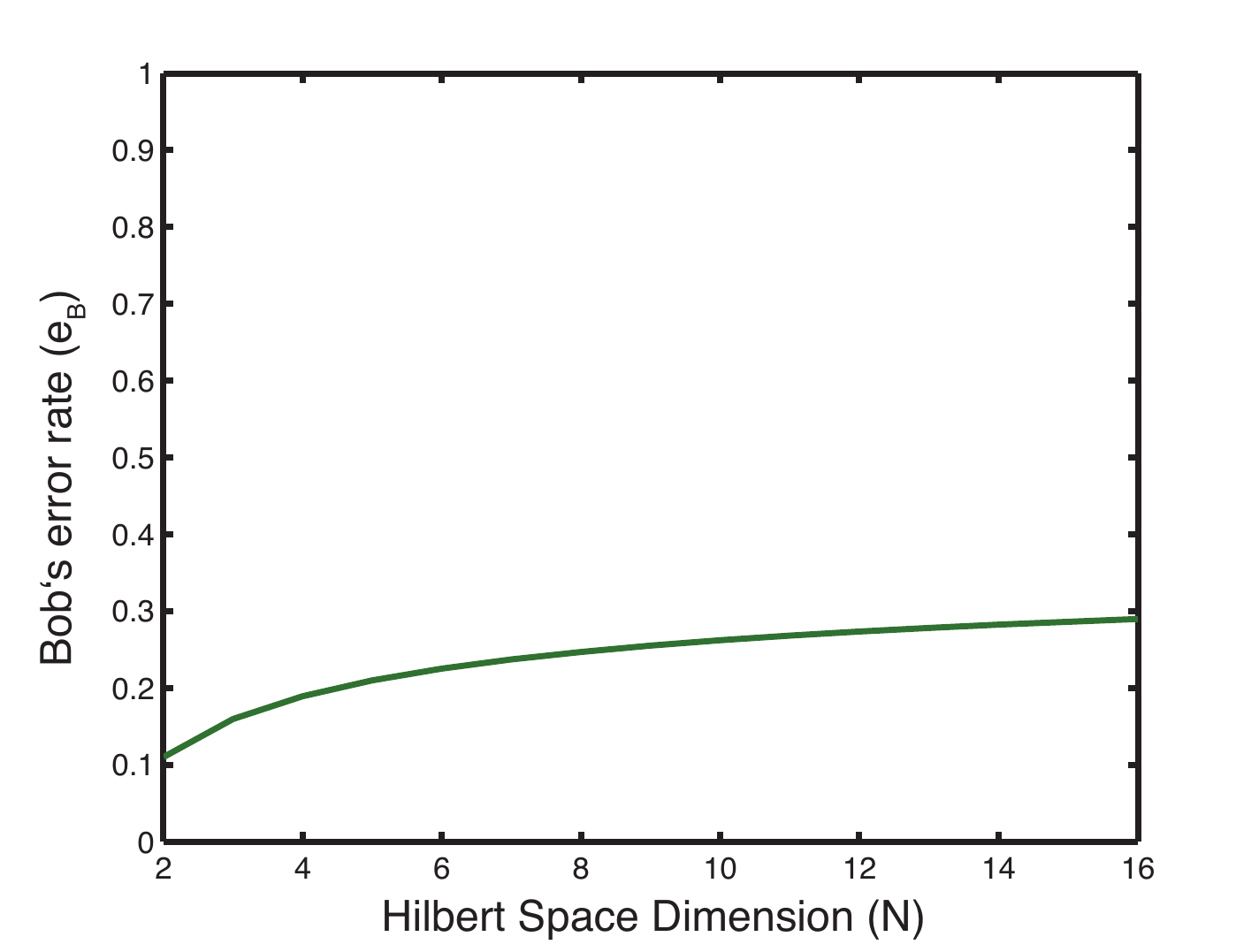}
  \caption{Bob's allowed error rate for finite coherent eavesdropping attacks as a function of system dimension $N$. The allowed error rate is independent of the number of MUBs used.}\label{coherent}
\end{figure}

In general, the error bound for an intercept-resend attack in an $N$-dimensional system with $M$ MUBs in given by \cite{Bourennane:2002uo}:

\be e_{\textrm{B}}(N,M)=\bigg(1-\frac{1}{M}\bigg)\bigg(1-\frac{1}{N}\bigg).\ee

\noindent Using this equation, we plot the error bound for an intercept-resend eavesdropping attack as a function of system dimension for $M=2$ MUBs (dashed blue line) and the maximum of $M=N+1$ MUBs (solid red line) in Fig.~\ref{errorrateIR}. It is clear that the allowed error rate goes up markedly with system dimension. For very large $N$, the error rate goes to 0.5 and 1.0 asymptotically for these two cases. Clearly, using more MUBs is beneficial for security, but has an adverse effect on data rates.

While the allowed error rate can be quite high for intercept-resend attacks on high-dimensional QKD systems, a stricter bound is imposed on the error rate in the case of finite coherent eavesdropping attacks. In these attacks, Eve coherently manipulates a finite number of qudits in order to gain information about the key \cite{Scarani:2009kj}. While the details of such attacks are outside the scope of this article, the error introduced by them on a QKD system follows the inequality \cite{Bourennane:2002uo}:

\be (1-e_{\textrm{B}})\log(e_{\textrm{B}})+e_{\textrm{B}}\log\bigg(\frac{e_{\textrm{B}}}{N-1}\bigg)>-\frac{1}{2}\log(N).\ee

\noindent Notice that in contrast with intercept-resend attacks, the error rate for coherent attacks depends only on system dimension $N$ and is independent of the number of MUBs, $M$. This equation can be numerically solved to produce values of the error bound $e_{\textrm{B}}$ as a function of system dimension $N$. We used the ``FindInstance" function in Mathematica to find values of the error bound, which are plotted in Fig.~\ref{coherent}. As can be seen, the allowed error rate for coherent attacks is indeed much stricter than that allowed for intercept-resend attacks. However, even in this case, there is a clear increase in the allowed error rate for larger system dimensions, $N$. For example, the allowed error rate for 16 modes is equal to 0.29, as opposed to 0.11 for 2 modes. This serves as ample motivation for using a high-dimensional encoding scheme for QKD such as that of OAM. 

\subsection{Generating OAM and ANG Modes}

Since OAM modes have a helical phase, a straightforward way of generating beams carrying OAM is by using a phase plate whose optical thickness varies in a similar fashion. These so called ``spiral phase plates" are commercially available today but are quite expensive, costing upwards of a thousand dollars per plate. This is because of the high precision required to manufacture them. In order to create an $\ell=\pm1$ spiral phase plate, for example, one must create a refractive index variation in glass that varies as $\lambda\theta/\pi$, where $\theta$ is the azimuthal position. Due to this manufacturing difficulty, there has been increased interest in the development of other techniques for generating OAM modes. Here we describe the technique of holography, which is the one we use in our lab.

\begin{figure}[t!]
\centering\includegraphics[scale=0.63]{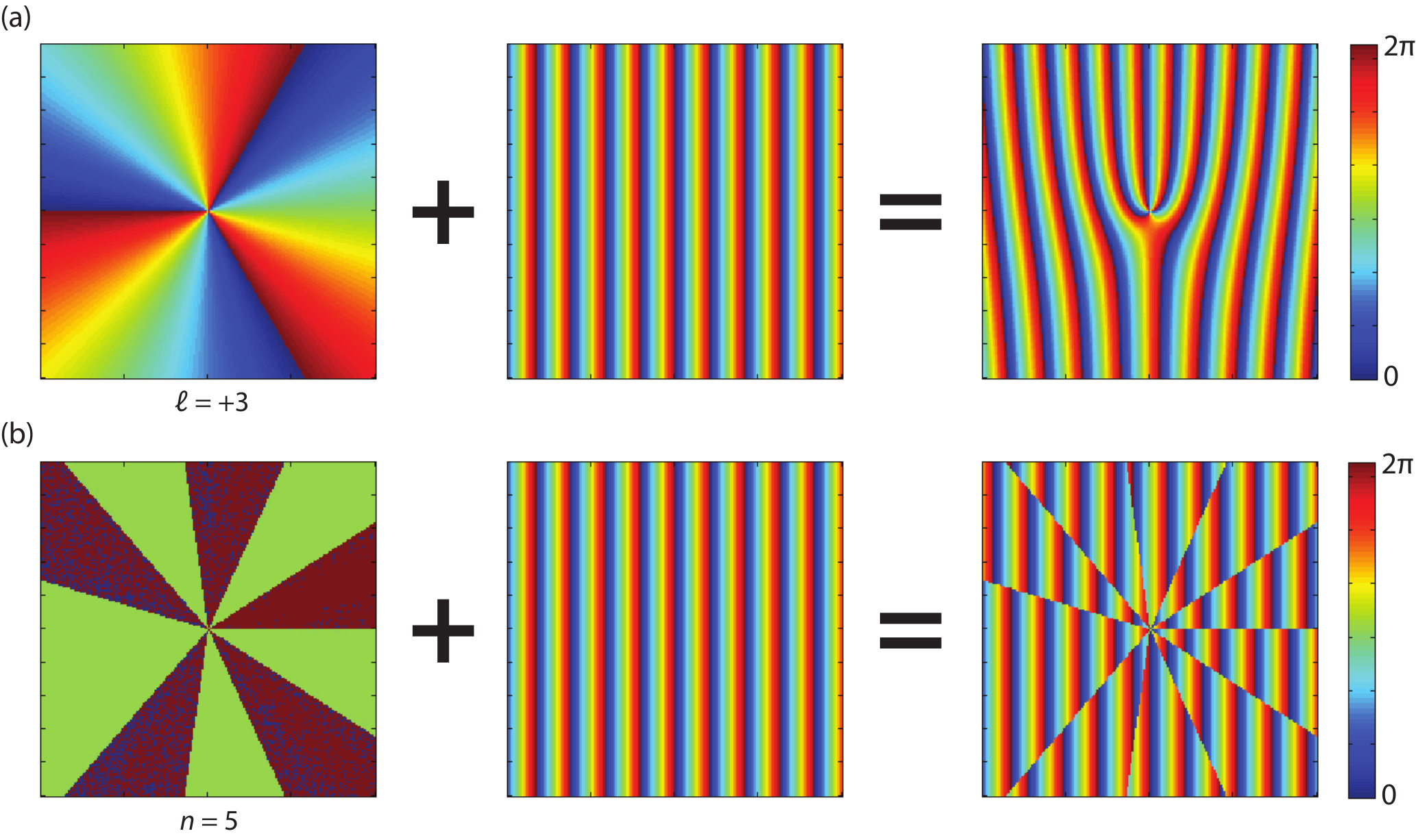}
  \caption{Phase profiles showing the addition of (a) an OAM mode with $\ell=+3$ with a plane wave mode generating a forked hologram and (b) an ANG mode with $n=5$ ($N=11$) with the same plane wave mode generating an ANG hologram. Holograms like these are implemented on spatial light modulators (SLMs) to generate arbitrary superpositions of OAM and ANG modes.}\label{Holograms}
\end{figure}

As explained in Section \ref{chap:QGI}, a hologram is formed by interfering two wavefronts. One of these wavefronts is usually a plane wave incident at a particular angle, and is conventionally referred to as the reference beam. The second wavefront can take on any structure. The interference pattern between these two wavefronts is written onto a holographic material, which is then developed to form a permanent hologram. In Section \ref{chap:QGI}, we used such a hologram as an image sorter. In this process, the second (more complicated) wavefront is sent through the developed hologram, producing a wavefront propagating in the direction of the original reference beam. One can easily flip this procedure around and use the same hologram to create the second complicated wavefront. This is carried out by sending a plane wave at the exact angle of the original reference beam, which interferes with the hologram to create the original, complicated wavefront. 

\begin{figure}[t!]
\centering\includegraphics[scale=0.5]{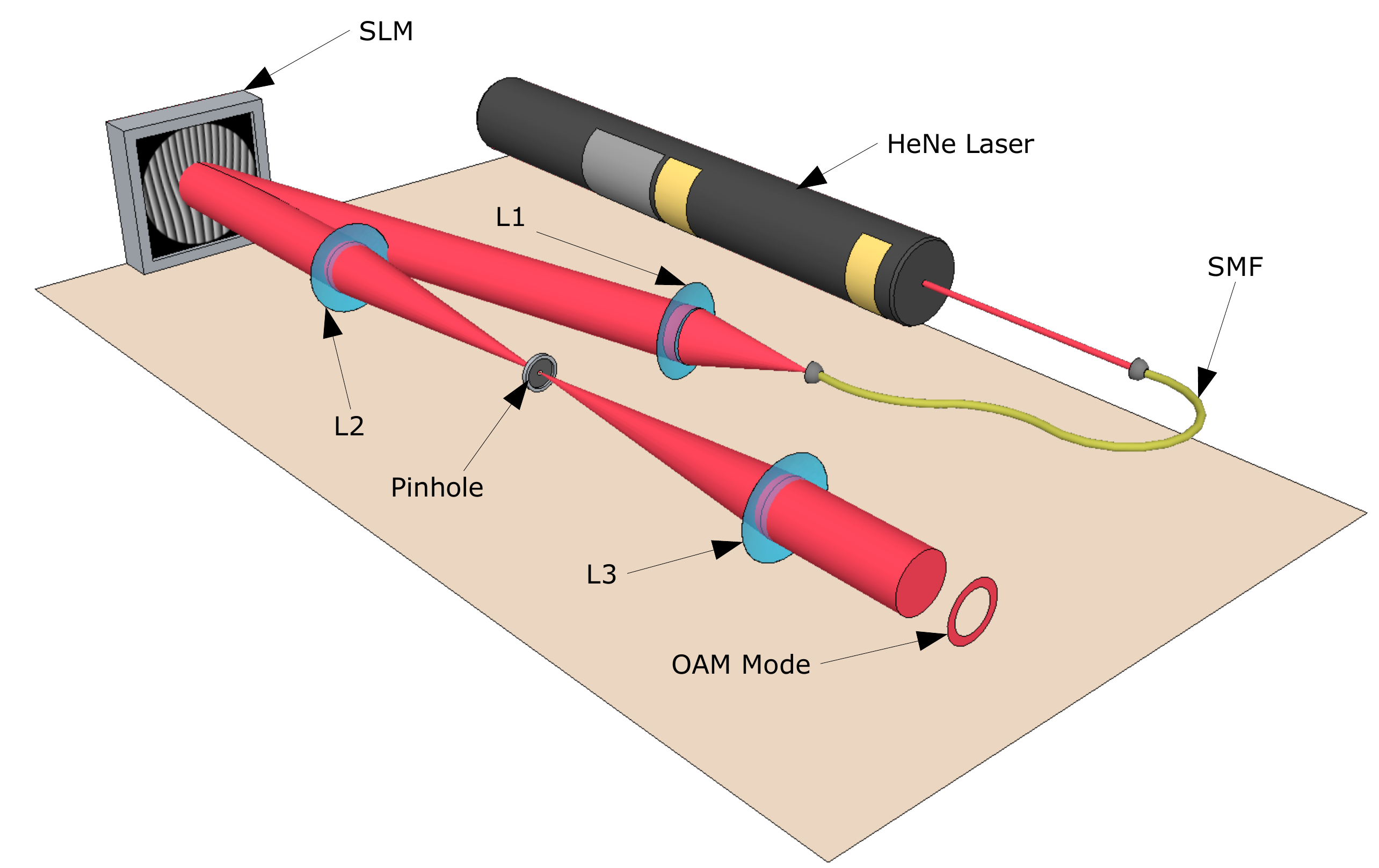}
  \caption{System used for generating an arbitrary superposition of OAM modes. A spatially filtered and collimated HeNe laser beam is incident on a ``forked" diffraction hologram implemented on an SLM. A $4f$ system of lenses (L2 and L3) along with a pinhole is used to remove background noise from the SLM. The OAM mode is obtained in the image plane of the SLM.}\label{OAMgen}
\end{figure}

In this manner, we can create a hologram that can be used for generating an arbitrary superposition of OAM modes. In Fig.~\ref{Holograms}(a), we show the phase profiles of an $\ell=+3$ OAM mode, a plane wave, and the their sum (mod $2\pi$). The OAM mode phase winds around the center of the beam with three $2\pi$ jumps, as expected. The phase of the plane wave mode resembles a linear grating. The combined phase shows a peculiar phase structure at its center. This is commonly referred to as a ``forked" hologram. First proposed in 1992 \cite{Heckenberg:1992vn}, it is a standard method for generating beams with phase singularities or OAM beams. The number of dislocations in the fork corresponds to the azimuthal quantum number of the OAM mode. For a negative OAM mode, the fork is upside down. When a plane wave is incident on such a hologram at the angle of the original plane wave or ``blaze" of the hologram, an OAM mode with $\ell=3$ is generated propagating normal to the hologram. A similar hologram can be used for generating angular position (ANG) modes, which are simply a complex superposition of OAM modes that resemble a wedge rotating around the center of the beam (see Section \ref{chap:Key} for a detailed discussion). The set of ANG modes formed by combining 11 OAM modes is given by

\be \Theta_n=\frac{1}{\sqrt{11}}\sum_{\ell=-5}^5\Psi_{\ell}\exp{\bigg(\frac{i2\pi n\ell}{11}\bigg)}.\ee

\begin{figure}[t!]
\centering\includegraphics[scale = .75]{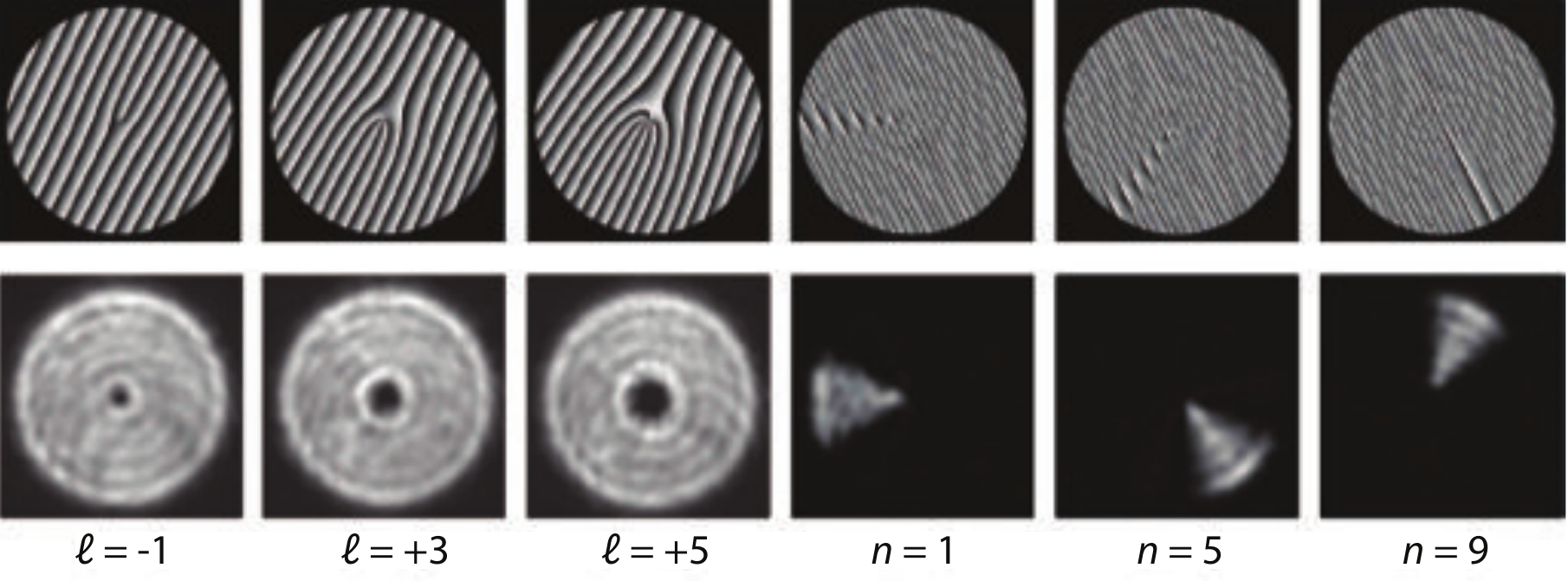}
\caption{Upper row: Holograms written onto the SLM to generate OAM and ANG modes. Lower row: CCD images of the corresponding modes generated. These are OAM mode numbers $\ell=-1$, 3, and 5, and ANG mode numbers $n=1$, 5, and 9 (Figure redrawn from Ref. \cite{Malik:2012ka}, copyright 2012 The Optical Society).}\label{SLMex}
\end{figure}

\noindent As can be seen above, this set is formed by coherently adding OAM modes with an azimuthal quantum number $\ell\leq\pm5$. One should note that the coefficient for each OAM mode in this superposition is equal, which is what makes the basis of ANG modes mutually unbiased with respect to the OAM basis. Specifically, if an ANG mode photon is measured in the OAM basis, it has an equal probability to appear in any of the component OAM modes, and vice versa. A hologram used for generating an ANG mode with $n=5$ is shown in Fig.~\ref{Holograms}(b). 

In our experiment, we generate OAM and ANG modes by implementing such holograms on a Holoeye PLUTO phase-only spatial light modulator (SLM) in conjunction with a $4f$ system of lenses \cite{Arrizon:2007wl, Gruneisen:2008bu}. A schematic for this system is shown in Fig.~\ref{OAMgen}. A HeNe laser is spatially filtered through a single mode fiber (SMF) and collimated by a lens (L1). This collimated, Gaussian beam is incident on an SLM with a forked diffraction grating as shown in Fig.~\ref{Holograms}(a). The beam diffracts off the SLM and is sent through a $4f$ system of lenses (L2 and L3) with a pinhole in between the two lenses. The purpose of this pinhole is to pick out the first diffracted order of the blazed hologram. The reason for doing so is that the zeroth order contains a lot of noise in the form of unwanted reflections from the SLM. A primary contributor to this is the periodic gap between SLM pixels, which also acts like a diffraction grating. By blazing the hologram in both $x$ and $y$ and picking off the first-order of diffracted light in the Fourier plane, we eliminate this background noise \cite{Arrizon:2007wl}. Figure \ref{SLMex} shows some of the holograms we use in our setup and CCD images of the OAM and ANG modes generated by them.

\subsection{Sorting OAM and ANG Modes}

One of the key hurdles to using OAM modes to perform QKD has been the need of a method of efficiently sorting single photons carrying OAM modes. This problem has eluded the scientific community for over a decade. A standard method for measuring the OAM content of a photon has been to project out each OAM mode. This procedure simply involves using the forked diffraction grating backwards (similar to the image sorter in Section \ref{chap:QGI}). An OAM mode with $\ell=+2$ incident on a forked diffraction grating with $\ell=-2$ will generating a plane wave (with $\ell=0$) traveling in a specific direction. When sent through a lens, this plane wave will produce a peak at $p=0$. Any other OAM mode with $\ell'\neq+2$ sent through the same forked diffraction grating will \textit{not} produce a plane wave, instead generating an OAM mode with $\ell''=\ell'-2$. This OAM mode, when Fourier transformed by a lens, will have a null at $p=0$ (the Fourier transform of an OAM mode is also an OAM mode). By placing a small aperture or fiber at the focus of this lens, a forked hologram can be used to test for a particular OAM mode. However, this procedure destroys the photon under test, and is thus limited to an efficiency of $1/N$, where $N$ is the number of OAM modes to be measured.

\begin{figure}[b!]
\centering\includegraphics[scale=0.48]{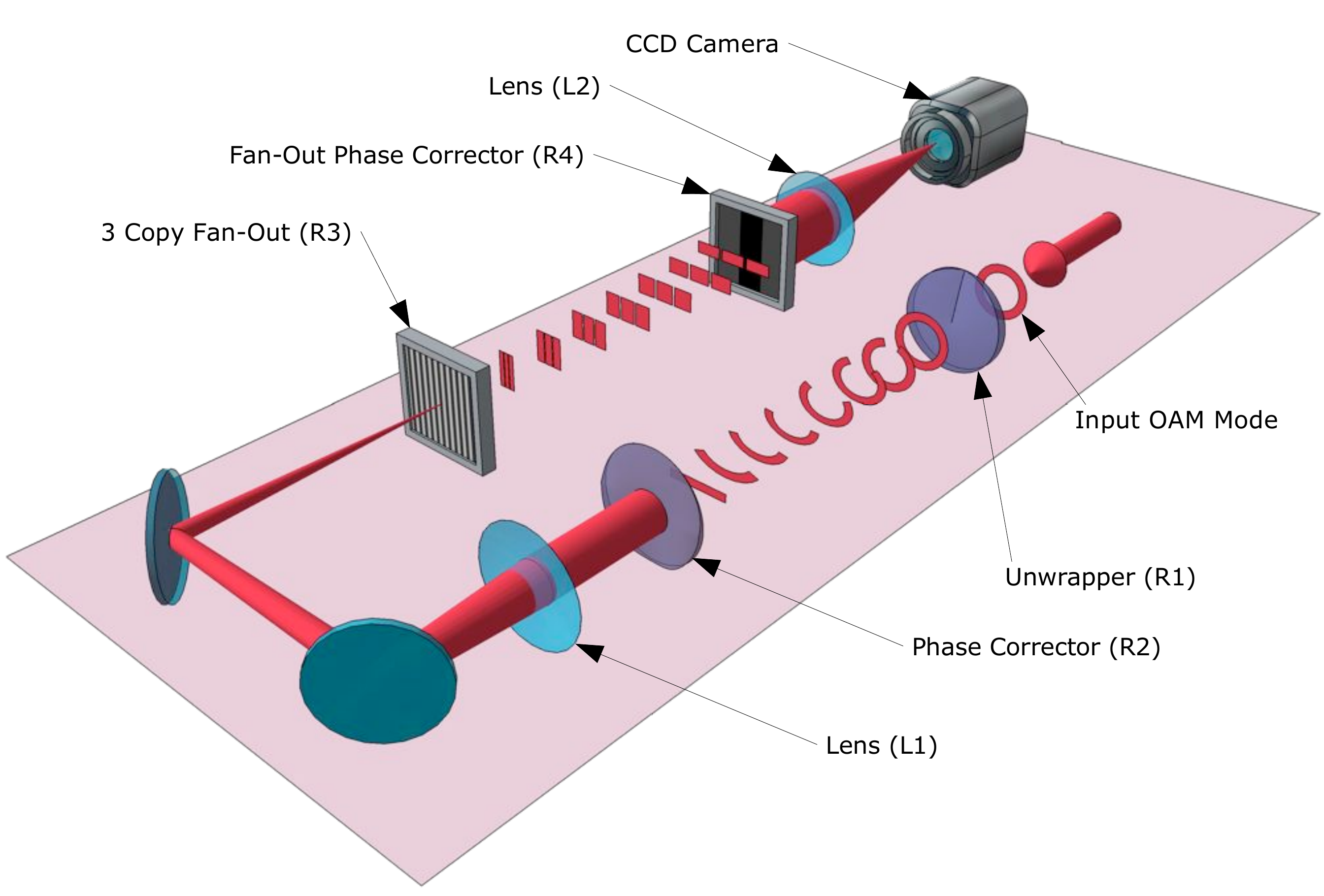}
  \caption{A schematic illustrating the OAM mode sorting procedure. The unwrapper element (R1) unwraps the helical phase of an input OAM mode, transforming it into a finite-sized plane wave mode with a tilt. The phase corrector element (R2) removes residual aberrations introduced during the mode transformation. A lens (L1) converts the tilted plane wave modes into somewhat spatially separated position modes. A fan-out element (R3) implemented on an SLM creates 3 adjacent copies of the finite plane wave mode, albeit with a phase offset between them. A final fan-out phase corrector element (R4) removes this phase offset between the copies. A lens (L2) then focus these larger plane wave modes into well separated position modes at the CCD camera (Figure adapted from Ref. \cite{Mirhosseini:2013em}).}\label{FanOut}
\end{figure}

The first method for efficiently sorting photons carrying OAM used a set of rotated Dove prisms in the arms of a Mach-Zehnder interferometer (MZI) \cite{Leach:2002wy}. In this interferometric method, photons with even $\ell$ were obtained from one MZI port, while photons with odd $\ell$ were obtained from the other. By cascading such OAM ``parity" checking MZIs, and cleverly inserting OAM parity shifting spiral phase plates (or holograms) between them, this method could, in principle, be used to efficiently sort single photons carrying OAM. However, the problems associated with this method are almost obvious --- besides the problem of scaling to large OAM dimensions, the use of many optical components would reduce the efficiency of the method by absorption of photons. 

\begin{figure}[t!]
\centering\includegraphics[scale=.9]{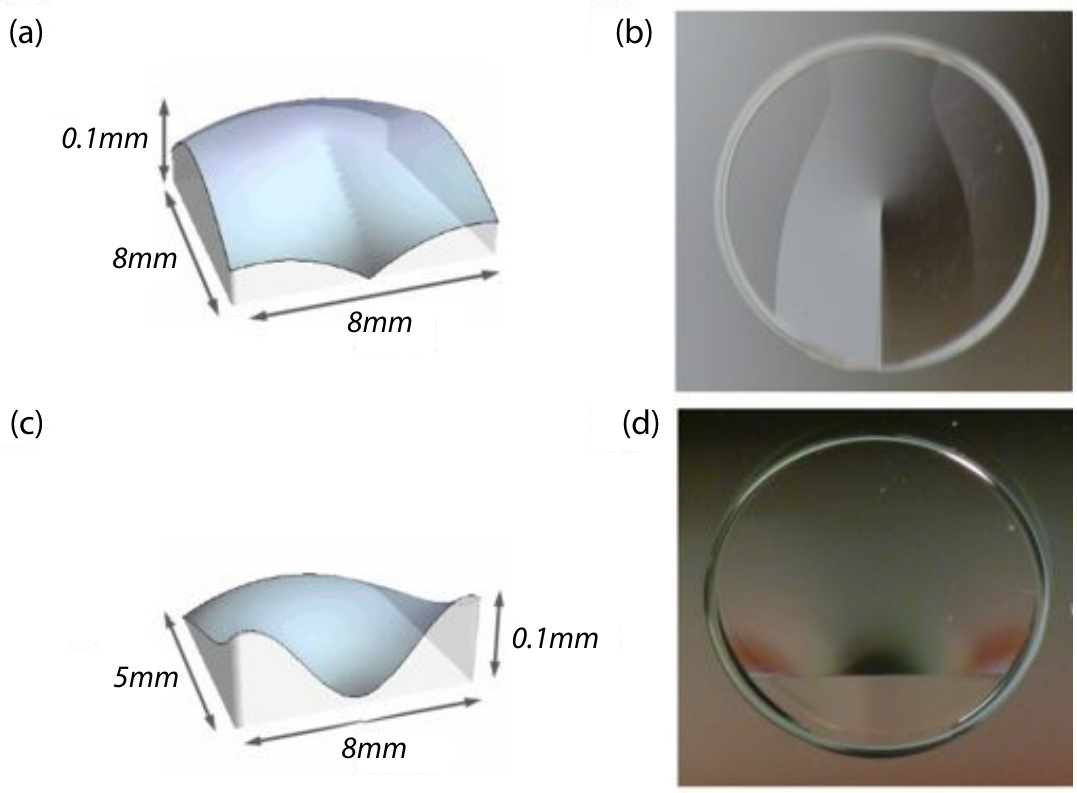}
  \caption{(a) and (c) Schematics showing the optical thickness of the two elements R1 and R2 as a function of position. (b) and (d) Photographs of the two elements R1 and R2 used in our setup, machined out of PMMA (Figure redrawn from Ref. \cite{Lavery:2012hi}).}\label{PMMA}
\end{figure}

Clearly, a more elegant solution was required, which was introduced recently by a method that uses a geometric transformation to convert an OAM mode with an azimuthal phase variation $e^{i\ell\phi}$ to a tilted plane wave mode with a position phase variation $e^{i\ell x}$ \cite{Berkhout:2010cb}. The tilt of the plane wave mode is proportional to the OAM quantum number $\ell$ of the OAM mode. In this manner, OAM modes can be sorted by first converting them into tilted plane waves, and then Fourier transforming the plane waves into separated position modes. Two custom refractive elements \cite{Lavery:2012hi} are used to optically map polar coordinates $(r,\varphi)$ in the input plane to rectilinear coordinates in the output plane $(x,y)$ via the log-polar mapping $x=a(\varphi \bmod{2\pi})$ and $y=-a\ln (r/b)$. Here, $a$ and $b$ are scaling constants that define the size of the converted mode \cite{Bryngdahl:1974tu}. The first element, the {\it unwrapper} (R1), maps intensities according to the coordinate transformation. A second element, the {\it phase corrector} (R2), corrects a residual aberration. Thus, optical waves with helical phase fronts are transformed into tilted plane waves, which can be sorted at the focus of a lens. This process is illustrated in Fig.~\ref{FanOut} for one input OAM mode. 

The two custom elements used in our setup were diamond machined with a Nanotech 3 axis ultra precision lathe in combination with a Nanotech NFTS6000 fast tool servo \cite{Lavery:2012hi}. The program used for the machining was written with DIFFSYS, which is a commercially available software. The program converted the input data given by a set of Cartesian coordinates ($x,y,z$) into files usable by the lathe and servo tools. The optical thickness of the first, unwrapping element can be written as a function of $(x,y)$ as

\be Z_1(x,y) = \frac{a}{f(n-1)}\bigg[y\,\textrm{arctan}(y/x)-x\ln(\sqrt(x^2+y^2)/b)+x-\frac{1}{2a}(x^2+y^2)\bigg].\ee 

\noindent Here, $f$ is the focal length of the lens integrated into both elements. This lens performs the Fourier transform operation that is required between the unwrapping and phase correcting procedures \cite{Berkhout:2010cb}. The two free parameters, $a$ and $b$, dictate the size and position of the transformed beam. The optical thickness of the second, phase correcting element can be similarly written as

\be Z_2(x,y) = -\frac{ab}{f(n-1)}\bigg[\exp\bigg(-\frac{u}{a}\bigg)\cos\bigg(\frac{v}{a}\bigg)-\frac{1}{2ab}(u^2+v^2)\bigg].\ee

\noindent Here, $u$ and $v$ are spatial Cartesian coordinates in the output plane. The distance between these two elements must be exactly $f$, and the elements must be aligned precisely along the same optical axis. For this reason, they are mounted in a cage system with fine position and rotation controls. A schematic of the optical thickness in 3D as well as photographs of the elements used in our setup is shown in Fig.~\ref{PMMA}.

While this method is substantially better at sorting OAM modes than previous methods, it is still limited to working approximately 80\% of the time \cite{Lavery:2012hi}. In other words, for a photon with OAM $\ell\hbar$, there exists an approximately 20\% probability of detecting it with OAM $m\hbar,\, m\neq\ell$. This is because the ``unwrapped" plane wave has a finite extent, which results in a diffraction limited spot at the focus of a lens. These spots have about 20\% overlap with neighboring spots, and hence about 20\% crosstalk. Clearly, this is not good enough for QKD, as any errors must be attributed to an eavesdropper. If we get an error 20\% of the time, this already places a strict bound on the allowed environmental error our system can handle. Further, it reduces the benefits of going to a higher dimensional state space. 

\begin{figure}[t!]
\centering\includegraphics[scale=0.48]{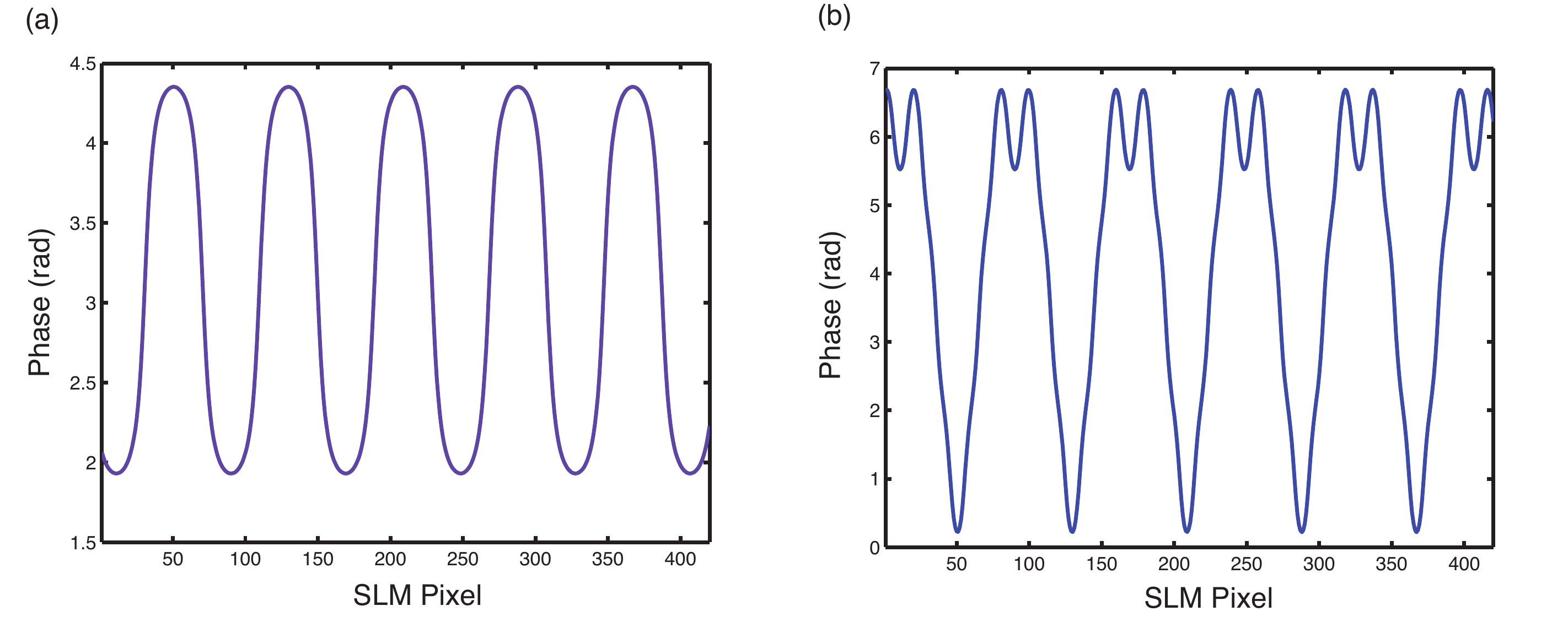}
  \caption{One-dimensional phase profiles of the fan-out holograms used for creating (a) 3 copies and (b) 9 copies. The profiles show a section (420 pixels) of the SLM.}\label{FanoutX}
\end{figure}

In two recent papers \cite{Mirhosseini:2013em, OSullivan:2012gj}, we showed that the technique of Berkhout et al. \cite{Berkhout:2010cb} can be combined with a holographic beam-splitting technique to sort OAM modes with only about 5\% crosstalk. The principle behind our method is straightforward --- by generating multiple, adjacent copies of the transformed plane wave mode, we increase its effective size. When this larger plane wave is sent through a lens, it is Fourier transformed into a smaller spot than before. More specifically, the spot size is reduced by $N$, where $N$ is the number of copies. The \textit{fan-out} element introduced in Ref. \cite{Prongue:1992uj} is a phase grating designed to diffract an incoming beam into $N$ uniformly spaced orders, each having the same spatial profile and equal energy. For perfect beam splitting, an optical element has to transform an incoming plane wave into a field distribution given by

\be
U(x,y)= \sum_{m=1}^{N} A_m e^{i\phi_m}e^{-i 2\pi s_m x/\lambda},
\ee

\begin{figure}[t!]
\centering
\includegraphics[width=13cm]{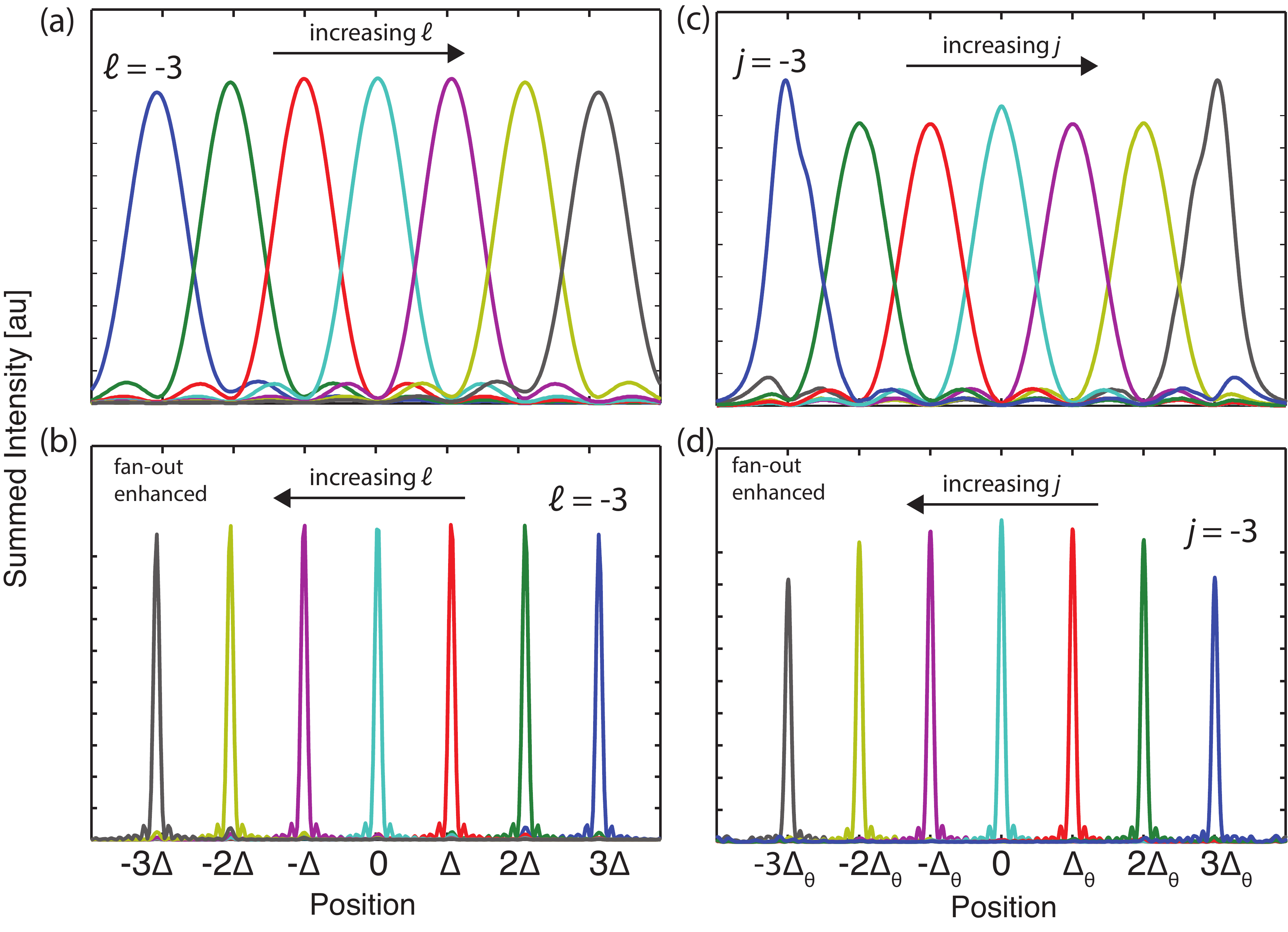}
\caption{Simulation results comparing (a) the output from the OAM sorter with (b) the output from the fan-out-enhanced OAM sorter for 7 input OAM modes and comparing (c) the output from the ANG sorter with (d) the output from the fan-out-enhanced ANG sorter for 7 input ANG modes. Different colors correspond to different modes. The number of copies produced by the fan-out element is 9 (Figure redrawn from Ref. \cite{OSullivan:2012gj}, copyright 2012 The Optical Society).}
\label{OAMANG}
\end{figure}

\noindent where \(A_m\)  is the amplitude, \(\phi_m\) is the phase, and \(s_m\) is the angle of propagation of the \(N\) copies. The fan-out element (R3) is the optimal design in the family of phase-only holograms which can approximately achieve this task \cite{Romero:2007vi}. Generally, the fan-out element introduces a relative phase \(\phi_m\) between the different copies. These are removed with a \textit{phase-correcting} element (R4) in the Fourier plane of the fan-out element (Fig.~\ref{FanOut}). The multiple copies are then Fourier transformed with a lens to a narrower spot than before. This process is illustrated in Fig.~\ref{FanOut} for a 3 copy fan-out. Using the specific values of \(A_m\) and \(\phi_m\) given in Refs. \cite{Prongue:1992uj,Romero:2007vi}, we can achieve an efficiency of more than 99\% while splitting the beam into nine copies. The one-dimensional phase profiles of the 3 and 9 copy fan-out holograms are plotted in Fig.~\ref{FanoutX}(a) and (b) respectively.

\begin{figure}[t!]
\centering
\includegraphics[width=15cm]{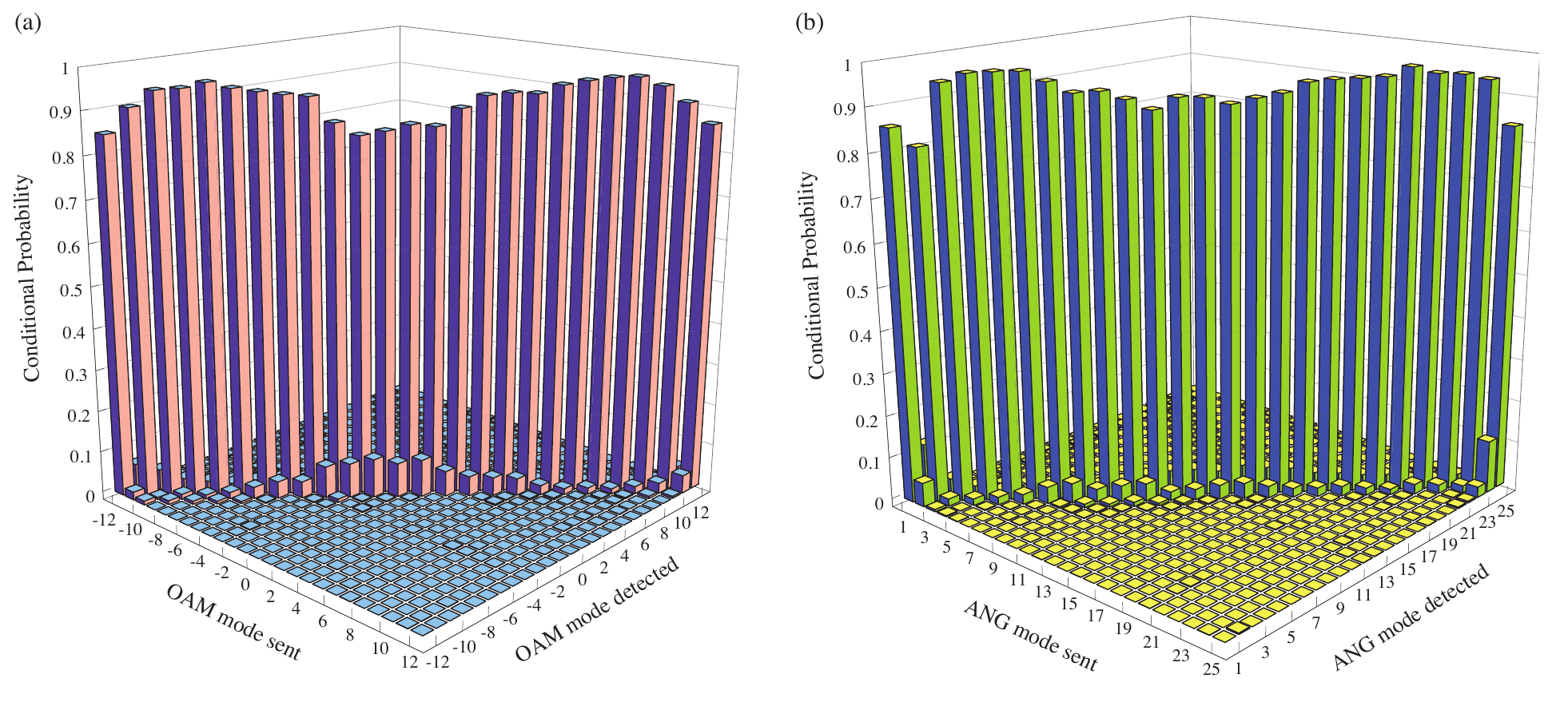}
\caption{Experimental results showing the intermodal crosstalk of a fan-out enhanced OAM and ANG mode sorter. The sorter is tested for 25 OAM modes and 25 ANG modes (Figure adapted from Ref. \cite{Mirhosseini:2013em}).}
\label{Xtalkmatrix}
\end{figure}

The same fan-out procedure can be used for sorting ANG modes as well \cite{Mirhosseini:2013em, OSullivan:2012gj}. In this case, the plane of the first phase correcting element (R2) is imaged onto the plane of the fan-out (R3). Figure \ref{OAMANG} shows simulation results comparing the fan-out enhanced OAM and ANG mode sorter with the previous versions of the sorter without the fan-out \cite{Berkhout:2010cb,Lavery:2012hi}. The decrease in the lateral size of the sorted position modes, and hence the crosstalk, is very clear. We have experimentally tested our sorting method for 25 OAM modes ($\ell=\pm16$) and 25 ANG modes. The crosstalk matrices for both of these cases are shown in Fig.~\ref{Xtalkmatrix}(a) and (b). One can see how the sorting process starts to break down for OAM modes with large $\ell$. For a mode number of $N=25$, we were able to achieve a mutual information of 4.16 bits/pulse in the ANG basis and 4.18 bits/pulse in the OAM basis. The ideal mutual information for 25 modes is equal to $\log_2(25)=4.64$ bits/pulse, which goes to show how close we are to the theoretical limit. In our test, we made measurements at high light levels using a HeNe laser and a Canon 5D Mark III camera. In principle, this can be extended to the single photon level with an appropriate multi-pixel single photon detector.

\subsection{Proposed High-Dimensional QKD Systems}

In this section, we describe two types of OAM-based QKD systems that could be built using the procedures for generating and sorting OAM and ANG modes explained above. The first is based on a high-dimensional version of the BB84 protocol \cite{Bennett:1984wv}, which uses two pairs of orthogonal polarization states in two mutually unbiased bases (MUBs) for encoding. The second is a high-dimensional variant of the Ekert protocol \cite{Ekert:1991kl}, which relies on the quantum correlations between two polarization-entangled photons for security. We also describe the progress we are making towards implementing the BB84-based OAM-QKD system in our lab. In the next section, we discuss the limitations of our current system. 

\subsubsection{BB84 OAM-QKD with Weak Coherent Pulses}
 
\begin{figure}[t]
\centering\includegraphics[scale = .8]{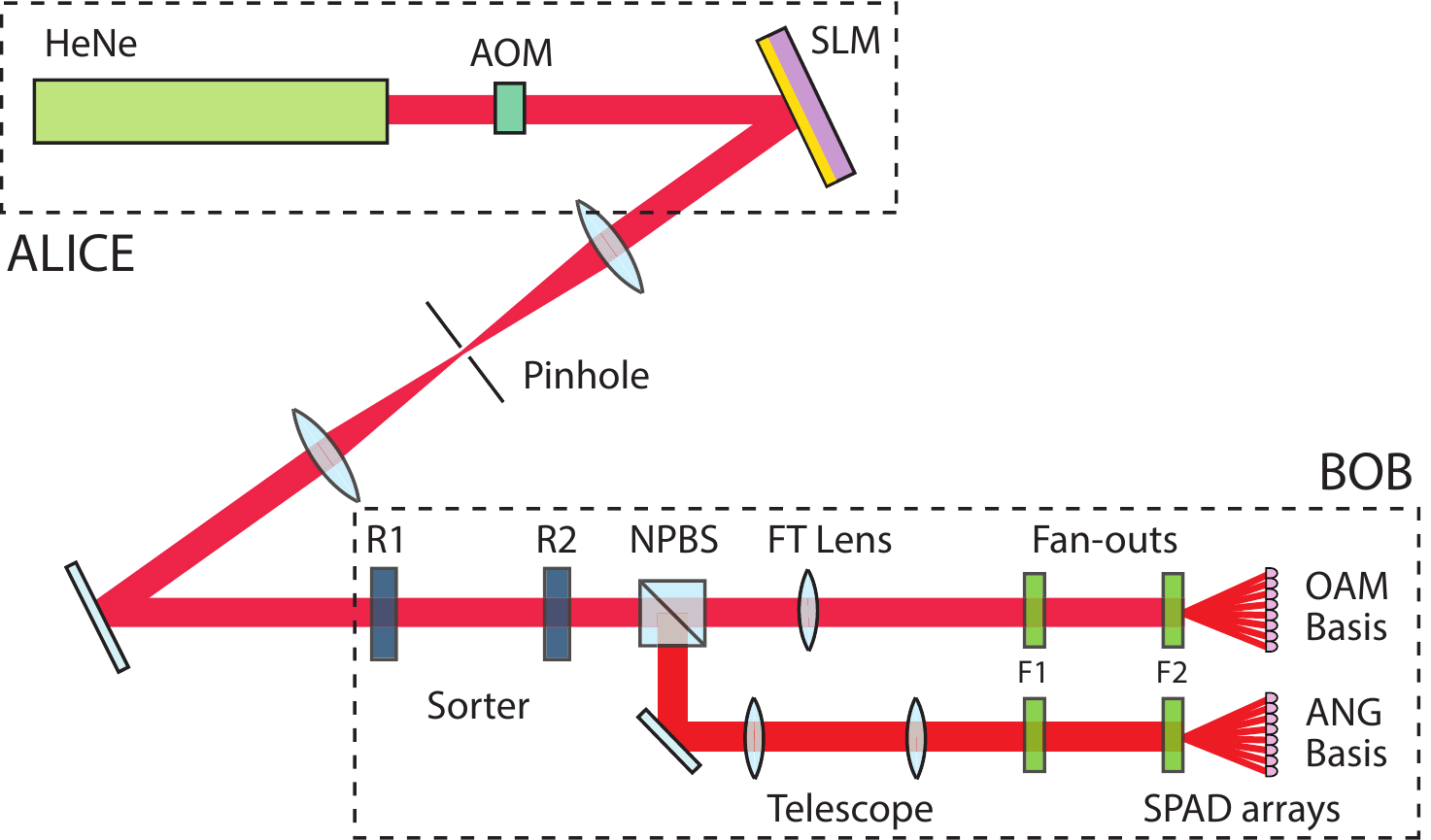}
\caption{Proposed high-dimensional QKD setup using the BB84 protocol. Our source is a HeNe laser operating at 633 nm that is modulated by an acousto-optic modulator (AOM) to carve out pulses containing an average number of photons dictated by the decoy state protocol. These pulses are tailored into OAM and ANG modes by a spatial light modulator (SLM) and $4f$ system. R1 and R2 are custom refractive elements used to transform the OAM and ANG modes into plane wave modes. F1 and F2 are fan-out and phase correcting elements used to enhance the sorting process. A beam splitter (NPBS) acts as a passive basis selector between the OAM and ANG bases. The transformed modes are detected with arrays of single photon avalanche detectors (SPADs).}\label{OAMBB84}
\end{figure}

As explained earlier in this section, using more than two dimensions for encoding in QKD also increases the number of possible MUBs one can use. Using more than two MUBs results in increased security, but a reduced key generation rate. For this reason, we are restricting ourselves to the two high-dimensional MUBs of OAM and ANG, introduced in Section \ref{chap:Key}. A schematic of our proposed QKD system is shown in Fig.~\ref{OAMBB84}. We use a HeNe laser modulated by an acousto-optical modulator (AOM) as our source. By adjusting the duration of the driving pulse, we can use the AOM to carve out pulses of light containing less than one photon on average. Due to the Poissonian statistics followed by coherent states, a highly attenuated laser pulse will always contain more than one photon with some probability. This opens up such a system to eavesdropping using photon-number splitting (PNS) attacks \cite{Lutkenhaus:2002jy}. In the simplest version of the PNS attack, an eavesdropper can insert a beam splitter into the channel and probabilistically split off a photon from pulses containing more than one photon. As all photons in the same pulse encode the same qubit, Eve can gain information about the qubit without destroying it or revealing herself. It is important to note that in general, multi-photon pulses do not necessarily undermine the security of a QKD system. However, they do limit the key generation rate, as more bits must be discarded during the privacy amplification process \cite{SchmittManderbach:2008vm}. 

Recently, a variation to the BB84 protocol was proposed which uses a simple technique to counter PNS attacks. In this technique, known as the decoy state protocol \cite{Lo:2005cd}, Alice prepares an additional set of ``decoy" states by randomly varying the number of photons in each pulse. She also randomly chooses which pulses will be used as signal states and which as decoy states. Thus, both the signal and the decoy states consist of pulses containing a varying distribution of average photon number that is known to Alice. The security lies in the fact that given a single $n$-photon pulse, Eve has no way of knowing whether it originated as a signal or decoy. Thus, any attempt by Eve to remove photons from a pulse will occur with the same probability for a signal as well as a decoy state. However, since these two kinds of states have different photon number statistics, the effect of removing a photon is different on both. By sharing the decoy state information after the sifting process, and measuring the ratio of the number of detection events to the number of signals originally sent for each kind of state, Alice and Bob can detect any PNS attacks by Eve with a high probability. This protocol has been implemented with many different intensities of decoy states \cite{Zhao:2006dc,SchmittManderbach:2007jj}. However, the protocol using two states---the vacuum and weak decoy state---has been shown to be optimal \cite{Ma:2005bd}.

In our proposed QKD system, we modulate the intensities of our pulses according to this protocol. The AOM is used to carve out pulses with varying intensities. Following the AOM, a spatial light modulator (SLM), a pinhole, and a $4f$ system of lenses are used for impressing OAM or ANG mode information onto each pulse (Fig.~\ref{OAMBB84}). The $4f$ system also images the SLM onto Bob's first detection plane at R1. Bob uses the sorting procedure explained earlier in this section to measure a pulse either in the OAM or ANG basis. The non-polarizing beam splitter (NPBS) acts as a passive selector of Bob's measurement basis, randomly measuring pulses in either the OAM or ANG basis. The same mode transforming elements (R1 and R2) are initially used to transform both OAM and ANG modes. Following the NPBS, two sets of fan-out elements (F1 and F2) carry out the beam-copying process for each basis. For the OAM basis, the output of R2 is Fourier transformed by a lens (FT Lens) onto the fan-out element F1. For the ANG basis, the output of R2 is imaged by a telescope system onto the fan-out element F1. Following the phase-correcting elements (F2), two sets of single photon avalanche detector (SPAD) arrays are used for detecting the photon states.

After Bob completes his measurements in the OAM and ANG bases, our high-dimensional protocol follows the standard steps of the BB84 protocol. Alice and Bob share their encoding and measurement basis choices with each other over a public channel. Using this information, they sift out the states where Bob did not measure in the preparation basis used by Alice. Following this, Alice and Bob perform the procedures of error correction \cite{Makkaveev:2005tf} and privacy amplification \cite{Huttner:uz}. Both these procedures merit detailed discussion. However, these topics are outside the scope of this article. After privacy amplification, Alice and Bob will share a secure key with enhanced security and an increased generation rate via the use of a high-dimensional Hilbert space. An experimental implementation of such a QKD protocol that achieved 2.1 bits/photon with 7 OAM and ANG modes was recently published on the arXiv by our group \cite{OAMQKD}.

\subsubsection{Ekert OAM-QKD with Entangled Photons}

The second proposed high-dimensional QKD system is based on an extension of the Ekert protocol \cite{Ekert:1991kl} to a high-dimensional Hilbert space. As explained in Section \ref{chap:QTT}, the security of the Ekert protocol relies on the strong quantum correlations shared by two members of an entangled pair. An eavesdropper trying to access information in this protocol disturbs these correlations, which can be quantified through entanglement measures such as the CHSH inequality \cite{Clauser:1969ff} or the Schmidt number \cite{Ekert:1995vv}. Bennett and Brassard argued that the Ekert protocol was formally identical to the BB84 protocol, and thus entanglement was not necessary to perform QKD. While this is true, the Ekert protocol simply provides an alternative method to do QKD in a different architecture --- the source is spatially separated from Alice and Bob. Also, a subtle yet important difference is that there is no active state preparation in the Ekert protocol. Alice and Bob simply rely on the probabilistic nature of wavefunction collapse to assign a bit value to their measured state. For example, a $D$-polarized photon encountering a polarizing beam splitter probabilistically goes into either the $H$ or the $V$ port. This removes certain technological requirements from Alice, as she no longer needs to employ expensive equipment such as a series of Pockels cells in order to create specific polarization states. On the contrary, of course, the Ekert protocol does require a maximally polarization-entangled state, which is fast becoming available cheaply, and is used even at the undergraduate laboratory level \cite{Galvez:2005tt}.

\begin{figure}[t!]
\centering\includegraphics[scale = .9]{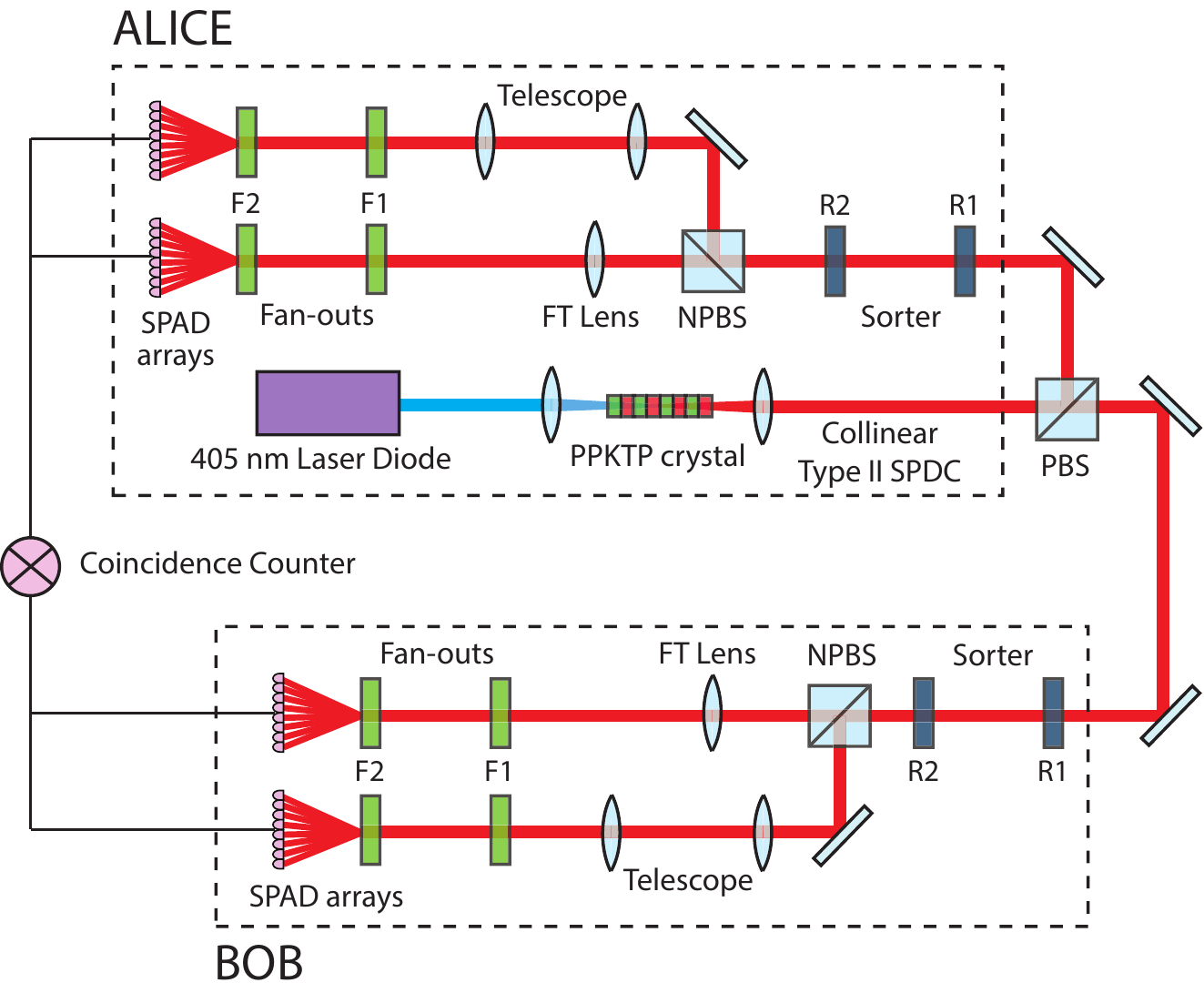}
\caption{Proposed high-dimensional QKD setup using the Ekert protocol. Our source is a diode laser operating at 405 nm that pumps a periodically-poled Potassium Titanyl Phosphate (PPKTP) crystal to generate type II downconverted photons at 810 nm exhibiting OAM entanglement. A polarizing beam splitter (PBS) separates the signal from the idler, directing one to Alice and the other to Bob. Alice and Bob use a similar detection setup as in the BB84 protocol (Fig.~\ref{OAMBB84}). R1 and R2 are custom refractive elements used to transform the OAM and ANG modes into plane wave modes. F1 and F2 are fan-out and phase correcting elements used to enhance the sorting process. A beam splitter (NPBS) acts as a passive basis selector between the OAM and ANG bases. The transformed modes are detected with arrays of single photon avalanche detectors (SPADs). Alice and Bob's SPAD arrays are connected with a coincidence counting circuit to ensure that only photons from the same entangled pair result in a signal.}\label{OAMEkert}
\end{figure}

To perform high-dimensional OAM-based QKD with the Ekert protocol, we require a bright source of photons maximally entangled in OAM \cite{Mair:2001ub}. The state of OAM-entangled photons generated in spontaneous parametric downconversion (SPDC) can be written as

\be \ket{\Psi} = \sum_{\ell=-\infty}^{\infty}c_{\ell}\ket{\ell}_A\ket{-\ell}_B. \ee

\noindent where $\ell$ is the azimuthal quantum number, $c_{\ell}$ is the probability amplitude, and $A$ and $B$ refer to the signal and idler photon, respectively. As can be seen, OAM-entangled photons are anti-correlated in OAM. Thus, if one photon of an entangled pair is measured to have an OAM of $+3\hbar$, its entangled partner photon must have an OAM of $-3\hbar$. For any realistic SPDC source, the OAM bandwidth (spiral bandwidth) does not extend to $\pm\infty$. This is because of the physical apertures in the system and finite size of the SPDC crystal. However, considerable work has been done on tailoring the OAM spectrum for use in quantum information \cite{Dada:2011dn,Pires:2010gy}. In an experimental realization, we plan on using a periodically-poled Potassium Titanyl Phosphate (PPKTP) crystal designed for degenerate, type-II, collinear SPDC. The PPKTP crystal is pumped by a laser diode at 405 nm to produced OAM-entangled photons at 810 nm. This source is based on an OAM-entanglement source used at IQOQI in Vienna \cite{Krenn:2013eq}.

The signal and idler photons are separated from one another by a polarizing beam splitter (PBS) and directed towards Alice and Bob, both of whom use a sorting procedure similar to the one described earlier for BB84-based OAM-QKD. Following the random measurement of their respective photon in either the OAM or the ANG basis, Alice and Bob use coincidence detection to ensure that their photons originated in the same entangled pair. Thus, OAM anti-correlations and ANG correlations between these two photons will ensure that Alice and Bob measure the opposite (for the OAM basis) or the same (for the ANG basis) state at the end. Just as in polarization-entanglement-based QKD, security must be proven by testing for high dimensional OAM-entanglement. This has been performed recently for an OAM dimensionality up to $\ell=11$ by violating the generalized Bell-type parameter $S_d$ by making projection measurements \cite{Dada:2011dn}. Interestingly, we can use our sorting method to calculate this very parameter directly. This is because the crosstalk terms (or the off-diagonal terms in the crosstalk matrix) obtained from the sorting process can be related to the measurements required to obtain $S_d$ \cite{Collins:2002ix,Dada:2011dn}. One should keep in mind that since our sorting process is not entirely perfect (approximately 5\% crosstalk), the entanglement measure will not be entirely accurate either. If the Bell-type parameter $S_d$ is found to be greater than 2, we know that the state is still entangled in OAM and no eavesdropper is present. 

\subsection{Limitations and Outlook}

\begin{figure}[t!]
\centering\includegraphics[scale = .8]{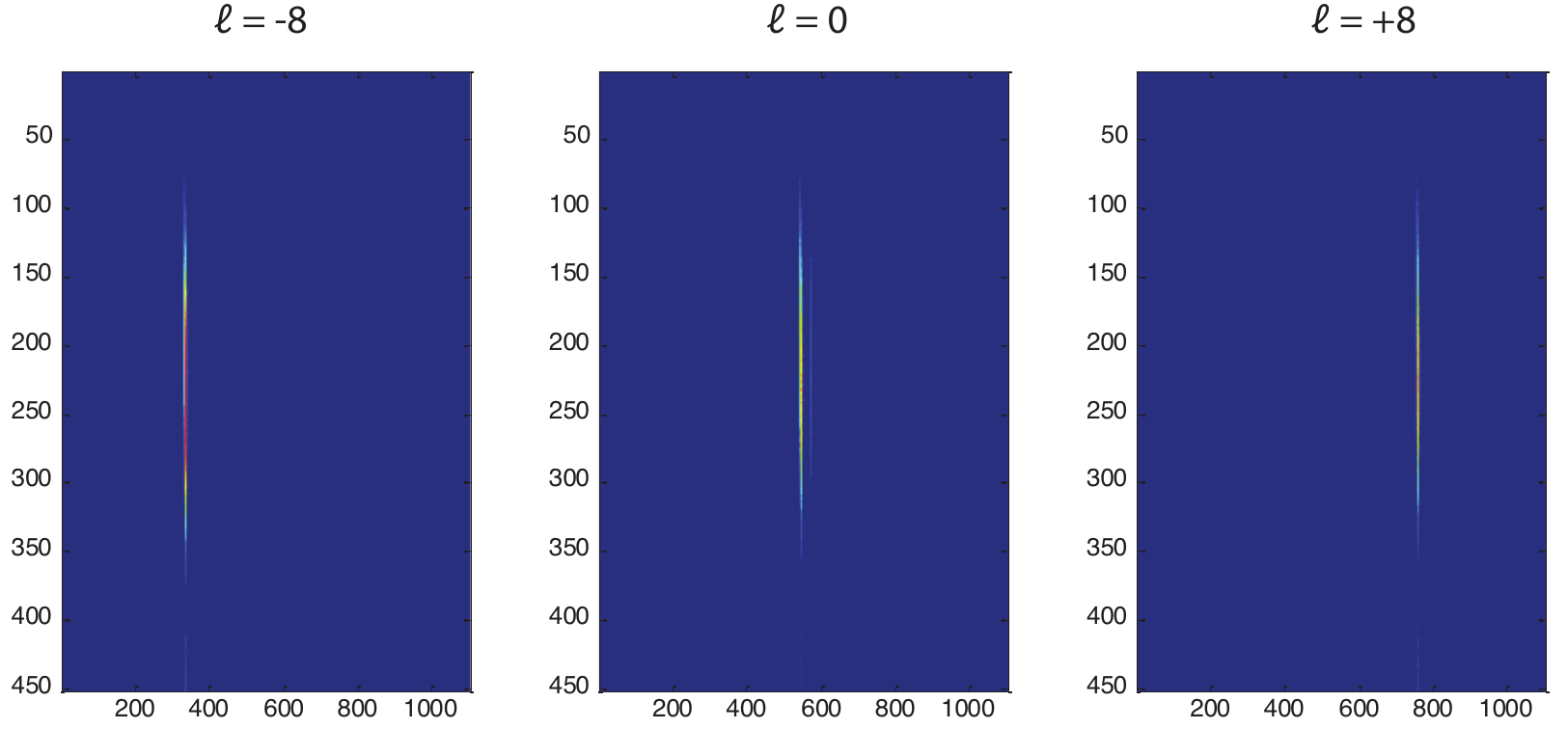}
\caption{A CCD image showing the mode structure of the transformed OAM mode, after passing through the OAM sorter, the fan-out holograms, and a Fourier transform lens. Three transformed OAM modes ($\ell=-8,0,+8$) are shown.}\label{FO_output}
\end{figure}

A chief limitation of our BB84 protocol-based OAM-QKD system is that an SLM is used for the generation of OAM and ANG modes. SLMs have a refresh rate of 60 Hz, which places a strict upper limit to how fast we can generate a key. For comparison, state of the art polarization-based systems have shown key generation rates exceeding 1 Mbit/s \cite{Dixon:2010ev}. Clearly, in order to compete with polarization-based QKD, we need a faster method of generating OAM and ANG modes. A promising option is to use digital micro-mirror devices (DMDs), which are cheaply available and can operate at up to 32 kHz speeds. DMDs are binary amplitude devices that, as the name suggests, rely on tiny mirrors to turn parts of a beam on and off. Using a DMD with a $4f$ system of lenses, one can convert an arbitrary amplitude pattern into a phase pattern \cite{Rodrigo:2006vw}. We have used such a device to generate OAM and ANG modes at speeds of up to 3.2 KHz \cite{Mirhosseini:2013go}. We are planning to replace the SLM currently being used for OAM and ANG state preparation with this method.

Another limitation of our current system lies in our detection system. While in the previous section we have proposed the use of SPAD arrays for detecting the transformed modes, the shape of these modes creates a unique challenge for their detection. As can be seen in Fig.~\ref{FO_output}, after passage through the OAM sorter, the fan-out elements and a Fourier transform lens, an OAM mode resembles a narrow line. Coupling this mode efficiently into a fiber will be a challenge and will perhaps require the use of cylindrical lenses. Optimization of the mode transformation process is also possible in order to obtain a mode that is more easily coupled into a fiber. The development of CCDs that work at the single photon level is progressing rapidly and will be key in the detection of such modes.


\section{Direct Measurement of a High-Dimensional Quantum State}
\label{chap:DM}

\subsection{Introduction}

Due in part to the no-cloning theorem \cite{Wootters:1982ex}, the measurement of a quantum state poses a unique challenge for experimentalists. Conventionally, a quantum state is measured through the indirect process of tomography \cite{DAriano:2003tw}, which requires significant post-processing times to reliably reconstruct the state \cite{Thew:2002tc}. For this reason, quantum tomography is an unfeasible method for measuring high-dimensional quantum states such as those of orbital angular momentum (OAM) \cite{Agnew:2011js}. Recently, an alternative method called ``direct measurement" was proposed that utilized sequential weak and strong measurements to directly characterize a quantum state, i.e. without any post-processing \cite{Lundeen:2011isa}. In this section, we review a recent experiment where we use this method to characterize a high-dimensional quantum state in the discrete basis of OAM \cite{Malik:2014bf}. Through weak measurements of orbital angular momentum and strong measurements of angular position, we measure the probability amplitudes of a pure quantum state with a dimensionality, $d=27$. Further, we use our method to study the relationship between the angular momentum operator and rotations of a quantum state in the natural basis of OAM \cite{Barnett:1990do}.

The act of measuring a quantum state disturbs it irreversibly, a phenomenon referred to as \textit{collapse} of the wavefunction. For example, precisely measuring the position of a single photon results in a photon with a broad superposition of momenta. Consequently, no quantum system can be fully characterized through a single measurement. An established method of characterizing a quantum state involves making a diverse set of measurements on a collection of identically prepared quantum states, followed by post-processing of the data. This process, known as quantum state tomography \cite{DAriano:2003tw}, is akin to its classical counterpart of imaging a three-dimensional object using two-dimensional projections. For a simple quantum system such as a polarization qubit, quantum tomography can be similarly visualized as making projections onto different axes of the Poincar\'{e} sphere in order to localize the state on the sphere \cite{Altepeter:2005dc}. A critical part of any real tomographic process is the analysis that follows this series of measurements---in order to obtain a physical quantum state, one must use lengthy numerical procedures to search over all the different state possibilities \cite{Banaszek:1999ho}. The time required for this post-processing step scales rather unfavorably with state dimension, and is catastrophically large for multipartite high-dimensional states \cite{Thew:2002tc,Agnew:2011js}.

Photons carrying orbital angular momentum (OAM) are one such example of a high\hyph dimensional quantum state that has come to the forefront recently \cite{MolinaTerriza:2007ig,Yao:2011ve,Fickler:2012hj}. The discrete, infinite dimensionality of the OAM Hilbert space provides a larger information capacity for quantum information systems \cite{Groeblacher:2005ec,Malik:2012ka}, as well as an increased tolerance to eavesdropping in quantum key distribution \cite{Bourennane:2002uo}. Photons entangled in OAM \cite{Mair:2001ub,Leach:2009tc,Leach:2010bm} are prime candidates not only for such high capacity, high security communication systems, but also for fundamental tests of quantum mechanics \cite{Dada:2011dn,Krenn:2014jy}. Thus, it is essential that fast, accurate, and efficient methods for characterizing such high-dimensional states be developed. Quantum state tomography of a pair of photons entangled in OAM, each with a dimensionality of $d=8$, was recently demonstrated---a process that took on the order of \textit{days} to complete \cite{Agnew:2011js}. 

In Section \ref{DM}, we reviewed a novel alternative to tomography called direct measurement. In this technique, the complex probability amplitude of a pure quantum state is directly obtained as an output of the measurement apparatus, bypassing the complicated post-processing step required in quantum tomography. In the first implementation of direct measurement \cite{Lundeen:2011isa}, the position of an ensemble of identically-prepared photons was weakly measured, which caused a minimal disturbance to their momentum. A subsequent strong measurement of their momentum revealed all the information necessary to characterize their state in the continuous bases of position and momentum. A recent experiment extended this idea to directly measuring the two-dimensional polarization state of a laser beam \cite{Salvail:2013bo}. Here, we apply this novel technique to characterize a photon in the discrete, infinite\hyph dimensional space of orbital angular momentum.

\subsection{Theoretical Description of Direct Measurement in the OAM basis}

In direct analogy to a photon's position and linear momentum, the angular position and OAM of a photon form a conjugate pair \cite{Yao:2006vc}. We can express the state of our photon as a superposition of states in the OAM or angular position basis as

\be \ket{\Psi}=\sum_{\ell}a_{\ell}\ket{\ell}\;\;\;\;\textrm{or}\;\;\;\;\ket{\Psi}=\sum_{\theta}b_{\theta}\ket{\theta},\label{states}\ee

\noindent where $a_{\ell}$ and $b_{\theta}$ are the complex probability amplitudes in the OAM and angular position basis respectively. 

By multiplying our state by a strategically chosen constant $c=\brac{\theta_0}{\ell}/\brac{\theta_0}{\Psi}$ and inserting the identity, we can expand our state as 

\be  c \ket{\Psi} = c \sum_{\ell} \ket{\ell}\brac{\ell}{\Psi} = \sum_{\ell} \ket{\ell} \frac{\brac{\theta_0}{\ell}\brac{\ell}{\Psi}}{\brac{\theta_0}{\Psi}}= \sum_{\ell}\avg{\pi_{\ell}}_{\textrm{w}}\ket{\ell}.\ee

\noindent Here we have introduced the quantity $\avg{\pi_{\ell}}_{\textrm{w}}$, which is proportional to the probability amplitude $a_{\ell}$ from Eq.~\ref{states}. This quantity, known as the \textit{weak value}, is defined as the average result of a weak measurement of a quantum state, followed by a strong measurement, or post-selection of another observable of the state \cite{Aharonov:1988fk, Dressel:2014ks}. In general, weak values can be complex and can lie significantly outside the eigenvalue range of the observables being measured \cite{Ritchie:1991gh,Aharonov:2010vx}. In our direct measurement technique, the OAM weak value $\avg{\pi_{\ell}}_{\textrm{w}}$ is equal to the average result obtained by making a weak measurement of a projector in the OAM basis ($\h{\pi}_{\ell} = \ket{\ell}\bra{\ell}$) followed by a strong measurement in the conjugate basis of angular position ($\theta=0$). In this manner, the scaled complex probability amplitudes $ca_\ell$ can be directly obtained by measuring the OAM weak value $\avg{\pi_{\ell}}_{\textrm{w}}$ for a finite set of $\ell$. Following this procedure, the constant $c$ can be eliminated by renormalizing the state $\ket{\Psi}$.

\be \ket{\Psi}=\frac{1}{c}\sum_{\ell}\avg{\pi_{\ell}}_{\textrm{w}}\ket{\ell}.\ee

In order to measure such weak values, previous demonstrations of direct measurement have utilized a two-system Hamiltonian, where the system of interest is coupled to a measurement pointer \cite{Lundeen:2011isa,Salvail:2013bo}. In this manner, the real and imaginary parts of the system weak value are obtained by measuring the change in the pointer's position and momentum respectively \cite{Lundeen:ta}. This is well illustrated in the experiment of Salvail et al. \cite{Salvail:2013bo}, where the authors coupled the polarization of a laser beam to its position through the use of a birefringent crystal. By measuring the shift in the beam's position and momentum, they were able to measure its real and imaginary polarization components. In contrast, we use the polarization of the photon as a measurement pointer \cite{Lundeen:2011isa}. By coupling a photon's OAM to its polarization, we perform weak measurements of OAM by rotating the polarization of the OAM mode to be measured by a small angle. After sequential weak and strong measurements are performed, the average change in the photon's linear and circular polarization is measured, which is proportional to the real and imaginary parts of the OAM weak value.

Here we derive the relationship between the OAM weak value $\avg{\pi_{\ell}}_{\textrm{w}}$ and expectation values of the $\h{\sigma}_x$ and $\h{\sigma}_y$ Pauli operators (Eq.~\ref{Pauli} in the text). The von Neumann formulation can be used to describe the coupling between the OAM (system) and polarization (pointer) observables \cite{Neumann:1955,Lundeen:ta}. The product Hamiltonian describing this interaction can be written as

\be\h{H}=-g\,\h{\pi}_{\ell}\cdot\h{S}_y=-\bigg(\frac{g\,\hbar}{2}\bigg)\,\h{\pi}_{\ell}\cdot\h{\sigma}_y,\ee

\noindent where $g$ is a constant indicating the strength of the coupling, $\h{\pi}_{\ell}$ is the projection operator in the OAM basis, and $\h{\sigma}_y$ is the Pauli spin operator in the $y$ direction. The measurement pointer is initially in a vertical polarization state

\be \ket{s_i} = \left[ \begin{array}{c}
0 \\
1 \\
\end{array} \right] \label{initial}\ee 

\noindent and the system is in an initial state $\ket{I}$. The initial system-pointer state is modified by a unitary interaction $\h{U}=\exp(-i\h{H}t/\hbar)$, which can be written using the product Hamiltonian above as

\be\h{U}=\exp\bigg(\frac{i\,gt\,\h{\pi}_{\ell}\cdot\h{\sigma}_y}{2}\bigg)=\exp\bigg(\frac{i\,\sin\alpha\,\h{\pi}_{\ell}\cdot\h{\sigma}_y}{2}\bigg)\ee

\noindent Here we have substituted $\sin\alpha$ in place of $gt$ as a coupling constant. This refers to the angle $\alpha$ by which we rotate the polarization of the OAM mode to be measured in our experiment. When $\alpha$ is small, the measurement is weak. In this case, we can express the operator $\h{U}$ as a Taylor series expansion truncated to first order in $\sin\alpha$. The initial state then evolves to

\bea \ket{\Psi(t)} &=& (1-\frac{i\h{H}t}{\hbar}-...)\ket{I}\ket{s_i}\nonumber\\
&=& \ket{I}\ket{s_i}+\frac{i\,\sin\alpha}{2}\h{\pi}_{\ell}\ket{I}\h{\sigma}_y\ket{s_i}
\eea

\noindent We can express the strong measurement as a projection into a final state $\ket{F}$:

\be \bra{F}\h{U}\ket{I}\ket{s_i}=\brac{F}{I}\ket{s_i}+\frac{i\,\sin\alpha}{2}\bra{F}\h{\pi}_{\ell}\ket{I}\h{\sigma}_y\ket{s_i}\ee

\noindent We can then divide by $\brac{F}{I}$ to get the final pointer polarization state:

\bea \ket{s_f} &=& \ket{s_i}+\frac{i\,\sin\alpha}{2}\frac{\bra{F}\h{\pi}_{\ell}\ket{I}}{\brac{F}{I}}\h{\sigma}_y\ket{s_i}\nonumber\\
&=& \ket{s_i}+\frac{i\,\sin\alpha}{2}\avg{\pi_{\ell}}_{\textrm{w}}\h{\sigma}_y\ket{s_i}
\eea

\noindent Notice that the weak value $\avg{\pi_{\ell}}_{\textrm{w}}=\bra{F}\h{\pi}_{\ell}\ket{I}/\brac{F}{I}$ appears in the above equation. Using this expression for the final state of the pointer, we can calculate the expectation value of $\h{\sigma}_x$ as follows:

\bea \bra{s_f}\h{\sigma}_x\ket{s_f} &=& \cancel{\bra{s_i}\h{\sigma}_x\ket{s_i}} +\frac{i\,\sin\alpha}{2}\bigg[\avg{\pi_{\ell}}_{\textrm{w}}\bra{s_i}\h{\sigma}_x\h{\sigma}_y\ket{s_i}-\avg{\pi_{\ell}}_{\textrm{w}}^\dag\bra{s_i}\h{\sigma}_y\h{\sigma}_x\ket{s_i}\bigg]\nonumber\\
\eea

\noindent Using the substitution $\avg{\pi_{\ell}}_{\textrm{w}}$ = Re\{$\avg{\pi_{\ell}}_{\textrm{w}}$\} + $i$Im\{$\avg{\pi_{\ell}}_{\textrm{w}}$\} and the initial state $\ket{s_i}$ from Eq.~\ref{initial}, the above equation can simplified further:

\bea \bra{s_f}\h{\sigma}_x\ket{s_f} &=&  \frac{i\,\sin\alpha}{2}\bigg[\textrm{Re}\{\avg{\pi_{\ell}}_{\textrm{w}}\}\bra{s_i}\h{\sigma}_x\h{\sigma}_y-\h{\sigma}_y\h{\sigma}_x\ket{s_i}\nonumber\\
&&+ \;i\,\textrm{Im}\{\avg{\pi_{\ell}}_{\textrm{w}}\}\bra{s_i}\cancel{\h{\sigma}_x\h{\sigma}_y+\h{\sigma}_y\h{\sigma}_x}\ket{s_i}\bigg]\nonumber\\
&=& -\sin\alpha\,\textrm{Re}\{\avg{\pi_{\ell}}_{\textrm{w}}\}\bra{s_i}\h{\sigma}_z\ket{s_i}\nonumber\\
&=& \sin\alpha\,\textrm{Re}\{\avg{\pi_{\ell}}_{\textrm{w}}\}
\eea

\noindent Similarly, we can calculate the expectation value of $\h{\sigma}_y$ as follows:

\bea 
\bra{s_f}\h{\sigma}_y\ket{s_f} &=& \cancel{\bra{s_i}\h{\sigma}_y\ket{s_i}} +\frac{i\,\sin\alpha}{2}\bigg[\avg{\pi_{\ell}}_{\textrm{w}}\bra{s_i}\h{\sigma}_y\h{\sigma}_y\ket{s_i}-\avg{\pi_{\ell}}_{\textrm{w}}^\dag\bra{s_i}\h{\sigma}_y\h{\sigma}_y\ket{s_i}\bigg]\nonumber\\
&=&  \frac{i\,\sin\alpha}{2}\bigg[\textrm{Re}\{\avg{\pi_{\ell}}_{\textrm{w}}\}\bra{s_i}\cancel{\h{\sigma}_y\h{\sigma}_y-\h{\sigma}_y\h{\sigma}_y}\ket{s_i}\nonumber\\
&&+\; i\,\textrm{Im}\{\avg{\pi_{\ell}}_{\textrm{w}}\}\bra{s_i}\h{\sigma}_y\h{\sigma}_y+\h{\sigma}_y\h{\sigma}_y\ket{s_i}\bigg]\nonumber\\
&=& -\sin\alpha\,\textrm{Im}\{\avg{\pi_{\ell}}_{\textrm{w}}\}\bra{s_i}\h{\sigma}_y^2\ket{s_i}\nonumber\\
&=& -\sin\alpha\,\textrm{Im}\{\avg{\pi_{\ell}}_{\textrm{w}}\}
\eea

\noindent Thus, we see that the real and imaginary parts of the OAM weak value $\avg{\pi_{\ell}}_{\textrm{w}}$ are proportional to the expectation values of the $\h{\sigma}_x$ and $\h{\sigma}_y$ Pauli operators (see Eq.~\ref{Pauli}):

\bea \avg{\pi_{\ell}}_{\textrm{w}} &=& \textrm{Re}\{\avg{\pi_{\ell}}_{\textrm{w}}\} + i\,\textrm{Im}\{\avg{\pi_{\ell}}_{\textrm{w}}\}\nonumber\\
&=& \frac{1}{\sin\alpha}\big[\bra{s_f}\h{\sigma}_x\ket{s_f}-i\bra{s_f}\h{\sigma}_y\ket{s_f}\big]
\eea


\subsection{Experimental Weak Measurement of OAM}

\begin{figure}[t!]
\centering\includegraphics[scale=.55]{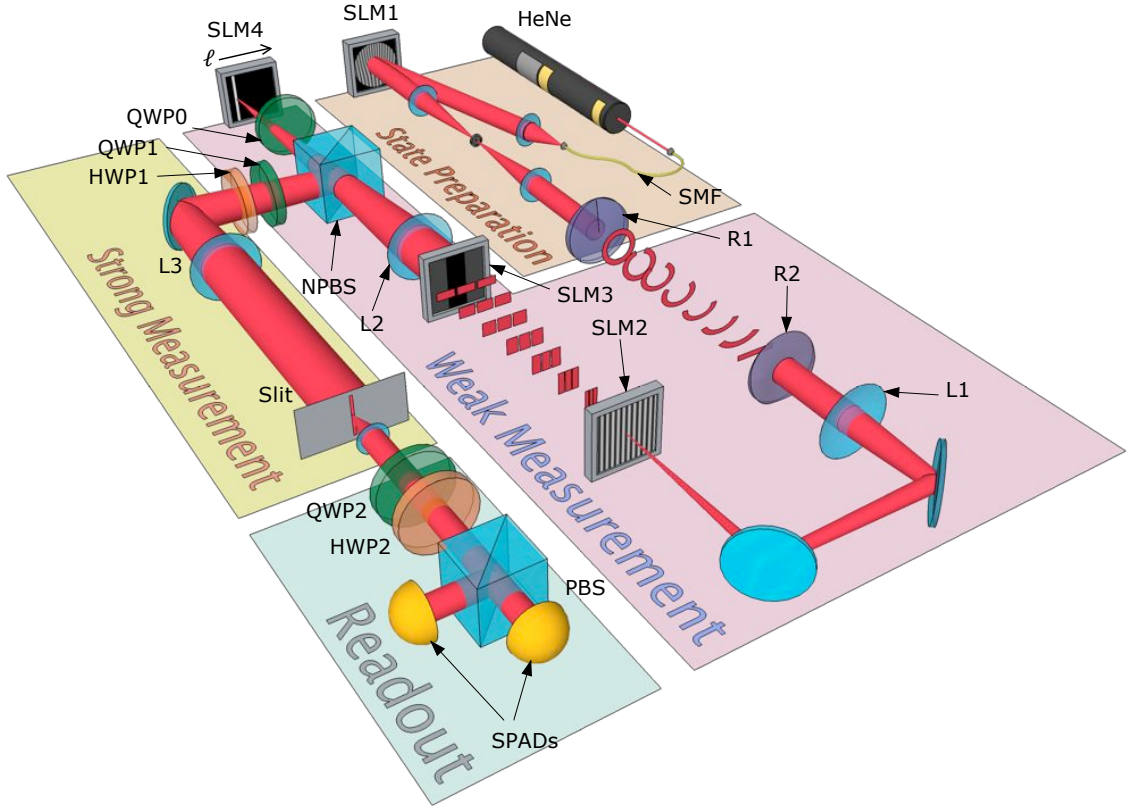}
\caption{Direct measurement of a high-dimensional quantum state. \textbf{State Preparation:} A quantum state in an arbitrary superposition of orbital angular momentum (OAM) modes is prepared by impressing phase information onto photons from an attenuated HeNe laser. \textbf{Weak Measurement:} A particular OAM mode is weakly measured by rotating its polarization by a small angle. This is accomplished by first separating the component OAM modes of the photon via a geometric transformation and then performing the polarization rotation. This process is depicted in the figure for one OAM mode. \textbf{Strong Measurement:} The angular position of the photon is strongly measured by using a slit to post-select states with an angle $\theta=0$. \textbf{Readout:} The OAM weak value $\avg{\pi_{\ell}}_{\textrm{w}}$ is obtained by measuring the change in the photon polarization in the linear and circular polarization bases (Figure adapted from Ref. \cite{Malik:2014bf}).}\label{DMsetup}
\end{figure}

In order to perform a weak measurement of OAM at the single photon level, one must first spatially separate the OAM components of the single photon. Only then can one rotate the polarization of the OAM mode to be measured by a small angle, which constitutes the weak measurement. Recently, we proposed a technique to efficiently separate the OAM components of a single photon \cite{Berkhout:2010cb,Mirhosseini:2013em}, which is discussed in detail in Section \ref{chap:OAMQKD}. Here, we implement this mode sorter technique using four phase-only elements (Fig.~\ref{DMsetup}, R1, R2, SLM2, and SLM3) to separate the OAM components with less than 10\% overlap. The first two elements, R1 and R2, are refractive holograms made out of Poly-methyl methacrylate (PMMA) that are used to map polar coordinates $(r,\theta)$ to rectilinear coordinates $(x,y)$ through the log-polar mapping $x=a(\theta\,\textrm{mod}\,2\pi)$ and $y=-a\ln{(r/b)}$ \cite{Lavery:2012hi}. This results in the transformation of an OAM mode with azimuthal phase variation $e^{i\ell\theta}$ to a momentum mode with position phase variation $e^{i\ell x/a}$. These momentum modes are then Fourier transformed by the lens L1 to position modes. At this stage, the component OAM modes of the photon still have an overlap of about 20\%. This is due to the finite size of the transformed momentum mode, which is bounded by the function $\textrm{rect}(x/2\pi a)$. A simple way to decrease the overlap and hence the size of the position mode is to simply increase the size of the momentum mode (while maintaining the phase ramp across it). We create three adjacent copies of the momentum mode by implementing a fan-out hologram and phase-corrector on SLM2 and SLM3 \cite{Prongue:1992uj}, also previously introduced in Section \ref{chap:OAMQKD}. The one-dimensional phase profile of the 3 copy fan-out hologram used here is shown in Fig.~\ref{FanoutX}. in After passing through another lens L2, this results in well-separated OAM modes having less than 10\% overlap on average with neighboring components.

In the next step, we rotate the polarization of the OAM mode to be weakly measured by an angle, $\alpha=\pi/9$. In contrast to the dynamic method used by Lundeen et al \cite{Lundeen:2011isa} in which they physically moved a half-wave plate (HWP) sliver through the beam, we use a static, programmable technique. A phase-only SLM acts as a variable phase retarder with individually addressable pixels. By sandwiching such an SLM between two quarter-wave plates (QWPs) whose extraordinary axes are aligned at $\pi/4$ radians to the SLM axis, one can rotate the polarization of any part of the beam through an arbitrary angle\cite{Davis:2000ur}. As shown in Fig.~\ref{DMsetup}, we use this technique to rotate the polarization of the OAM mode to be weakly measured. Since we use SLM4 in reflection, only one quarter-wave plate (QWP0) is needed. However, QWP1 and HWP1 are used to remove any ellipticity introduced by reflection through the non-polarizing beamsplitter (NPBS).  A strong measurement of angular position is performed by a 10 $\mu$m slit placed in the Fourier plane of lens L3. Since the plane of the slit is conjugate to the plane (SLM4) where the OAM modes are spatially separated, a measurement of linear position by the slit is equivalent to a measurement of angular position. 

The average change in the photon's linear and circular polarization is proportional to Re$\avg{\pi_{\ell}}_{\textrm{w}}$ and Im$\avg{\pi_{\ell}}_{\textrm{w}}$ respectively. As shown in the previous section, for an initially vertically-polarized state, the OAM weak value is given by

\be\avg{\pi_{\ell}}_{\textrm{w}}=\frac{1}{\sin\alpha}\bigg(\bra{s_f}\h{\sigma}_x\ket{s_f}-i\bra{s_f}\h{\sigma}_y\ket{s_f}\bigg),\label{Pauli}\ee

\noindent where $\alpha$ is the rotation angle, $\sigma_x$ and $\sigma_y$ are the Pauli operators in the $x$ and $y$ directions, and $\ket{s_f}$ is the final polarization state of the photon. In order to measure the expectation values of $\sigma_x$ and $\sigma_y$, we transform to the linear and circular polarization bases with QWP2 and HWP2, and measure the two Stokes parameters with a polarizing beamsplitter (PBS) and two single-photon avalanche detectors (SPADs). In this manner, we directly obtain the scaled complex probability amplitudes $ca_{\ell}$ by scanning the weak measurement through $\ell$ values of $\pm13$. Although the OAM of a photon exists in a discrete, unbounded Hilbert space, we are not able to go beyond a dimensionality of $d=27$ as our mode transformation technique begins to break down for higher OAM modes.

\subsection{Measuring the Wavefunction in the OAM Basis}

To test our method, we generate single photon states by strongly attenuating a HeNe laser to the single photon level. These photons are then tailored into a high-dimensional quantum state by impressing a specific OAM distribution on them with SLM1 and a 4$f$ system of lenses (Fig.~\ref{DMsetup}) \cite{Arrizon:2007wl}. We create a sinc-distribution of OAM using a wedge-shaped mask on the SLM. Analogous to how a rectangular aperture diffracts light into a sinc-distribution of linear momenta, a single photon diffracting through an angular aperture of width $\Delta\theta$ results in a quantum state with a sinc-distribution of OAM probability amplitudes \cite{Yao:2006vc} 

\be a_{\ell} = k\,\textrm{sinc}\bigg(\frac{\Delta\theta\ell}{2}\bigg).\label{sinc}\ee

\begin{figure}[t!]
\centering\includegraphics[scale=.60]{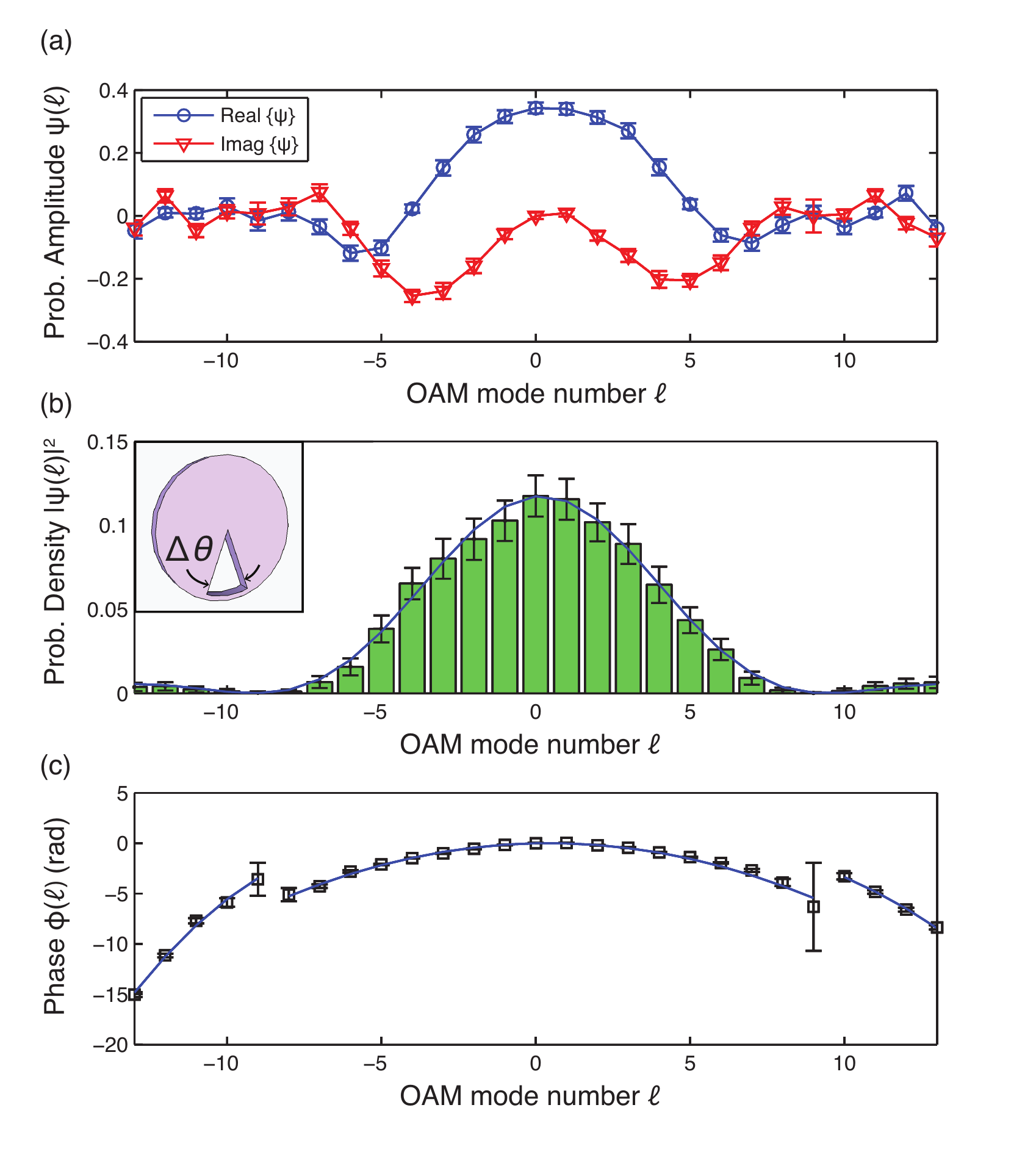}
\caption{Experimental data showing direct measurement of a 27-dimensional quantum state in the OAM basis. The state is created by sending single-photons through an angular aperture of width $\Delta\theta = 2\pi/9$ rad (inset of (b)). (a) The measured real (blue circles) and imaginary parts (red triangles) of the wavefunction, (b) the calculated probability density $\abs{\Psi(\ell)}^2$, and (c) the calculated phase $\phi(\ell)$ are plotted as functions of the OAM quantum number $\ell$ up to a dimensionality of $\ell=\pm13$. $\pi$-phase jumps occur when the probability amplitude is negative (not seen in the probability density). Theoretical fits to the probability density and phase are plotted as a blue line (Figure adapted from Ref. \cite{Malik:2014bf}).}\label{results1}
\end{figure}

\noindent The first nulls of this OAM distribution lie at OAM values $\ell=\pm2\pi/\Delta\theta$. Using an angular aperture of width $2\pi/9$ rad (inset of Fig.~\ref{results1}(b)), we create such an ensemble of identical single photons and perform the direct measurement procedure on them. The measured real and imaginary parts of the wavefunction are plotted in Fig.~\ref{results1}(a) as a function of $\ell$. From these, we calculate the probability density $\abs{\Psi(\ell)}^2$ and the phase $\phi(\ell)$, which are plotted in Figs. \ref{results1}(b) and (c). The width of the sinc-squared fit to the probability density is measured to be $9.26\pm0.21$, which is very close to the value of 9 predicted from theory. The measured phase of the OAM distribution in Fig.~\ref{results1}(c) has a tilted quadratic shape that is acquired from propagation through the system. Interestingly, $\pi$-phase jumps appear at the two minima of the sinc-squared probability density. This is because the sinc distribution of probability amplitudes in Eq. \eqref{sinc} changes sign from positive to negative at these two points. Additionally, the phase error is large when the amplitude goes to zero. This is because the noise due to the background and detector dark counts overwhelms our signal in this regime. Theoretical fits to the phase and probability density are plotted as blue lines.

\subsection{The Angular Momentum Operator as a Generator of Rotations}

Here, we use our technique to study the effect of rotation on a single photon carrying a range of angular momenta. Rotation of a quantum state by an angle $\theta_0$ can be expressed by the unitary operator $\h{U}=\textrm{exp}(i\h{L}_z\theta_0)$, where $\h{L}_z$ is the angular momentum operator. Operating on our quantum state $\ket{\Psi}$ with $\h{U}$, we get

\be \ket{\Psi'} = \h{U}\ket{\Psi}=\sum_\ell k\,\textrm{sinc}\bigg(\frac{\Delta\theta\ell}{2}\bigg)e^{i\ell\theta_0}\ket{\ell}.\ee

\begin{figure}[t!]
\centering\includegraphics[scale=.60]{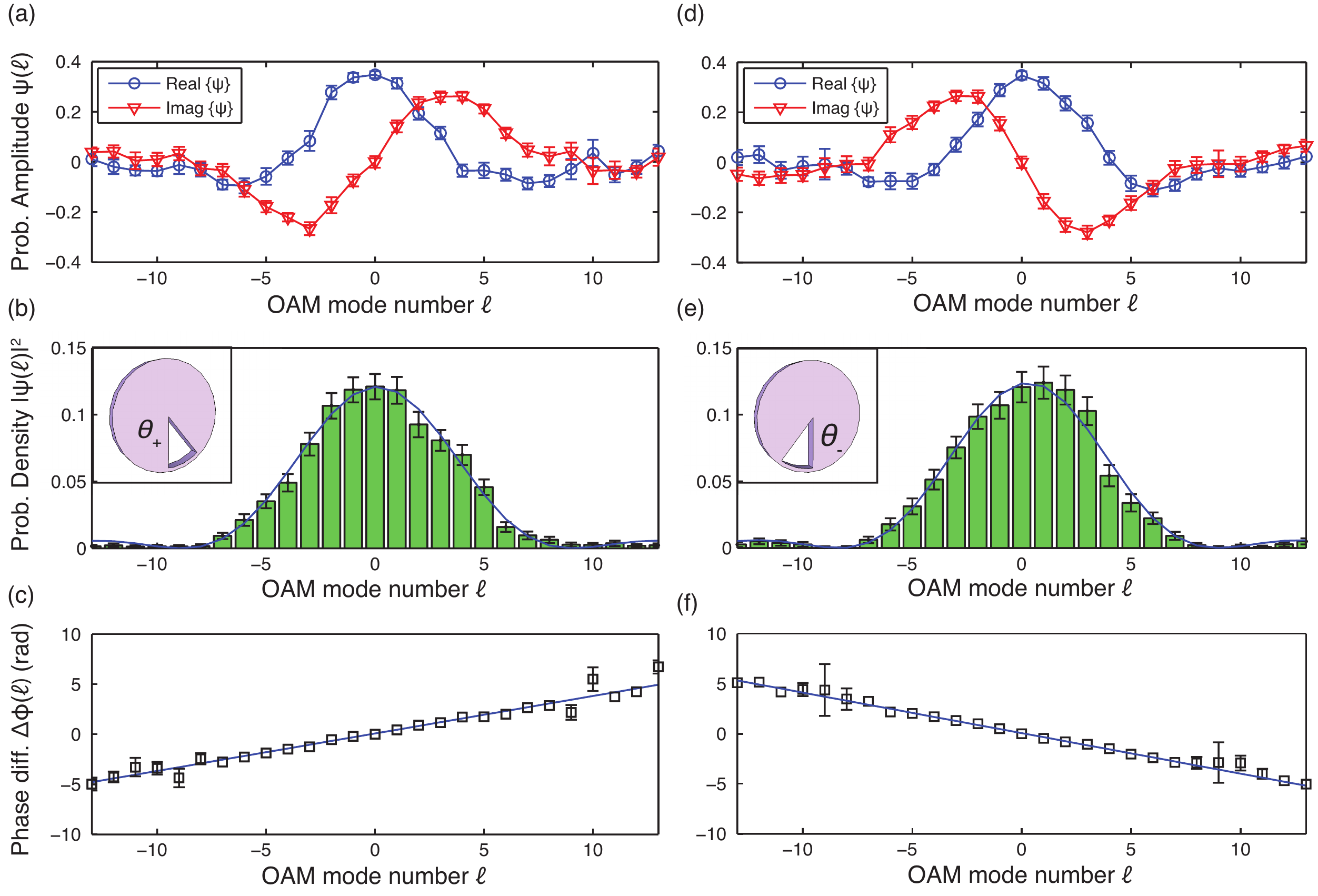}
\caption{Experimental data showing the direct measurement of a high-dimensional quantum state rotated by an angle $\theta_{\pm} = \pm\pi/9$ rad (insets). (a) and (d) The measured real (blue circles) and imaginary parts (red triangles) of the rotated wavefunctions. (b) and (e) The calculated probability densities $\abs{\Psi(\ell)_{\pm}}^2$. (c) and (f) The phase difference $\Delta\phi_{\pm}(\ell)$ between the calculated phase and the phase of the unrotated case. Theoretical fits to the probability densities and phases are plotted as blue lines. Error bars larger than the symbols are shown (Figure adapted from Ref. \cite{Malik:2014bf}).}\label{results2}
\end{figure}

\noindent Thus, the \textit{rotation} of a state vector by an angle $\theta_0$ manifests as an $\ell$\textit{-dependent phase} $e^{i\ell\theta_0}$ in the OAM basis. For this reason, the angular momentum operator is called the generator of rotations in quantum mechanics under the paraxial approximation \cite{Barnett:1990do}. In order to generate such a linear OAM-dependent phase, we create a rotated wavefunction by rotating our angular aperture by an angle $\theta_+=\pi/9$ rad (inset of Fig.~\ref{results2}(b)). Then, we perform the direct measurement procedure as explained in the previous section. The real and imaginary parts of the rotated wavefunction as a function of $\ell$ (Fig.~\ref{results2}(a)) are measured. The probability density and phase of the wavefunction are calculated from these measured values and plotted in Figs. \ref{results2}(b) and (c). For clarity, we subtract the phase of the zero rotation case (Fig.~\ref{results1}(c)) from our measured values of phase, so the effect of rotation is clear. Barring experimental error, the amplitude does not change significantly from the unrotated case (Fig.~\ref{results1}(b)). However, the phase of the single-photon OAM distribution exhibits a distinct $\ell$-dependent phase ramp with a slope of $0.373\pm0.007$ rad/mode. This is in close agreement with theory, which predicts the phase to have a form $\phi(\ell)=\pm\pi\ell/9$, corresponding to a phase ramp with a slope of $0.35$ rad/mode. Errors in slope are calculated by the process of chi-square minimization. This process is repeated for a negative rotation angle $\theta_-=-\pi/9$ rad, which results in a mostly unchanged probability density, but an $\ell$-dependent phase ramp with a negative slope of $-0.404\pm0.007$ rad/mode (Figs. \ref{results2} (d)-(f)).

These results clearly illustrate the relationship between phase and rotation in the OAM basis in that every $\ell$-component acquires a phase proportional to the azimuthal quantum number $\ell$. The measured slopes in both cases are slightly larger than those expected from theory due to errors introduced in the geometrical transformation that is used to spatially separate the OAM modes. The mode sorting process is extremely sensitive to misalignment, and a very small displacement of the transforming elements R1 and R2 can propagate as a phase error.

\subsection{Summary and Outlook}

To summarize, we have measured the complex probability amplitudes that characterize the wavefunction in the high-dimensional bases of orbital angular momentum and angular position. Using our technique, we have also measured the effects of rotation on a quantum state in the OAM basis. The rotation manifests as an OAM mode-dependent phase and provides a clean visualization of the relationship between the angular momentum operator and rotations in quantum mechanics. 

While we have directly measured pure states of OAM, this method can be extended to perform measurements of mixed, or general quantum states \cite{Lundeen:2012db,Salvail:2013bo}. By scanning the strong measurement of angular position as well, one can measure the Dirac distribution, which is informationally equivalent to the density matrix of a quantum state \cite{Dirac:1945wx,Chaturvedi:2006kz}. Furthermore, by extending this technique to two photons, photons entangled in OAM can be measured. In this case, one would need to perform independent weak and strong measurements on each photon, followed by a two-photon coincidence-detection scheme for the polarization measurement. 

Direct measurement offers distinct advantages over conventional methods of quantum state characterization such as tomography. This method does not require a global reconstruction, a step that involves prohibitively long processing times for high-dimensional quantum states such as those of OAM. Consequently, the quantum state is more accessible in that it can be measured locally as a function of OAM quantum number $\ell$, as in our experiment. These advantages may open up avenues for measuring quantum states directly in the middle of dynamically changing systems such as quantum circuits and free-space quantum communication links.


\section{Conclusions}
\label{chap:Conc}

In the immensely technological world of today, security is perhaps something we take for granted. However, security pervades through our quotidian tasks such as withdrawing money from the ATM and checking email on our smartphones. The machine gun-toting bank robbers of yesteryear have become a figment of Hollywood imagination, being replaced instead by individuals sitting behind a desk with an internet connection. The threat of cyber crimes such as identity theft and account hacking have never been more real, and nations are starting to realize that such concerns are not just limited to the individual. Large amounts of money have been spent by countries in the past for developing complex encryption algorithms that can be cracked by a sophisticated hacker. 

Over the past thirty years, the simplest limitation of a quantum state has led to the development of elegant technologies that allow unconditionally secure communication. The fact that one cannot create a copy of a quantum state allows one to use it as ``digital bait." A hacker who intercepts a secure quantum communication channel will disturb the delicate quantum states used in the channel, thus revealing his or her presence. The technologies reviewed in this article form part of this quantum revolution in security. We have applied ideas borrowed from quantum key distribution protocols to the field of optical imaging and surveillance. This promises a form of security for active imaging systems such as lidar that has not been seen before. Quantum ghost imaging has long been destined to the lab bench due the to impracticality of measuring an entire image at the single photon level. By extending this scheme to the identification of a set of images, we have made quantum ghost imaging easier to apply in the real world.

Quantum key distribution (QKD) is the bastion of secure quantum technologies, with commercial short-range QKD systems available today. One branch of research in this field is exploring the engineering extension of polarization-based QKD to longer and longer distances, such as ground to satellite. We have taken a parallel approach in this article, instead exploring alternative methods of encoding for QKD. QKD protocols using polarization states for encoding are limited to how much information they can send per photon, as well as how much eavesdropping error they can tolerate. By using the discrete, infinite dimensional state space of orbital angular momentum (OAM) for encoding, we can build a QKD system that promises a vastly increased information capacity and a significantly higher tolerance to error than conventional polarization-based QKD. In this article, we have discussed the development of technologies that allow us to generate and measure single photons carrying superpositions of OAM. Further, we have explored our ability to use such states to perform communication in a real-world setting with atmospheric turbulence. The development of working OAM-based QKD systems is the next research goal in our lab, and we are making fast progress towards achieving it.

The accurate characterization of a quantum state is important for fields as diverse as information science, physical chemistry, and foundational physics. A quantum state is conventionally measured by the process of quantum state tomography. Recently, an alternative to tomography was presented that used sequential weak and strong measurements in order to completely characterize a quantum state. In contrast with tomography, this ``direct measurement" method does not involve a time consuming post-processing step. In this article, we have reviewed the first application of the direct measurement technique for measuring a quantum state in a discrete, high-dimensional state space such as that of OAM. This serves as a significant advance for this technique, which has been previously used to characterize a photon in the continuous basis of position-momentum and a classical beam in the two-dimensional basis of polarization. The technique of direct measurement is especially advantageous when used for measuring high-dimensional quantum states. This is because the post-processing time required for the tomography of such states is prohibitive. Our experiment serves as the first application of direct measurement that sets it clearly apart from tomography in this regard. Further, it displays the potential this technique has for measuring quantum states directly in the middle of dynamic quantum processes.

The quantum technologies presented in this article serve to advance the state-of-the-art of research in the fields of secure quantum communication, quantum imaging, and quantum state characterization. In addition, by interfacing between these fields, they show the potential that exists for the exchange of ideas and techniques across these disciplines. It is likely that quantum technologies today are the beginning of a technological revolution that will encompass more than just security. 


\section*{Acknowledgments}

We would like to thank Mohammad Mirhosseini, Omar Maga\~{n}a-Loaiza, Malcolm O'Sullivan, Prof.~Zhimin Shi, Dr.~Heedeuk Shin, Brandon Rodenberg, Prof.~Anand Jha, Dr.~Petros Zerom, Prof.~Jonathan Leach, Dr.~Martin Lavery, and Prof.~Miles Padgett for their direct contributions to many of the experiments that are reviewed in this article. Additionally, we would like to thank Dr.~Justin Dressel, Dr.~Filippo Miatto, and Prof.~Andrew Jordan for many enlightening discussions on quantum weak values. This work was supported by the DARPA InPho program. M.M.~acknowledges funding from the European Commission through a Marie Curie fellowship. R.W.B.~acknowledges support from the Canada Excellence Research Chairs program.

\bibliography{Thesis_custom,all2}

\begin{thebibliography}{100}
\newcommand{\enquote}[1]{``#1''}

\bibitem{Wootters:1982ex}
W.~K. Wootters and W.~H. Zurek, \enquote{{A single quantum cannot be cloned},}
  Nature \textbf{299}, 802--803 (1982).

\bibitem{Bennett:1984wv}
C.~Bennett and G.~Brassard, \enquote{{Quantum cryptography: Public key
  distribution and coin tossing},} in \enquote{Proceedings of IEEE
  International Conference on Computers, Systems, and Signal Processing,}
  (Bangalore, 1984), p. 175.

\bibitem{Boto:2000eg}
A.~N. Boto, P.~Kok, D.~S. Abrams, S.~L. Braunstein, C.~P. Williams, and J.~P.
  Dowling, \enquote{{Quantum Interferometric Optical Lithography: Exploiting
  Entanglement to Beat the Diffraction Limit},} Phys. Rev. Lett. \textbf{85},
  2733--2736 (2000).

\bibitem{Nagata:2007jo}
T.~Nagata, R.~Okamoto, J.~L. O'Brien, K.~Sasaki, and S.~Takeuchi,
  \enquote{{Beating the standard quantum limit with four-entangled photons},}
  Science \textbf{316}, 726--729 (2007).

\bibitem{Plato:Republic}
Plato and B.~Jowett, \emph{{Plato's The Republic}} (New York: The Modern
  Library, 1941).

\bibitem{cat_website}
Rhys, \enquote{{Schr\"{o}dinger's Cat},}
  http://goodcanadiankid.com/schrodingers-cat/  (2012).

\bibitem{RomeroIsart:2009iv}
O.~Romero-Isart, M.~L. Juan, R.~Quidant, and J.~I. Cirac, \enquote{{Toward
  quantum superposition of living organisms},} New J. Phys. \textbf{12}, 033015
  (2010).

\bibitem{Schrodinger:1935ub}
E.~Schr{\"o}dinger, \enquote{{Die gegenw{\"a}rtige Situation in der
  Quantenmechanik},} Naturwissenschaften \textbf{23}, 823--828 (1935).

\bibitem{Bell:1964wu}
J.~Bell, \enquote{{On the einstein-podolsky-rosen paradox},} Physics
  \textbf{1}, 195--200 (1964).

\bibitem{Clauser:1969ff}
J.~Clauser, M.~Horne, A.~Shimony, and R.~Holt, \enquote{{Proposed Experiment to
  Test Local Hidden-Variable Theories},} Phys. Rev. Lett. \textbf{23}, 880--884
  (1969).

\bibitem{Aspect:1982ja}
A.~Aspect, J.~Dalibard, and G.~Roger, \enquote{{Experimental Test of Bell's
  Inequalities Using Time- Varying Analyzers},} Phys. Rev. Lett. \textbf{49},
  1804--1807 (1982).

\bibitem{Ekert:1991kl}
A.~K. Ekert, \enquote{{Quantum cryptography based on Bell{\textquoteright}s
  theorem},} Phys. Rev. Lett. \textbf{67}, 661--663 (1991).

\bibitem{Bouwmeester:1997wk}
D.~Bouwmeester, J.-W. Pan, K.~Mattle, M.~Eibl, H.~Weinfurter, and A.~Zeilinger,
  \enquote{{Experimental quantum teleportation},} Nature \textbf{390}, 575--579
  (1997).

\bibitem{Boyd:2011eh}
R.~W. Boyd and J.~P. Dowling, \enquote{{Quantum lithography: status of the
  field},} Quantum Inf Process \textbf{11}, 891--901 (2011).

\bibitem{Pittman:1995jb}
T.~Pittman, Y.~Shih, D.~Strekalov, and A.~Sergienko, \enquote{{Optical imaging
  by means of two-photon quantum entanglement},} Phys. Rev. A \textbf{52},
  R3429--R3432 (1995).

\bibitem{Monken:1998vy}
C.~Monken, P.~Ribeiro, and S.~Padua, \enquote{{Transfer of angular spectrum and
  image formation in spontaneous parametric down-conversion},} Phys. Rev. A
  \textbf{57}, 3123--3126 (1998).

\bibitem{Walborn:2007ix}
S.~P. Walborn and C.~Monken, \enquote{{Transverse spatial entanglement in
  parametric down-conversion},} Phys. Rev. A \textbf{76}, 062305 (2007).

\bibitem{Beth:1936vh}
R.~A. Beth, \enquote{{Mechanical detection and measurement of the angular
  momentum of light},} Phys. Rev. \textbf{50}, 115 (1936).

\bibitem{Allen:1992by}
L.~Allen, M.~Beijersbergen, R.~Spreeuw, and J.~P. Woerdman, \enquote{{Orbital
  angular momentum of light and the transformation of Laguerre-Gaussian laser
  modes},} Phys. Rev. A \textbf{45}, 8185--8189 (1992).

\bibitem{Aharonov:1988fk}
Y.~Aharonov, D.~Z. Albert, and L.~Vaidman, \enquote{{How the result of a
  measurement of a component of the spin of a spin-1/2 particle can turn out to
  be 100},} Phys. Rev. Lett. \textbf{60}, 1351--1354 (1988).

\bibitem{Dressel:2014ks}
J.~Dressel, M.~Malik, F.~M. Miatto, A.~N. Jordan, and R.~W. Boyd,
  \enquote{{Colloquium: Understanding quantum weak values: Basics and
  applications},} Rev. Mod. Phys. \textbf{86}, 307--316 (2014).

\bibitem{Ritchie:1991gh}
N.~Ritchie, J.~Story, and R.~Hulet, \enquote{{Realization of a measurement of a
  {\textquoteleft}{\textquoteleft}weak
  value{\textquoteright}{\textquoteright}},} Phys. Rev. Lett. \textbf{66},
  1107--1110 (1991).

\bibitem{Giovannetti:2004jg}
V.~Giovannetti, S.~Lloyd, and L.~Maccone, \enquote{{Quantum-Enhanced
  Measurements: Beating the Standard Quantum Limit},} Science \textbf{306},
  1330--1336 (2004).

\bibitem{Dutton:2010fj}
Z.~Dutton, J.~H. Shapiro, and S.~Guha, \enquote{{LADAR resolution improvement
  using receivers enhanced with squeezed-vacuum injection and phase-sensitive
  amplification},} Journal of the Optical Society of America B \textbf{27}, A63
  (2010).

\bibitem{Steane:1999fn}
A.~Steane, \enquote{{Quantum computing},} Rep. Prog. Phys. \textbf{61},
  117--173 (1999).

\bibitem{Bennett:2000kl}
C.~H. Bennett and D.~P. DiVincenzo, \enquote{{Quantum information and
  computation},} Nature \textbf{404}, 247--255 (2000).

\bibitem{Gisin:2002zz}
N.~Gisin, G.~Ribordy, W.~Tittel, and H.~Zbinden, \enquote{{Quantum
  cryptography},} Rev. Mod. Phys. \textbf{74}, 145--195 (2002).

\bibitem{Lo:2005cd}
H.-K. Lo, X.~Ma, and K.~Chen, \enquote{{Decoy State Quantum Key Distribution},}
  Phys. Rev. Lett. \textbf{94}, 230504 (2005).

\bibitem{Agnew:2012gs}
M.~Agnew, J.~Leach, and R.~Boyd, \enquote{{Observation of entanglement
  witnesses for orbital angular momentum states},} Eur. Phys. J. D \textbf{66},
  156 (2012).

\bibitem{Strekalov:1995ub}
D.~Strekalov, A.~Sergienko, and D.~Klyshko, \enquote{{Observation of two-photon
  {\textquotedblleft}Ghost{\textquotedblright} interference and diffraction},}
  Phys. Rev. Lett. \textbf{74}, 3600 (1995).

\bibitem{Bennink:2002jr}
R.~Bennink, S.~Bentley, and R.~W. Boyd, \enquote{{"Two-photon" coincidence
  imaging with a classical source},} Phys. Rev. Lett. \textbf{89}, 113601
  (2002).

\bibitem{Gatti:2004ec}
A.~Gatti, E.~Brambilla, M.~Bache, and L.~Lugiato, \enquote{{Ghost Imaging with
  Thermal Light: Comparing Entanglement and Classical Correlation},} Phys. Rev.
  Lett. \textbf{93}, 093602 (2004).

\bibitem{Bennink:2004df}
R.~Bennink, S.~Bentley, R.~W. Boyd, and J.~Howell, \enquote{{Quantum and
  classical coincidence imaging},} Phys. Rev. Lett. \textbf{92}, 033601 (2004).

\bibitem{Ferri:2005hq}
F.~Ferri, D.~Magatti, A.~Gatti, M.~Bache, E.~Brambilla, and L.~Lugiato,
  \enquote{{High-resolution ghost image and ghost diffraction experiments with
  thermal light},} Phys. Rev. Lett. \textbf{94}, 183602 (2005).

\bibitem{Meyers:2011cv}
R.~E. Meyers, K.~S. Deacon, and Y.~Shih, \enquote{{Turbulence-free ghost
  imaging},} Appl. Phys. Lett. \textbf{98}, 111115 (2011).

\bibitem{Dixon:2011wq}
P.~Dixon, G.~Howland, K.~Chan, M.~O'Sullivan, B.~Rodenburg, N.~Hardy,
  J.~Shapiro, D.~Simon, A.~Sergienko, and R.~W. Boyd, \enquote{{Quantum ghost
  imaging through turbulence},} Phys. Rev. A \textbf{83}, 051803 (2011).

\bibitem{Katz:2009fv}
O.~Katz, Y.~Bromberg, and Y.~Silberberg, \enquote{{Compressive ghost imaging},}
  Appl. Phys. Lett. \textbf{95}, 131110 (2009).

\bibitem{Bromberg:2009ca}
Y.~Bromberg, O.~Katz, and Y.~Silberberg, \enquote{{Ghost imaging with a single
  detector},} Phys. Rev. A \textbf{79}, 053840 (2009).

\bibitem{MaganaLoaiza:2013jq}
O.~S. Maga{\~n}a-Loaiza, M.~Malik, G.~A. Howland, J.~C. Howell, and R.~W. Boyd,
  \enquote{{Compressive object tracking using entangled photons},} Appl. Phys.
  Lett. \textbf{102}, 231104 (2013).

\bibitem{Brida:2010jw}
G.~Brida, M.~Genovese, and I.~R. Berchera, \enquote{{Experimental realization
  of sub-shot-noise quantum imaging},} Nat. Phot. \textbf{4}, 227--230 (2010).

\bibitem{Lemos:2014uw}
G.~B. Lemos, V.~Borish, G.~D. Cole, S.~Ramelow, R.~Lapkiewicz, and
  A.~Zeilinger, \enquote{{Quantum Imaging with Undetected Photons},} arXiv p.
  1401.4318 (2014).

\bibitem{Ferrie:2013gf}
C.~Ferrie and J.~Combes, \enquote{{Weak Value Amplification is Suboptimal for
  Estimation and Detection},} Phys. Rev. Lett. \textbf{112}, 040406 (2014).

\bibitem{Jordan:2014jv}
A.~N. Jordan, J.~Mart{\'\i}nez-Rinc{\'o}n, and J.~C. Howell,
  \enquote{{Technical Advantages for Weak-Value Amplification: When Less Is
  More},} Phys. Rev. X \textbf{4}, 011031 (2014).

\bibitem{Altepeter:2005dc}
J.~Altepeter, E.~Jeffrey, and P.~Kwiat, \enquote{{Photonic state tomography},}
  Adv Atom Mol Opt Phy \textbf{52}, 105--159 (2005).

\bibitem{Lundeen:2011isa}
J.~S. Lundeen, B.~Sutherland, A.~Patel, and C.~Stewart, \enquote{{Direct
  measurement of the quantum wavefunction},} Nature \textbf{474}, 188--191
  (2011).

\bibitem{Lundeen:2012db}
J.~S. Lundeen and C.~Bamber, \enquote{{Procedure for direct measurement of
  general quantum states using weak measurement},} Phys. Rev. Lett.
  \textbf{108}, 070402 (2012).

\bibitem{Salvail:2013bo}
J.~Z. Salvail, M.~Agnew, A.~S. Johnson, E.~Bolduc, J.~Leach, and R.~W. Boyd,
  \enquote{{Full characterization of polarization states of light via direct
  measurement},} Nat. Phot. \textbf{7}, 316--321 (2013).

\bibitem{Haapasalo:2011en}
E.~Haapasalo, P.~Lahti, and J.~Schultz, \enquote{{Weak versus approximate
  values in quantum state determination},} Phys. Rev. A \textbf{84}, 052107
  (2011).

\bibitem{Malik:2012js}
M.~Malik, O.~S. Maga{\~n}a-Loaiza, and R.~W. Boyd, \enquote{{Quantum-secured
  imaging},} Appl. Phys. Lett. \textbf{101}, 241103--241103--4 (2012).

\bibitem{Roome:1990th}
S.~J. Roome, \enquote{{Digital radio frequency memory},} Electronics {\&}
  communication engineering journal \textbf{2}, 147--153 (1990).

\bibitem{Goodman:Fourier}
J.~Goodman, \emph{{Introduction to Fourier Optics}} (Roberts \& Company, 2005).

\bibitem{Broadbent:2009tq}
C.~Broadbent, P.~Zerom, H.~Shin, J.~Howell, and R.~W. Boyd,
  \enquote{{Discriminating orthogonal single-photon images},} Phys. Rev. A
  \textbf{79}, 033802 (2009).

\bibitem{Malik:2010cr}
M.~Malik, H.~Shin, M.~N. O'Sullivan, P.~Zerom, and R.~W. Boyd,
  \enquote{{Quantum ghost image identification with correlated photon pairs},}
  Phys. Rev. Lett. \textbf{104}, 163602 (2010).

\bibitem{Abouraddy:2002vi}
A.~Abouraddy, B.~Saleh, and A.~Sergienko, \enquote{{Entangled-photon Fourier
  optics},} J. Opt. Soc. Am. A. \textbf{19}, 1174 (2002).

\bibitem{He:2008ud}
G.~He, X.~Wang, D.~Li, and J.~Hu, \enquote{{A high-speed image sensing
  technique with adjustable frame rate based on an ordinary CCD},}
  Optik-International Journal for Light and Electron Optics \textbf{119},
  548--552 (2008).

\bibitem{Turin:1960tw}
G.~Turin, \enquote{{An introduction to matched filters},} IEEE Trans. on Inf.
  Theo. \textbf{6}, 311--329 (1960).

\bibitem{Morris:1984vk}
G.~Morris, \enquote{{Image correlation at low light levels: a computer
  simulation},} Applied Optics \textbf{23}, 3152--3159 (1984).

\bibitem{An:1999ui}
X.~An, D.~Psaltis, and G.~Burr, \enquote{{Thermal fixing of 10,000 holograms in
  LiNbO3 : Fe},} Applied Optics \textbf{38}, 386--393 (1999).

\bibitem{Hiskett:2006kt}
P.~A. Hiskett, D.~Rosenberg, C.~G. Peterson, R.~J. Hughes, S.~Nam, A.~E. Lita,
  A.~J. Miller, and J.~E. Nordholt, \enquote{{Long-distance quantum key
  distribution in optical fibre},} New J. Phys. \textbf{8}, 193--193 (2006).

\bibitem{Ursin:2007jj}
R.~Ursin, F.~Tiefenbacher, T.~Schmitt-Manderbach, H.~Weier, T.~Scheidl,
  M.~Lindenthal, B.~Blauensteiner, T.~Jennewein, J.~Perdigues, P.~Trojek,
  B.~{\"O}mer, M.~F{\"u}rst, M.~Meyenburg, J.~Rarity, Z.~Sodnik, C.~Barbieri,
  H.~Weinfurter, and A.~Zeilinger, \enquote{{Entanglement-based quantum
  communication over 144 km},} Nature Physics \textbf{3}, 481--486 (2007).

\bibitem{Bourennane:2002uo}
M.~Bourennane, A.~Karlsson, G.~Bjork, N.~Gisin, and N.~Cerf, \enquote{{Quantum
  key distribution using multilevel encoding: security analysis},} J. Phys.
  A-Math. Gen. \textbf{35}, 10065--10076 (2002).

\bibitem{Groeblacher:2005ec}
S.~Groblacher, T.~Jennewein, A.~Vaziri, G.~Weihs, and A.~Zeilinger,
  \enquote{{Experimental quantum cryptography with qutrits},} New J. Phys.
  \textbf{8}, 75 (2006).

\bibitem{Cerf:2002fp}
N.~Cerf, M.~Bourennane, A.~Karlsson, and N.~Gisin, \enquote{{Security of
  Quantum Key Distribution Using d-Level Systems},} Phys. Rev. Lett.
  \textbf{88}, 127902 (2002).

\bibitem{Paterson:2005wj}
C.~Paterson, \enquote{{Atmospheric turbulence and orbital angular momentum of
  single photons for optical communication},} Phys. Rev. Lett. \textbf{94},
  153901 (2005).

\bibitem{Tyler:2009wz}
G.~Tyler and R.~W. Boyd, \enquote{{Influence of atmospheric turbulence on the
  propagation of quantum states of light carrying orbital angular momentum},}
  Optics Letters \textbf{34}, 142--144 (2009).

\bibitem{Smith:2006ek}
B.~Smith and M.~Raymer, \enquote{{Two-photon wave mechanics},} Phys. Rev. A
  \textbf{74}, 062104 (2006).

\bibitem{Gbur:2008fb}
G.~Gbur and R.~K. Tyson, \enquote{{Vortex beam propagation through atmospheric
  turbulence and topological charge conservation},} J. Opt. Soc. Am. A.
  \textbf{25}, 225 (2008).

\bibitem{Roux:2011dr}
F.~Roux, \enquote{{Infinitesimal-propagation equation for decoherence of an
  orbital-angular-momentum-entangled biphoton state in atmospheric
  turbulence},} Phys. Rev. A \textbf{83}, 053822 (2011).

\bibitem{Malik:2012ka}
M.~Malik, M.~N. O'Sullivan, B.~Rodenburg, M.~Mirhosseini, J.~Leach, M.~P.~J.
  Lavery, M.~J. Padgett, and R.~W. Boyd, \enquote{{Influence of atmospheric
  turbulence on optical communications using orbital angular momentum for
  encoding},} Optics Express \textbf{20}, 13195--13200 (2012).

\bibitem{Rodenburg:2013dn}
B.~Rodenburg, M.~Mirhosseini, M.~Malik, O.~S. Maga{\~n}a-Loaiza, M.~Yanakas,
  L.~Maher, N.~K. Steinhoff, G.~A. Tyler, and R.~W. Boyd, \enquote{{Simulating
  thick atmospheric turbulence in the lab with application to orbital angular
  momentum communication},} New. J. Phys. \textbf{16}, 033020 (2014).

\bibitem{Krenn:2014wj}
M.~Krenn, R.~Fickler, M.~Fink, J.~Handsteiner, M.~Malik, T.~Scheidl, R.~Ursin,
  and A.~Zeilinger, \enquote{{Twisted light communication through turbulent air
  across Vienna},} arXiv p. 1402.2602 (2014).

\bibitem{Shannon:1948zz}
C.~E. Shannon, \enquote{{A mathematical theory of communication},} Bell
  Syst.Tech.J. \textbf{27}, 379--423 (1948).

\bibitem{Nielsen:2004}
M.~Nielsen and I.~Chuang, \emph{{Quantum Computation and Quantum Information}}
  (Cambridge University Press, 2004).

\bibitem{Wootters:1989ba}
W.~K. Wootters and B.~D. Fields, \enquote{{Optimal state-determination by
  mutually unbiased measurements},} Annals of Physics \textbf{191}, 363--381
  (1989).

\bibitem{Bruss:1998gg}
D.~Bru{\ss}, \enquote{{Optimal Eavesdropping in Quantum Cryptography with Six
  States},} Phys. Rev. Lett. \textbf{81}, 3018--3021 (1998).

\bibitem{Scarani:2009kj}
V.~Scarani, H.~Bechmann-Pasquinucci, N.~J. Cerf, M.~Dusek, N.~Luetkenhaus, and
  M.~Peev, \enquote{{The security of practical quantum key distribution},} Rev.
  Mod. Phys. \textbf{81}, 1301--1350 (2009).

\bibitem{Heckenberg:1992vn}
N.~R. Heckenberg, R.~McDuff, C.~P. Smith, and A.~G. White, \enquote{{Generation
  of optical phase singularities by computer-generated holograms},} Optics
  Letters \textbf{17}, 221--223 (1992).

\bibitem{Arrizon:2007wl}
V.~Arrizon, U.~Ruiz, R.~Carrada, and L.~A. Gonzalez, \enquote{{Pixelated phase
  computer holograms for the accurate encoding of scalar complex fields},} J.
  Opt. Soc. Am. A. \textbf{24}, 3500--3507 (2007).

\bibitem{Gruneisen:2008bu}
M.~T. Gruneisen, W.~A. Miller, R.~C. Dymale, and A.~M. Sweiti,
  \enquote{{Holographic generation of complex fields with spatial light
  modulators: application to quantum key distribution},} Applied Optics
  \textbf{47}, A32 (2008).

\bibitem{Leach:2002wy}
J.~Leach, M.~J. Padgett, S.~M. Barnett, and S.~Franke-Arnold,
  \enquote{{Measuring the orbital angular momentum of a single photon},} Phys.
  Rev. Lett. \textbf{88}, 257901 (2002).

\bibitem{Berkhout:2010cb}
G.~C.~G. Berkhout, M.~P.~J. Lavery, J.~Courtial, M.~W. Beijersbergen, and M.~J.
  Padgett, \enquote{{Efficient Sorting of Orbital Angular Momentum States of
  Light},} Phys. Rev. Lett. \textbf{105}, 153601 (2010).

\bibitem{Lavery:2012hi}
M.~P.~J. Lavery, D.~J. Robertson, G.~C.~G. Berkhout, G.~D. Love, M.~J. Padgett,
  and J.~Courtial, \enquote{{Refractive elements for the measurement of the
  orbital angular momentum of a single photon},} Optics Express \textbf{20},
  2110--2115 (2012).

\bibitem{Bryngdahl:1974tu}
O.~Bryngdahl, \enquote{{Geometrical transformations in optics},} JOSA
  \textbf{64}, 1092--1099 (1974).

\bibitem{Mirhosseini:2013em}
M.~Mirhosseini, M.~Malik, Z.~Shi, and R.~W. Boyd, \enquote{{Efficient
  separation of the orbital angular momentum eigenstates of light},} Nat.
  Commun. \textbf{4:2781} (2013).

\bibitem{OSullivan:2012gj}
M.~N. O'Sullivan, M.~Mirhosseini, M.~Malik, and R.~W. Boyd,
  \enquote{{Near-perfect sorting of orbital angular momentum and angular
  position states of light},} Optics Express \textbf{20}, 24444--24449 (2012).

\bibitem{Prongue:1992uj}
D.~Prongue, H.~P. Herzig, R.~Dandliker, and M.~T. Gale, \enquote{{Optimized
  Kinoform Structures for Highly Efficient Fan-Out Elements},} Applied Optics
  \textbf{31}, 5706--5711 (1992).

\bibitem{Romero:2007vi}
L.~A. Romero and F.~M. Dickey, \enquote{{Theory of optimal beam splitting by
  phase gratings. I. One-dimensional gratings},} J. Opt. Soc. Am. A.
  \textbf{24}, 2280--2295 (2007).

\bibitem{Lutkenhaus:2002jy}
N.~L{\"u}tkenhaus and M.~Jahma, \enquote{{Quantum key distribution with
  realistic states: photon-number statistics in the photon-number splitting
  attack},} Journal of Optics \textbf{4}, 44--44 (2002).

\bibitem{SchmittManderbach:2008vm}
T.~Schmitt-Manderbach, \enquote{{Long distance free-space quantum key
  distribution},} Ph.D. thesis,
  Ludwig{\textendash}Maximilians{\textendash}Universitat Munchen (2008).

\bibitem{Zhao:2006dc}
Y.~Zhao, B.~Qi, X.~Ma, H.-K. Lo, and L.~Qian, \enquote{{Experimental Quantum
  Key Distribution with Decoy States},} Phys. Rev. Lett. \textbf{96}, 070502
  (2006).

\bibitem{SchmittManderbach:2007jj}
T.~Schmitt-Manderbach, H.~Weier, M.~F{\"u}rst, R.~Ursin, F.~Tiefenbacher,
  T.~Scheidl, J.~Perdigues, Z.~Sodnik, C.~Kurtsiefer, J.~Rarity, A.~Zeilinger,
  and H.~Weinfurter, \enquote{{Experimental Demonstration of Free-Space
  Decoy-State Quantum Key Distribution over 144~km},} Phys. Rev. Lett.
  \textbf{98}, 010504 (2007).

\bibitem{Ma:2005bd}
X.~Ma, B.~Qi, Y.~Zhao, and H.-K. Lo, \enquote{{Practical decoy state for
  quantum key distribution},} Phys. Rev. A \textbf{72}, 012326 (2005).

\bibitem{Makkaveev:2005tf}
A.~Makkaveev, S.~Molotkov, D.~Pomozov, and A.~Timofeev, \enquote{{Practical
  error-correction procedures in quantum cryptography},} J Exp Theor Phys
  \textbf{101}, 230--252 (2005).

\bibitem{Huttner:uz}
B.~Huttner and A.~K. Ekert, \enquote{{Information Gain in Quantum
  Eavesdropping},} J. Mod. Opt. \textbf{41}, 2455--2466 (1994).

\bibitem{OAMQKD}
M.~Mirhosseini, O.~Maga\~{n}a Loaiza, M.~O'Sullivan, B.~Rodenburg, M.~Malik,
  D.~J. Gauthier, and R.~W. Boyd, \enquote{{High-dimensional quantum
  cryptography with twisted light},} arXiv p. 1402.7113 (2014).

\bibitem{Ekert:1995vv}
A.~Ekert and P.~Knight, \enquote{{Entangled quantum systems and the Schmidt
  decomposition},} Am. J. Phys. \textbf{63}, 415--423 (1995).

\bibitem{Galvez:2005tt}
E.~Galvez, C.~Holbrow, M.~Pysher, J.~Martin, N.~Courtemanche, L.~Heilig, and
  J.~Spencer, \enquote{{Interference with correlated photons: Five quantum
  mechanics experiments for undergraduates},} Am. J. Phys. \textbf{73}, 127
  (2005).

\bibitem{Mair:2001ub}
A.~Mair, A.~Vaziri, G.~Weihs, and A.~Zeilinger, \enquote{{Entanglement of the
  orbital angular momentum states of photons},} Nature \textbf{412}, 313--316
  (2001).

\bibitem{Dada:2011dn}
A.~C. Dada, J.~Leach, G.~S. Buller, M.~J. Padgett, and E.~Andersson,
  \enquote{{Experimental high-dimensional two-photon entanglement and
  violations of generalized Bell inequalities},} Nature Physics \textbf{7},
  677--680 (2011).

\bibitem{Pires:2010gy}
H.~D.~L. Pires, H.~C.~B. Florijn, and M.~P. van Exter, \enquote{{Measurement of
  the Spiral Spectrum of Entangled Two-Photon States},} Phys. Rev. Lett.
  \textbf{104}, 020505 (2010).

\bibitem{Krenn:2013eq}
M.~Krenn, R.~Fickler, M.~Huber, R.~Lapkiewicz, W.~Plick, S.~Ramelow, and
  A.~Zeilinger, \enquote{{Entangled singularity patterns of photons in
  Ince-Gauss modes},} Phys. Rev. A \textbf{87}, 012326 (2013).

\bibitem{Collins:2002ix}
D.~Collins, N.~Gisin, N.~Linden, S.~Massar, and S.~Popescu, \enquote{{Bell
  Inequalities for Arbitrarily High-Dimensional Systems},} Phys. Rev. Lett.
  \textbf{88}, 040404 (2002).

\bibitem{Dixon:2010ev}
A.~R. Dixon, Z.~L. Yuan, J.~F. Dynes, A.~W. Sharpe, and A.~J. Shields,
  \enquote{{Continuous operation of high bit rate quantum key distribution},}
  Appl. Phys. Lett. \textbf{96}, 161102 (2010).

\bibitem{Rodrigo:2006vw}
P.~J. Rodrigo, I.~R. Perch-Nielsen, and J.~Gl{\"u}ckstad, \enquote{{High-speed
  phase modulation using the RPC method with a digital micromirror-array
  device},} Optics Express \textbf{14}, 5588--5593 (2006).

\bibitem{Mirhosseini:2013go}
M.~Mirhosseini, O.~S. Maga{\~n}a-Loaiza, C.~Chen, B.~Rodenburg, M.~Malik, and
  R.~W. Boyd, \enquote{{Rapid generation of light beams carrying orbital
  angular momentum},} Optics Express \textbf{21}, 30204--30211 (2013).

\bibitem{DAriano:2003tw}
G.~M. D'Ariano, M.~G.~A. Paris, and M.~F. Sacchi, \enquote{{Quantum
  tomography},} Advances in Imaging and Electron Physics \textbf{128}, 205--308
  (2003).

\bibitem{Thew:2002tc}
R.~T. Thew, K.~Nemoto, A.~G. White, and W.~J. Munro, \enquote{{Qudit
  quantum-state tomography},} Phys. Rev. A \textbf{66}, 012303 (2002).

\bibitem{Agnew:2011js}
M.~Agnew, J.~Leach, M.~McLaren, F.~S. Roux, and R.~W. Boyd,
  \enquote{{Tomography of the quantum state of photons entangled in high
  dimensions},} Phys. Rev. A \textbf{84}, 062101 (2011).

\bibitem{Malik:2014bf}
M.~Malik, M.~Mirhosseini, M.~P.~J. Lavery, J.~Leach, M.~J. Padgett, and R.~W.
  Boyd, \enquote{{Direct measurement of a 27-dimensional
  orbital-angular-momentum state vector},} Nat. Commun. \textbf{5:3115} (2014).

\bibitem{Barnett:1990do}
S.~M. Barnett and D.~T. Pegg, \enquote{{Quantum theory of rotation angles},}
  Phys. Rev. A \textbf{41}, 3427--3435 (1990).

\bibitem{Banaszek:1999ho}
K.~Banaszek, G.~D{\textquoteright}Ariano, M.~Paris, and M.~Sacchi,
  \enquote{{Maximum-likelihood estimation of the density matrix},} Phys. Rev. A
  \textbf{61}, 010304 (1999).

\bibitem{MolinaTerriza:2007ig}
G.~Molina-Terriza, J.~P. Torres, and L.~Torner, \enquote{{Twisted photons},}
  Nature Physics \textbf{3}, 305--310 (2007).

\bibitem{Yao:2011ve}
A.~Yao and M.~J. Padgett, \enquote{{Orbital angular momentum: origins, behavior
  and applications},} Adv. Opt. Photon. \textbf{3}, 161--204 (2011).

\bibitem{Fickler:2012hj}
R.~Fickler, R.~Lapkiewicz, W.~N. Plick, M.~Krenn, C.~Schaeff, S.~Ramelow, and
  A.~Zeilinger, \enquote{{Quantum Entanglement of High Angular Momenta},}
  Science \textbf{338}, 640--643 (2012).

\bibitem{Leach:2009tc}
J.~Leach, B.~Jack, J.~Romero, M.~Ritsch-Marte, R.~W. Boyd, A.~K. Jha, S.~M.
  Barnett, S.~Franke-Arnold, and M.~J. Padgett, \enquote{{Violation of a Bell
  inequality in two-dimensional orbital angular momentum state-spaces},} Optics
  Express \textbf{17}, 8287--8293 (2009).

\bibitem{Leach:2010bm}
J.~Leach, B.~Jack, J.~Romero, A.~K. Jha, A.~M. Yao, S.~Franke-Arnold, D.~G.
  Ireland, R.~W. Boyd, S.~M. Barnett, and M.~J. Padgett, \enquote{{Quantum
  Correlations in Optical Angle-Orbital Angular Momentum Variables},} Science
  \textbf{329}, 662--665 (2010).

\bibitem{Krenn:2014jy}
M.~Krenn, M.~Huber, R.~Fickler, R.~Lapkiewicz, S.~Ramelow, and A.~Zeilinger,
  \enquote{{Generation and confirmation of a (100 $\times$ 100)-dimensional
  entangled quantum system},} PNAS \textbf{111}, 6243 (2014).

\bibitem{Yao:2006vc}
E.~Yao, S.~Franke-Arnold, J.~Courtial, S.~Barnett, and M.~J. Padgett,
  \enquote{{Fourier relationship between angular position and optical orbital
  angular momentum},} Optics Express \textbf{14}, 9071--9076 (2006).

\bibitem{Aharonov:2010vx}
Y.~Aharonov, S.~Popescu, and J.~Tollaksen, \enquote{{A time-symmetric
  formulation of quantum mechanics},} Physics Today \textbf{63}, 27--33 (2010).

\bibitem{Lundeen:ta}
J.~S. Lundeen and K.~J. Resch, \enquote{{Practical measurement of joint weak
  values and their connection to the annihilation operator},} Physics Letters A
  \textbf{334}, 337--344 (2005).

\bibitem{Neumann:1955}
J.~von Neumann, \emph{{Mathematical Foundations of Quantum Mechanics}}
  (Princeton University, Princeton, NJ, 1955).

\bibitem{Davis:2000ur}
J.~Davis, D.~McNamara, D.~Cottrell, and T.~Sonehara, \enquote{{Two-dimensional
  polarization encoding with a phase-only liquid-crystal spatial light
  modulator},} Applied Optics \textbf{39}, 1549--1554 (2000).

\bibitem{Dirac:1945wx}
P.~Dirac, \enquote{{On the analogy between classical and quantum mechanics},}
  Rev. Mod. Phys. \textbf{17}, 195--199 (1945).

\bibitem{Chaturvedi:2006kz}
S.~Chaturvedi, E.~Ercolessi, G.~Marmo, G.~Morandi, N.~Mukunda, and R.~Simon,
  \enquote{{Wigner{\textendash}Weyl correspondence in quantum mechanics for
  continuous and discrete systems{\textemdash}a Dirac-inspired view},} J. Phys.
  A-Math. Gen. \textbf{39}, 1405--1423 (2006).

\end{thebibliography}
\bibliographystyle{osajnl}

\end{document}